\documentclass[12pt]{article}

\newlength{\abstractwidth}
\usepackage{amsmath, nccmath}
\usepackage{amsfonts}
\usepackage{amssymb}
\usepackage{latexsym}

\usepackage{color}
\usepackage{graphicx}
\usepackage{tikz}
\usepackage{dsfont}

\usetikzlibrary{arrows,shapes,positioning}
\usetikzlibrary{decorations.markings}
\usepackage[rightcaption]{sidecap}
\tikzstyle arrowstyle=[scale=1]
\tikzstyle directed=[postaction={decorate,decoration={markings,
    mark=at position .65 with {\arrow[arrowstyle]{stealth}}}}]
\tikzstyle reverse directed=[postaction={decorate,decoration={markings,
    mark=at position .65 with {\arrowreversed[arrowstyle]{stealth};}}}]
\usetikzlibrary{positioning}

\usepackage{hyperref}
\definecolor{darkred}{rgb}{0.8,0.1,0.1}
\hypersetup{colorlinks=true, linkcolor=darkred, citecolor=blue, linktoc=page}

\renewcommand{\thefootnote}{\fnsymbol{footnote}}
\renewcommand{\thanks}[1]{\footnote{#1}}
\newcommand{\starttext}{
\setcounter{footnote}{0}
\setcounter{section}{0}
\renewcommand{\thefootnote}{\arabic{footnote}}}

\newcommand{\bea}{\begin{eqnarray}}
\newcommand{\eea}{\end{eqnarray}}
\newcommand{\be}{\begin{eqnarray}}
\newcommand{\ee}{\end{eqnarray}}
\newcommand{\bma}{\begin{matrix}}
\newcommand{\ema}{\cr\end{matrix}}
\newcommand{\<}{\langle}
\renewcommand{\>}{\rangle}

\def\mthree{{\cal L}}

\def\cA{{\cal A}}
\def\cB{{\cal B}}
\def\cC{{\cal C}}
\def\cD{{\cal D}}
\def\cE{{\cal E}}
\def\cF{{\cal F}}
\def\cG{{\cal G}}

\def\cI{{\cal I}}
\def\cJ{{\cal J}}
\def\cK{{\cal K}}
\def\cL{{\cal L}}
\def\cM{{\cal M}}
\def\cN{{\cal N}}
\def\cO{{\cal O}}
\def\cP{{\cal P}}

\def\cR{{\cal R}}

\def\cV{{\cal V}}
\def\cW{{\cal W}}

\def\cZ{{\cal Z}}

\def\mA{\mathfrak{A}}
\def\mB{\mathfrak{B}}

\def\mJ{\mathfrak{J}}

\def\mT{\mathfrak{T}}

\def\mt{\mathfrak{t}}

\def\mw{\mathfrak{w}}

\def\ZZ{{\mathbb Z}}
\def\RR{{\mathbb R}}

\def\CC{{\mathbb C}}

\def\Re{{\rm Re \,}}
\def\Im{{\rm Im \,}}
\def\tr{{\rm tr}}
\def\Tr{{\rm Tr}}
\def\det{{\rm det \,}}

\def\half{{1\over 2}}
\def\thalf{{\tfrac{1}{2}}}
\def\p{\partial}

\def\trop{^{\rm tr}}
\def\sgn{{\rm sgn}}

\def\a{\alpha}
\def\b{\beta}
\def\g{\gamma}

\def\tet{\vartheta}
\def\ep{\varepsilon}
\def\om{\omega}

\def\no{\nonumber}
\def\sm{\smallskip}

\begin{document}
\starttext
\setcounter{footnote}{0}

\begin{flushright}
2020 June 24 \\
revised July 3 \\
UUITP-16/20
\end{flushright}

\bigskip

\begin{center}

{\Large \bf Two-loop superstring five-point amplitudes I }

\vskip 0.1in

{ \bf Construction via chiral splitting and pure spinors   }

\vskip 0.4in

{{  \bf Eric D'Hoker$^{(a)}$, Carlos R. Mafra${}^{(b)}$,  Boris Pioline$^{(c)}$, Oliver Schlotterer$^{(d)}$}
\footnote{E-mail: {\tt \small dhoker@physics.ucla.edu}; {\tt \small c.r.mafra@soton.ac.uk}; \\ 
\color{black}{} \hskip 0.7in
{\tt \small pioline@lpthe.jussieu.fr};
{\tt \small oliver.schlotterer@physics.uu.se}.}}

\vskip 0.1in

 ${}^{(a)}$ {\sl Mani L. Bhaumik Institute for Theoretical Physics}\\
 { \sl Department of Physics and Astronomy }\\
{\sl University of California, Los Angeles, CA 90095, USA}\\

\vskip 0.1in

${}^{(b)}$ {\sl STAG Research Centre and Mathematical Sciences,} \\ {\sl  University of Southampton, Highfield, Southampton SO17 1BJ, UK}

\vskip 0.1in

 ${}^{(c)}$ { \sl Laboratoire de Physique Th\'eorique et Hautes Energies}\\
{\sl CNRS and Sorbonne Universit\'e, UMR 7589}\\
{\sl Campus Pierre et Marie Curie, 4 Place Jussieu 75252 Paris, France}

\vskip 0.1in

 ${}^{(d)}$ { \sl Department of Physics and Astronomy,} \\ {\sl Uppsala University, 75108 Uppsala, Sweden}

\begin{abstract}
The full two-loop amplitudes for five massless states in Type~II and Heterotic superstrings are constructed in terms of convergent integrals over the genus-two moduli space of compact Riemann surfaces and integrals of Green functions and Abelian differentials on the surface. The construction combines elements from the BRST cohomology of the pure spinor formulation and from chiral splitting with the help of loop momenta and homology invariance.  The $\alpha' \to 0$ limit of the resulting superstring amplitude is shown to be in perfect agreement with the previously known amplitude computed in Type~II supergravity. Investigations of the $\alpha'$ expansion of the Type~II amplitude and
comparisons with predictions from S-duality are relegated to a first companion paper. A construction from first principles in the RNS formulation of the genus-two amplitude with five external NS states is relegated to a second companion paper.

\end{abstract}

\end{center}

\newpage

\setcounter{tocdepth}{2} 
\tableofcontents

\newpage

\baselineskip=16pt
\setcounter{equation}{0}
\setcounter{footnote}{0}

\section{Introduction}
\label{sec:1}
\setcounter{equation}{0}

The perturbative evaluation of superstring amplitudes in the Ramond-Neveu-Schwarz (RNS) formulation proceeds systematically from first principles (see for example \cite{Friedan:1985ge,DHoker:1988pdl,DHoker:2002hof,Witten:2012bh} and references therein). Space-time supersymmetry is achieved in the RNS formulation by assembling the separate contributions from the NS and R sectors and integrating over super moduli which includes a sum over spin structures. By contrast, the pure spinor formulation \cite{Berkovits:2000fe,Berkovits:2004px, Berkovits:2005bt} requires only an integral over bosonic moduli and
is manifestly supersymmetric. It provides a streamlined approach to the evaluation of  multi-particle superstring amplitudes with arbitrary external massless states (see for example \cite{Mafra:2011nv, Mafra:2018qqe}  and references therein). However, for genus three and greater, the pure spinor formulation faces the complication of a composite $b$-ghost whose presence is required to produce a suitable measure on moduli space. Various problems associated with the $b$-ghost and with the integration over pure spinor zero modes remain incompletely resolved to date.

\sm

While the explicit calculation of higher-genus amplitudes in superstring theory is of interest in its
own right, it is also mainly motivated by  the systematic study of the low energy effective interactions induced by string theory and the derivation of associated non-renormalization theorems, as well as by the exploration of the hidden structures of scattering amplitudes in quantum field theory through the  $\alpha' \rightarrow 0$ limit, such as the perturbative relations between gauge theories and supergravity. Another, more mathematical  motivation is to gain a better  understanding of the higher-genus modular forms that enter multi-loop string amplitudes.

\sm

The focus of this paper will be on genus-two amplitudes. In the RNS
formulation, amplitudes receive contributions from even and odd spin structure
sectors. The measure for the even spin structure sector was evaluated in
\cite{DHoker:2001kkt,DHoker:2001qqx,DHoker:2001foj,DHoker:2001jaf} with the
help of the canonical holomorphic projection of the genus-two  even spin
structure super moduli space onto moduli space. An alternative derivation of
the measure using algebraic geometry methods was given more recently in
\cite{Witten:2012ga,Witten:2013tpa}.  The genus-two amplitude for four
external NS bosons was evaluated for both the Type II and Heterotic strings
and is given by convergent integrals over the moduli space of genus-two
compact Riemann surfaces, and integrals over each surface of combinations of
Green functions in \cite{D'Hoker:2005jc,DHoker:2005dys}. The absolute 
normalization of the Type IIB amplitude and a comparison of its low energy
expansion with the implications from S-duality were obtained in
\cite{D'Hoker:2005ht} with further results derived in \cite{DHoker:2013fcx,DHoker:2014oxd}. 
A general formulation for the even spin structure part of the genus-two
amplitude for an arbitrary number of NS states was given using Dolbeault
cohomology in \cite{DHoker:2007csw}, but no explicit formulas for amplitudes
with more than 4 external states have been obtained in the RNS
formulation yet.

\sm

The genus-two results for four massless states in Type II were  reproduced soon after the RNS calculations
using the pure spinor formulation, and extended to obtain the
amplitudes involving external R states and thus external fermions
\cite{Berkovits:2005df}. Agreement with the results from RNS was verified in
\cite{Berkovits:2005ng}, including the precise normalization of the amplitude
\cite{Gomez:2010ad}. The pure spinor prescription was also applied to
genus-two amplitudes with five external states in \cite{Gomez:2015uha} and to
genus-three amplitudes with four external states in \cite{Gomez:2013sla}. In
both cases, finite expressions consistent with S-duality were obtained for the 
leading terms in the low energy expansion of these amplitudes. While for the genus-two amplitude
with five external states the full expression will be derived below, the
divergences in the zero-mode integrals of the bosonic ghosts pose difficulties
when attempting the same for the genus-three amplitude.

\sm

In the present paper, we shall construct the genus-two amplitudes for five
massless external states of the supergravity multiplet for Type II superstrings, and the
supergravity or the super Yang-Mills multiplet for Heterotic strings. The extension to Type I superstrings is expected to follow from our construction as well but will not be considered in any detail here. We shall follow the prescription neither of the RNS formulation nor of the pure spinor formulation. Instead we shall combine ingredients of
both formulations with properties of the corresponding maximal supergravity
amplitudes. Specifically, we shall use the vertex operator BRST cohomology
(see \cite{Mafra:2014oia} and references therein) from the pure spinor
formulation, and import the chiral splitting procedure and 
homology invariance properties of chiral amplitudes which were developed in the context of 
the RNS formulation \cite{DHoker:1988pdl,DHoker:1989cxq}. 

\sm

It will turn out that the construction via a combination of these ingredients 
produces unique amplitudes in the above theories in terms of integrals over
the moduli space of compact Riemann surfaces and, for each surface, integrals
over combinations of Green functions and meromorphic Abelian differentials.
The integrals are convergent after analytic continuation in the external
momenta, as is familiar from genus-one amplitudes \cite{DHoker:1994gnm}.

\sm

Our key result is the construction of the chiral amplitude ${\cal K}_{(5)}$ which is a function of external momenta, chiral polarization vectors and spinors, loop momenta, and a complex analytic dependence on vertex operator  points and moduli of the underlying compact Riemann surface $\Sigma$.  The integration
of the pairing of left and right chiral amplitudes over loop momenta, vertex operator points, and moduli 
gives the physical amplitude for five external states in the supergravity multiplet. For example, the Type II amplitudes take the form, 
\bea
{\cal A}_{(5)} = \int  \langle  {\cal K}_{(5)} \overline{  \tilde {\cal K}_{(5)}} \rangle_0 \, |\cI_{(5)}|^2
\label{teaser.1}
\eea
The integral encompasses moduli, vertex points, and loop momenta and includes the chiral Koba-Nielsen factor $\cI_{(5)}$,  as will be explained in detail in the sequel. Furthermore, the bracket $ \langle  \ldots \rangle_0$ denotes the prescription of the pure spinor formalism \cite{Berkovits:2000fe} to integrate over spinor zero modes, which extracts the power  of $\theta^5 \tilde \theta^5$ from the enclosed superfields. The chiral amplitude $\cK_{(5)}$ in (\ref{teaser.1}) will be  determined in a basis of holomorphic five-forms on $\Sigma^5$,
\bea
 {\cal K}_{(5)} =  \Delta(3,4) \Delta(5,1) \omega_I(2){\cal K}^I_{1,2,3|4,5} + {\rm cycl}(1,2,3,4,5)
  \label{teaser.2}
\eea
where $\Delta(i,j)$ is the bi-holomorphic combination of holomorphic one-forms $\omega_{1,2}$,
\bea
\Delta(i,j)= \omega_1(z_i)  \omega_2(z_j) - \omega_2(z_i)  \omega_1(z_j)
\eea
familiar from \cite{D'Hoker:2005jc,DHoker:2005dys}. All the dependence on the external polarization vectors and 
spinors is captured by the coefficients ${\cal K}^I_{1,2,3|4,5}$ which are scalar functions on $\Sigma^5$,
\begin{align}
{\cal K}^I_{1,2,3|4,5} &= 2\pi p_m^I T^m_{1,2,3|4,5} 
- g^I_{2,3} \, T_{23,1|4,5} - g^I_{2,1} \, T_{21,3|4,5}  - g^I_{3,1} \, T_{31,2|4,5}  \notag \\
&- g_{2,4}^I \, S_{2;4|5|1,2}  - g_{3,4}^I \, S_{3;4|5|2,1}  - g_{1,4}^I \, S_{1;4|5|2,3}  \label{teaser.3} 
 \\
&- g_{2,5}^I \, S_{2;5|4|3,1}  - g_{3,5}^I \, S_{3;5|4|2,1}  - g_{1,5}^I \, S_{1;5|4|2,1}  \notag
\end{align}
The dependence on the loop momenta $p_m^I$ is explicit in \eqref{teaser.3}, while the dependence on vertex
positions and moduli enters  through the following combinations of theta functions,
\bea
g_{i,j}^I = \frac{ \partial }{\partial \zeta_I} \ln \vartheta[\nu](\zeta|\Omega) 
\qquad \hbox{ for } \qquad \zeta_I = \int^{z_i}_{z_j} \omega_I
\eea
The choice of odd spin structure $\nu$ is immaterial as long as it is the same for all terms in (\ref{teaser.3}). 
The kinematic factors  $T^m_{1,2,3|4,5}, T_{23,1|4,5}, S_{2;4|5|1,2}$ in pure spinor superspace will be 
developed below, giving access to arbitrary combinations of external states from the massless supersymmetry 
multiplets. These kinematic factors  are independent of moduli, vertex points, and loop momenta.

\sm

Our construction of the chiral amplitude ${\cal K}_{(5)}$ in this paper
does not proceed directly from first principles, and it is therefore important to carry out 
consistency checks to confirm its validity.  A first check consists in showing
that those terms of the chiral amplitude which have singularities at
coincident vertex points agree with the OPEs derived from first principles in 
\cite{Gomez:2015uha}. A second check consists of comparing the
$\alpha'\to 0$ limit of the Type II superstring amplitudes with the
predictions from the corresponding maximal supergravity calculations. 
Both checks will be carried out in this paper and demonstrate perfect agreement.

\sm

As further checks, the investigation of the low energy expansion of the amplitude
for five external states in Type II string theory and the comparison with
predictions from S-duality, carried out in \cite{Gomez:2015uha} to lowest
order, will be extended to higher orders in a companion paper \cite{compone}. Finally, 
the genus-two amplitude for five external NS bosons will be evaluated through the RNS formalism in
another companion paper \cite{comptwo}, where its form will be compared with the amplitude obtained here.


\subsection*{Organization}

The remainder of this paper is organized as follows. In section \ref{sec:2} we
review and summarize the required key ingredients of the
non-minimal pure spinor formulation, its BRST cohomology, its zero-mode
counting, and its vertex operators, as well as the chiral splitting procedure
applied to pure spinors. Section~\ref{sec:3} briefly reviews selected aspects
of multi-loop computations in the pure spinor formalism and the derivation of
the amplitude with four external massless states. In section \ref{sec:4}, we
make use of BRST cohomology and chiral splitting to construct a chiral
amplitude with five external massless states. In section \ref{sec:4A} we shall 
recast this result in various alternative representations which make manifest 
Bose and Fermi symmetry, homology invariance,  BRST invariance, and short distance singularities.
 In section~\ref{sec:5} we
continue to use the results of chiral splitting to assemble left and right
moving chiral blocks into the full amplitudes for five external states in the
Type II and Heterotic strings.  In section~\ref{sec:6} we
check the worldline limit of our results to reproduce the loop integrand
of the two-loop five-point amplitude in supergravity.
In section \ref{sec:7} we conclude and offer a
perspective on some future directions of investigation. 

\sm

Various identities for the Clifford-Dirac algebra and pure spinors are collected in
appendix~\ref{sec:A}; basics ingredients of Riemann surfaces and their
function theory are summarized in appendix~\ref{sec:C}; a detailed derivation
of the chiral splitting procedure suitable for the pure spinor formulation is
presented in appendix~\ref{sec:D}; and the operator product expansions of the
pure spinor worldsheet fields are gathered in appendix~\ref{sec:B}.

\subsection*{Acknowledgments}

We are grateful to Piotr Tourkine for  helpful discussions on the tropical limit and to  Nicholas Geiser for useful comments on the manuscript. The research of ED is supported in part by NSF grant PHY-19-14412. BP and OS are grateful to UCLA and the Mani Bhaumik Institute for kind hospitality and creating a stimulating atmosphere during
initial stages of this work. CRM is supported by a University Research Fellowship from the Royal Society. OS is supported by the European Research Council under ERC-STG-804286 UNISCAMP.

\newpage

\section{Pure spinors and chiral splitting}
\setcounter{equation}{0}
\label{sec:2}

In this section we derive the basic building blocks for the five-point
amplitude in terms of the BRST cohomology of the pure spinor superstring and
the chiral splitting procedure.  The source of these building blocks may be
found in the non-minimal pure spinor superstring, whose formulation is suited
to two-loop calculations  in view of the presence of its $b$-ghost. Salient
features of the non-minimal pure spinor superstring may be found in
\cite{Berkovits:2005bt}. Throughout, we assume Euclidean signature 
both on the worldsheet and in target space.

\subsection{Worldsheet fields, action, and symmetries}
\label{sec:21}

The fields of the non-minimal pure spinor superstring on the worldsheet $\Sigma$  are the worldsheet scalar $x^m$ with $m=1,\cdots, 10$; the left-moving worldsheet scalars $ \theta ^\a, \lambda ^\a, \bar \lambda _\a, r_\a$ with $\alpha = 1, \cdots, 16$;  the left-moving worldsheet $(1,0)$-forms $p_\a, w_\a, \bar w^\a , s^\a$; and  their right-moving counterparts suitable either for the Type II or Heterotic strings.  Despite the notation, the fields $\lambda ^\a, \bar \lambda _\a$ and $w_\a, \bar w^\a$ are not complex conjugates of one another, but independent fields transforming under conjugate representations of the Lorentz group. Under the $SO(10)$ Lorentz group, the field $x^m$ transforms as a vector; $\theta ^\a, \lambda ^\a, \bar w^\a, s^\a$ transform as Weyl spinors in the {\bf 16} of $SO(10)$; and $p_\a, w_\a, \bar \lambda _\a, r_\a$ transform as Weyl spinors in the {\bf 16'}. The fields $\theta ^\a, p_\a$ are anti-commuting matter fields while $\lambda ^\a, \bar \lambda _\a, w_\a, \bar w^\a$ are commuting ghosts, and $ s^\a , r_\a$ are anti-commuting spinor ghosts. The pure spinor constraints on the ghost fields are,
\bea
\label{pure}
\lambda  \gamma ^m \lambda = \bar \lambda \gamma ^m \bar \lambda =\bar \lambda \gamma ^m r=0
\eea
These identities are invariant under $SO(10)$ and reduce the number of independent components of each field $\lambda ^\a, \bar \lambda _\a, r_\a$  from 16 to 11 in an $SO(10)$-invariant way.\footnote{This counting may be seen explicitly by decomposing the fields under the $U(5)$ maximal subgroup of $SO(10)$ under which the spinor representation {\bf 16} decomposes into the representations ${\bf 1} \oplus {\bf 5}^* \oplus {\bf 10}$ of $SU(5)$.  The constraints (\ref{pure}) are responsible for projecting out the representation ${\bf 5}^*$ from each field, leaving 11 independent components for each one of the fields $\lambda ^\a, \bar \lambda _\a$ and $r_\a$. Basic identities for the $16 \times 16$ Clifford-Dirac $\gamma$-matrices and pure spinor identities  are given in appendix~\ref{sec:A}.}

\sm

The action for $x^m$ and the left-moving worldsheet fields is given by,\footnote{Throughout, we shall set $\alpha'=2$ and use local complex coordinates $z, \bar z$ on $\Sigma$ with $\p=\p/\p z$, $\bar \p = \p / \p \bar z$.  The  fields $p_\a, w_\a, \bar w^\a, s^\a$ will denote the coefficients of the differential $dz$ of their corresponding $(1,0)$ form fields expressed in local coordinates. The coordinate volume form on $\Sigma$ is $d^2 \! z = { i \over 2} d z \wedge d\bar z$. When no confusion is expected to arise,  the integral of a $(1,1)$-form $v \, d^2z$ on $\Sigma$ will be denoted in shorthand by $\int _\Sigma v \, d^2\! z  \to  \int _\Sigma v$, while  the integral of a $(1,0)$ form $\omega \, dz$ along a curve $\cC$ will be denoted  $\int_\cC  \omega \, dz \to \int _\cC \omega$.}
\bea
\label{action}
I= { 1 \over 2 \pi} \int_\Sigma \left ( \half \p x^m \bar \p x_m + p_\a \bar \p \theta ^\a 
- w_\a \bar \p \lambda ^\a - \bar w^\a \bar \p \bar \lambda _\a + s^\a \bar \p r_\a \right )
\eea
The action $I$ is invariant under global Lorentz transformations of $SO(10)$. It is also invariant under global supersymmetry transformations which are generated by a constant spinor $\epsilon ^\a$,
\bea
\delta x^m = - \thalf \epsilon \gamma ^m  \theta 
\hskip 1in 
\delta \theta ^\a = \epsilon ^\a
\eea
The corresponding translation and supersymmetry currents are given by,
\bea
\label{dpi}
\Pi ^m & = & \p x^m + \thalf \theta \gamma ^m \p \theta 
\no \\
d_\a & = & p_\a - \thalf  \p x^m (\gamma _m \theta ) _\a -  \tfrac{1}{8} (\theta \gamma ^m \p \theta ) (\gamma _m \theta ) _\a
\eea
Both currents are invariant under supersymmetry.  The stress tensor is given by,
\bea
\label{stress}
T_{{\rm tot}}= - \thalf \p x^m \p x_m - p_\a \p \theta ^\a + w_\a \p \lambda ^\a 
+ \bar w ^\a \p \bar \lambda _\a - s^\a \p r_\a
\eea
 The matter fields $x^m, \theta ^\a, p_\a$ are unconstrained free fields while the ghost fields are subject to the pure spinor constraints (\ref{pure}).  It will often be convenient to use the field $d_\a$ instead of $p_\a$ by carrying out the field-dependent shift in (\ref{dpi}).  The 16 components of the spinor $d_\a$ are unconstrained. The operator product relations  are given in appendix~\ref{sec:B}.

 \subsubsection{Gauge symmetry of the ghost fields and gauge invariant composites}
 \label{sec:21}
 
 In view of the pure spinor constraints (\ref{pure}) on $\lambda ^\a, \bar \lambda _\a, r_\a$, their respective conjugates $w_\a, \bar w^\a, s^\a$ are subject to gauge transformations, 
\bea
\label{delw}
\delta w_\a & = & \Lambda _m (\gamma ^m \lambda ) _\a
\no \\
\delta \bar w^\a & = & \bar \Lambda _m (\gamma ^m \bar \lambda ) ^\a - \phi _m (\gamma ^m r)^\a
\no \\
\delta s^\a & = & \phi _m (\gamma ^m \bar \lambda)^\a
\eea
which leave the action $I$ invariant for arbitrary commuting $\Lambda _m, \bar \Lambda _m$ and  anti-commuting $ \phi_m$ functions on $\Sigma$. As a result, the number of fields $w_\a, \bar w^\a, s^\a$ modulo gauge transformations is reduced from 16 to 11 for each field.  Linear combinations  of $w_\a, \bar w^\a, s^\a$ 
(with $\lambda$ and $\bar \lambda$-valued coefficients)  that are  invariant under these gauge transformations are given by,
\begin{align}
N_{mn} & = \thalf w \gamma _{mn} \lambda &  J & = w\lambda 
\no \\
\bar N_{mn} & = \thalf (\bar w \gamma _{mn} \bar \lambda  - s \gamma _{mn} r) &  \bar J & = \bar w \bar \lambda - sr
\no \\
S_{mn} & = \thalf s \gamma _{mn} \bar \lambda & S & = s \bar \lambda  
\end{align}
The composites $N_{mn}, \bar N_{mn}$ are the $SO(10)$ currents of the ghost fields $\lambda ^\a, w_\a$,  $\bar \lambda _\a, \bar w^\a$, $s^\a, r_\a$, while $J, \bar J$ are $U(1)$ currents. The ghost number current is defined by,
\bea
J_{{\rm gh}} = w \lambda - \bar w \bar \lambda
\eea
so that $\lambda, \bar w$ have ghost number $+1$ and $w, \bar \lambda$ have ghost number $-1$ while all other fields, including the composites $\Pi ^m, d_\a$, and $ T_{{\rm tot}}$,  have zero ghost number. The partial stress tensors $ T_\lambda  = w \p \lambda$ and $T_{\bar \lambda}  = \bar w \p \bar \lambda- s\p r$ are also invariant but will not be needed here. 

\sm

 In view of the pure spinor constraints (\ref{pure}),
only 11 amongst the fields $(N_{mn},J)$ are linearly independent of one
another (with $\lambda$-valued coefficients), and similarly only 11 amongst
$(\bar N_{mn}, \bar J)$ and 11 amongst $(S_{mn}, S)$ are linearly independent
 (with $\bar \lambda$-valued coefficients).

\subsection{Chiral splitting}
\label{sec:22}

The spinor-valued fields in the non-minimal pure spinor formulation, $\theta
^\a, p_\a, \lambda ^\a, w_\a, \bar \lambda _\a, \bar w^\a$, $r_\a$, and
$s^\a$, are conformal primary fields whose correlators on a Riemann surface
$\Sigma$ of arbitrary genus $h$ are complex analytic on $\Sigma$ and on
moduli.  The vector-valued field $x^m$, however, is not a conformal primary
due to the presence of  translational zero modes. As a result the inverse of the scalar Laplacian 
on the space orthogonal to the zero mode depends on certain choices, including the volume form on $\Sigma$.
Choosing the volume form to be the canonical K\"ahler form of unit volume (with $Y^{IJ}$ denoting the entries
of the inverse of $Y=\Im \Omega$),\footnote{A summary of function theory on compact Riemann surfaces, including the definitions of meromorphic differentials, Jacobi theta-functions, and the prime form, is given in appendix \ref{sec:C}. Throughout, we shall use the Einstein convention for the summation over pairs
of repeated upper and lower indices $I,J=1,\cdots, h$, where $h$ is the genus, which we 
keep general in this section.}
\be
\label{defkappa}
\kappa(z) = \frac{i}{4} Y^{IJ} \om _I (z) \wedge \bar \om _J (z) = { i \over 2} \kappa _{z \bar z} dz \wedge d \bar z
\ee
the inverse of the scalar Laplacian on the space orthogonal to the zero mode gives the Arakelov Green function $\cG$ which satisfies, 
\be
 \partial_z \bar\partial_{\bar z} \cG(z,w|\Omega) = - \pi \delta^{(2)}(z,w)  + \pi \kappa_{z \bar z} (z) \, 
\hskip 0.6in
\int_\Sigma \cG(z,w |\Omega) \kappa(w)=0
\ee
The Arakelov Green function is globally well-defined,  symmetric in $z,w$, invariant under conformal transformations, and gives the two-point function of $x^m$ as follows $\< x^m (z) x^n(w)\>=\eta ^{mn} \cG(z,w|\Omega)$. The Arakelov Green function is related to the more familiar  ``string Green function",
\bea
\label{Green}
G(z,w|\Omega) = - \ln |E(z,w|\Omega)|^2 + 2 \pi Y^{IJ} \, \Big ( \Im \int _w ^z \om _I \Big ) \Big ( \Im \int _w ^z \om _J \Big )
\eea 
via a shift
\bea
\label{GreenA}
\cG(z,w|\Omega)  = G(z,w|\Omega) - \gamma(z|\Omega) - \gamma(w|\Omega) 
\eea
where
\be
\gamma(z|\Omega)= \int_{\Sigma} G(z,w|\Omega)\, \kappa(w) - \frac12  \int_{\Sigma\times \Sigma} 
\kappa(w)\, G(w,w' |\Omega)\, \kappa(w')
\label{gashift}
\ee
Unlike $\cG(z,w|\Omega)$, the string Green function $G(z,w|\Omega)$ depends on a choice of local coordinates, due to the fact that $E(z,w|\Omega)$ is a form of weight $(-\thalf, 0)$ in $z$ and $w$, and is not globally well-defined on $\Sigma$. However, the difference $\cG(z,w|\Omega)  - G(z,w|\Omega)$ cancels from correlators
upon imposing momentum conservation, so we may equally well use the two-point function 
$\< x^m (z) x^n(w)\>=\eta ^{mn} G(z,w|\Omega)$ in computing correlators of $x^m$. The use of the Arakelov Green function will be especially important when carrying out a low-energy expansion of the amplitudes and guarantees that individual terms are properly conformal invariant  \cite{DHoker:2017pvk,DHoker:2018mys}.

\sm

By contrast, the field $\p x^m(z)$ is a $(1,0)$ form and conformal primary field. Its correlators are meromorphic on $\Sigma$, as may be seen from the two-point function $\< \p x^m (z) \p x^n (w) \> = \eta ^{mn} \p_z \p_w
G(z,w|\Omega)=\eta ^{mn}\p_z \p_w \cG(z,w|\Omega)$ with,
\bea
\label{xx}
\p_z \p_w G(z,w|\Omega )  =  -  \p_z \p_w \ln E(z,w|\Omega) + \pi  Y^{IJ} \om _I (z) \om _J (w)
\eea
Note that neither the Green functions $G$, $\cG$ nor their derivatives $\p_z \p_w G$ are complex analytic in the moduli $\Omega$, as evident  from the presence of $Y^{IJ}$ in (\ref{xx}).

\sm

The chiral splitting procedure \cite{DHoker:1988pdl,DHoker:1989cxq,Verlinde:1987sd} introduces loop momenta to re-express conformal correlators of the $x^m$-field in terms of an integral over loop momenta whose integrand is a product of left and right chiral blocks.  Each chiral conformal block is complex analytic  in the vertex points on $\Sigma$ and in the moduli of $\Sigma$. Chiral conformal blocks have a universal monodromy behavior as the points are moved
around one another and/or moved around the homology cycles of $\Sigma$. The chiral splitting procedure is a key ingredient in the evaluation of the genus-two measure  and four-point amplitudes in the RNS formulation \cite{DHoker:2002hof, DHoker:2001kkt, D'Hoker:2005jc}.

\sm

The momentum  flowing through a simple closed cycle $\cC$ on $\Sigma$ is given by the integral along $\cC$ of the  space-time translation current  $\p x^m (z)$ and is dubbed the loop momentum through $\cC$.  On a surface of genus $h$, there are $h$ independent loop momenta, which we shall denote by $(p^I)^m$ with $I=1,\cdots, h$ (not to be confused with the spinor field $p_\a$ of (\ref{action})). The choice of their routing is not unique but may be fixed canonically to the cycles $\mA_I$ given a choice of canonical homology basis $\mA_I, \mB_I$,  
\bea
(p^I)^m = { 1 \over 2 \pi} \oint _{\mA_I}  \p x^m  \hskip 1in I=1,\cdots, h
\label{defloopmom}
\eea
The normalization is fixed to reproduce the momentum flowing through a cylinder.

\sm

The construction of the chiral blocks for the correlators of the field   $\p x^m$ and the exponential $e^{i k\cdot x}$ is formulated in terms of a set of effective rules, starting from a  generating function for $N$-point $x^m$ correlators (see appendix \ref{sec:D} for a detailed derivation), 
\bea
\label{genM}
\cJ= \int Dx \, \exp \left \{ - { 1 \over 4 \pi} \int  _\Sigma \p x \cdot \bar \p  x + \sum _{j=1}^N \Big ( i k_j \cdot x (z_j)
+ \ep _j \cdot  \p x (z_j) + \bar \eta _j \cdot \bar \p x (z_j) \Big ) \right \}
\quad
\eea
Throughout we shall assume that the incoming momenta $k_j$ and the polarization vectors $\ep_j$ and $\eta _j$ are complex-valued and satisfy $k_j^2=k _j \cdot \ep _j= k_j \cdot \eta _j=0$ for all $j=1,\cdots, N$ and that the total momentum $\sum_{j=1}^N k_j$ vanishes. We shall also assume that  the coefficients $\ep _j$ and $\eta_j$ are independent of one another so that,  at a given point $z_j$, either $\ep_j$ or $\eta_j$ or both may vanish independently.  The functional integral will be understood as a generating function for correlators which are linear in each $\ep _j \cdot \p x(z_j)$ and $\bar \eta _j \cdot \bar \p x(z_j)$ so that terms of quadratic order and higher in a given $\ep_j$ or $\eta_j$ will never be needed. 

\sm

It is shown in appendix \ref{sec:D} that $\cJ$ may be obtained as an integral over loop momenta $p_I^m$ of a pairing of chiral conformal blocks, 
\bea
\label{pairing0}
\cJ =  \delta  \Big(\sum_{j=1}^N k_j\Big) \int_{\RR^{10h}}  dp \,  \cB (z_i,  \ep_i, k_i, p^I |\Omega ) 
\,  \overline{\cB (z_i,\eta_i ,  -k_i^*, -p^I |\Omega )} 
\eea
where the chiral block is given by, 
\bea
\cB (z_i, \ep_i, k_i, p^I |\Omega ) & = &  
\cB_0(z_i,k_i,p^I | \Omega) \left \< \exp \sum _{j=1}^N 
\left \{  \ep _j \cdot \Big ( \p_z x_+  +   2 \pi  p^I \om_I \Big )   
+ i k _j \cdot  x_+ \right  \}(z_j) \right \>
\no \\
\cB_0 (z_i,k_i,p^I | \Omega) & = & Z (\Omega)  ^{-10} \exp \left \{  i \pi  \Omega _{IJ} p ^I  \cdot p ^J 
+  \sum_{j=1}^N  2 \pi  i p^I \cdot k_j  \int ^{z_j} _{z_0} \om _I   \right \}
\label{BandP}
\eea
Note that the dependence on the base point $z_0$ drops out by momentum conservation. The chiral scalar partition function $Z(\Omega) $ is holomorphic in $\Omega$. It may be evaluated using chiral bosonization \cite{Verlinde:1986kw} and is given explicitly in terms of $\tet$-functions for genus two in \cite{DHoker:2001jaf}, however its form will not be needed in this work.  The field  $x_+^m$ is an effective chiral scalar field whose Wick contraction rule is given by,
\bea
\label{wick}
\< x_+ ^m (z) \, x_+ ^n (w) \> = - \eta ^{mn} \ln E(z,w|\Omega)
\eea
Recall that the field $x_+^m$ is not a conformal primary field, a property which is reflected in the non-trivial monodromy of the above correlator as $z$ and $w$ are swapped and as they are moved around non-trivial homology cycles.

\subsubsection{Homology invariance}

The chiral field $x_+^m(z)$ and, as a result, the chiral blocks $\cB$ have non-trivial monodromy as a  point $z_i$ is taken around a homology cycle of the surface. The corresponding transformations are familiar from the chiral splitting procedure \cite{DHoker:1989cxq},
\bea
\label{Bmon}
\cB (z_i + \delta_{ij} \mA_J , \ep_i, k_i, p^I |\Omega ) & = & e^{2 \pi i p_J \cdot k_j} \cB (z_i, \ep_i, k_i, p^I |\Omega ) 
\no \\
\cB (z_i+ \delta_{ij} \mB_J , \ep_i, k_i, p^I |\Omega ) & = & \cB (z_i, \ep_i, k_i, p^I +  \delta^I_J\,  k_j  |\Omega ) 
\eea
These monodromy transformations are universal in the sense that they are the same for the chiral blocks of the bosonic string, the Type II string, and the Heterotic strings. In the RNS formulation, they hold for each spin structure separately \cite{DHoker:1989cxq}.

\sm

Alternatively, we may interpret the monodromy relations of (\ref{Bmon}) as an invariance under a suitable action of the homology group of $\Sigma$ on the chiral blocks, to which we shall refer as ``homology invariance" for short. To do so, we consider a representation $\cR$ of the homology group $H_1(\Sigma, \ZZ)$ acting on both the vertex points $z_j$ and the loop momenta $p^I$, defined by the following transformations on the chiral block $\cB$,
\bea
\label{Bmon1}
\cR(z_j , \mA_J) \, \cB (z_i, \ep_i, k_i, p^I |\Omega ) 
& =  & e^{- 2 \pi i p_J \cdot k_j} \, \cB (z_i + \delta_{ij} \mA_J , \ep_i, k_i, p^I |\Omega ) 
\no \\
\cR(z_j, \mB_J) \, \cB (z_i, \ep_i, k_i, p^I |\Omega ) 
& = & \cB (z_i+ \delta_{ij} \mB_J , \ep_i, k_i, p^I -  \delta^I_J\,  k_j |\Omega ) 
\eea
These transformations mutually commute for arbitrary pairs of $(j,J)$, in agreement with the Abelian nature of the homology group. The transformation laws of (\ref{Bmon}) are then equivalent to the invariance of $\cB$ under the action of $\cR$,
\bea
\cR(z_j,\mA_J) \cB = \cR(z_j, \mB_J) \cB=\cB
\eea
The full generating function $\cJ$ of (\ref{genM}), obtained by assembling the factors of left and right chirality is, of course, invariant under these transformations. Upon integration over loop momenta the resulting correlator is single-valued in the vertex points $z_i$ thanks to the translation invariance of the loop momentum integration measure $dp$ and its domain $\RR^{10h}$.

\subsubsection{Summary of the chiral splitting procedure}

The chiral splitting procedure may be summarized  by the following prescriptions,
\begin{enumerate}
\itemsep=-0.05in
\item Carrying out the following replacements,
\bea
\label{rules1}
e^{ik\cdot x} \to e^{i k \cdot x_+}
\hskip 1in 
\p x^m (z) \to  \p x_+ ^m (z) +  2 \pi (p^I)^m \om _I (z) 
\eea
\item Wick contracting the chiral field $x_+^m$ using  (\ref{wick});
\item Including the factor $\cB_0(z_i,k_i,p^I|\Omega)$ defined in (\ref{BandP});
\item Integrating over all loop momenta of the paired chiral blocks in  (\ref{pairing0}).
\end{enumerate}

Henceforth, we shall assume that these effective rules are used whenever the fields $\p x^m$ or $e^{i k \cdot x}$ occur. For example, to construct a chiral block involving the composite field $\Pi^m$ defined in (\ref{dpi}) we shall perform the following substitution,\footnote{Note that the field $\p x^m$ also enters in the relation between the fields $p_\a$ and $d_\a$ in (\ref{dpi}). Since throughout we will work exclusively in terms of the field $d_\a$, this occurrence of $\p x^m$ will be immaterial.}
\bea
\label{rules2}
\Pi ^m & \to & \p x_+^m  + \half \theta \gamma ^m \p \theta +  2 \pi (p^I)^m \om_I 
\eea
and then carry out the Wick contractions of the field $x_+^m$ using (\ref{wick}). To simplify notations until the evaluation of the chiral block is needed, however, we shall retain the notations $\p x^m$ and $e^{i k \cdot x}$ at intermediate stages of the evaluations. Henceforth the dependence on moduli through $\Omega$ will be understood but no longer exhibited.

\subsection{BRST transformations}

The BRST charge $Q$ of the non-minimal pure spinor formalism has ghost number 1 and  is given by \cite{Berkovits:2005bt},
\bea
Q = \oint  \big ( \lambda ^\a d_\a + \bar w^\a r_\a \big )
\eea
The operator product expansion of the worldsheet fields, given in appendix \ref{sec:B}, may be used to evaluate their BRST transformation, and we have,\footnote{Throughout, we shall use standard CFT notation and write $Qf$ instead of $[Q, f]$ or $\{ Q,f\}$ for the BRST transformation of a bosonic or fermionic  field $f$, respectively.} 
\begin{align}
\label{BRST}
Q \, x^m & =  \thalf \lambda \gamma ^m \theta         & \hskip 0.5in     Q \, \lambda ^\a &=0
\no \\
Q \, \theta ^\a & =  \lambda ^\a                                 & \hskip 0.5in      Q \, \bar \lambda _\a & = r_\a 
\no \\
Q \, d_\a & =  - ( \lambda \gamma ^m) _\a \Pi _m    & \hskip 0.5in     Q  \, r_\a & = 0
\no \\
Q \, \Pi ^m & =  \lambda \gamma ^m \p \theta          & \hskip 0.5in 
Q  N_{mn} & =  - \thalf ( d \gamma _{mn} \lambda)
\end{align}
With the help of the pure spinor constraints (\ref{pure}) it may be verified that the relation,  
\bea
Q^2=0
\eea 
is properly realized on all fields. The BRST transformations of $w_\a, \bar w^\a$, and $s^\a$ are not invariant under the gauge transformations (\ref{delw}) and will not be needed, other than in the gauge invariant combination $N_{mn}$. Throughout, the field $p_\a$ will be traded for the supersymmetry current $d_\a$, which is simply related to it by a  shift given in (\ref{dpi}). A convenient unified expression may be derived from (\ref{BRST}) for the BRST transformation of any local function $f(x,\theta)$, which depends only on $x$ and $\theta$ but not on their worldsheet derivatives, 
\bea
Q f(x, \theta ) = \lambda ^\a D_\a f (x,\theta)
\eea
where $D_\a$ is the super derivative defined by,
\bea 
\label{superD}
D_\a ={ \p \over \p \theta ^\a} + \half \gamma ^m _{\a \b} \theta ^\beta { \p \over \p x^m}
\hskip 1in 
\{ D_\a, D_\b \} = \gamma ^m _{\a \b} \p_m
\eea
where we use the standard notation $\p_m = \p / \p x^m$.

\subsection{Vertex operators}

Vertex operators for massless physical states are constructed from the plane wave solutions to the linearized 10-dimensional super-Yang-Mills and supergravity equations. The spinor part of the vertex operators is chirally split as it stands, and the chiral splitting for the $x^m$ field will be carried out in the subsequent section.  The chiral vertex operators involve chiral spinor fields and the 10-dimensional super Yang-Mills multiplet and are governed by the linearized 10-dimensional super Yang-Mills equations. The fields of the super-multiplet $(A_\a, A_m, W^\a, F_{mn})$ satisfy the following equations,\footnote{The field equations of linearized 10-dimensional  super Yang-Mills theory \cite{Witten:1985nt}  may be expressed in terms of the covariant derivatives $\cD_\a = D_\a+A_\a$ and $\cD_m = \p_m + A_m$ subject to gauge transformations $\delta A_\a  =  D_\a \mA$, $\delta A_m   =  \p_m \mA$, the Jacobi identities, and the superspace torsion constraint  $F_{\a \b} = \{ \cD_\a, \cD_\b \} - \gamma ^m _{\a \b} \cD_m=0$. The  field strengths  $F_{\a m}=[\cD_\a, \cD_m]$ and $F_{mn} = \p_m A_n - \p _n A_m$ satisfy (\ref{SYM})  with $F_{\a m} = (\gamma _m) _{\a \b} W^\b$.} 
\begin{align}
\label{SYM}
D_\a A_\b + D_\b A_\a & =  \gamma^m _{\a \b} A_m 
&
D_\a W^\b & =  \tfrac{1}{4} (\gamma ^{mn} ) _\a {}^\b F_{mn}
\no \\
D_\a A_m - \p_m A_\a & =  (\gamma _m)_{\a \b}  W^\b 
&
D_\a F_{mn} & =   (\p_m \gamma _n - \p_n \gamma _m)_{\a  \b}  W^\b
\end{align}
For later use, we record the field equation and Bianchi identity for $W^\a$, 
\bea
\label{fieldW}
\gamma ^m \p _m W^\a=0 \hskip 1in D_\a W^\a=0
\eea
The fields $A_m$, $W^\a$, and $F_{mn}$ may be expressed in terms of the field $A_\a$ which has odd grading. A plane wave solution with momentum $k$ is given  in the gauge $\theta ^\a A_\a=0$ 
by \cite{Harnad:1985bc,Grassi:2004ih,Policastro:2006vt}, 
\bea
A_\a (x,\theta) = \Big ( \thalf \ep ^m (\gamma _m \theta)_\a 
- \tfrac{1}{3} (\chi \gamma ^m \theta ) (\gamma _m \theta )_\a + \cdots \Big ) \, e^{i k\cdot x}
\label{thetexp}
\eea
where the ellipses stand for terms with higher powers of $\theta$. The parameters $\ep$ and $\chi$ are the polarization vector and spinor, respectively. For massless external states we have $k^2=0$ and $k\cdot \ep= k \cdot \gamma \chi=0$. The dependence of the SYM fields on $k, \ep, \chi$ will be suppressed  throughout. 

\sm

Vertex operators for physical massless states are built out of chiral vertex operators times their conjugates. A chiral vertex operator is a $(1,0)$ form on the worldsheet which is BRST invariant up to an exact differential. To construct such vertex operators, we begin by obtaining the BRST variations of the linearized SYM fields, 
\bea
\label{Q3}
Q \, A_\a  & = & \lambda ^\b D_\b A_\a  
\no \\
Q \, W^\a   & =  & \tfrac{1}{4} (\lambda \gamma ^{mn} )^\a F_{mn}
\no \\
Q \, A_m & =  & (\lambda \gamma _m W) + \lambda ^\b \p_m A_\b 
\no \\
Q \,  F_{mn} & = & (\lambda \gamma _n \p_m W) - (\lambda \gamma _m \p_n W)
\eea
Some immediate consequences for composites, to be of later use,  are as follows,
\bea
\label{Q4}
Q \, (\lambda \gamma ^m W)  & = &0
\no \\
Q \, (\lambda \gamma ^{mn} \gamma ^r W )  & = & - \tfrac{1}{4} (\lambda  \gamma ^{mnpqr} \lambda ) F_{pq}
\no \\
Q \, (\lambda \gamma ^{mnpqr} \lambda ) F_{mn} & = & 0
\eea 
The un-integrated vertex operator $V$ is a worldsheet $(0,0)$ form of ghost number 1 given by,
\bea 
V= \lambda ^\a A_\a (x, \theta)
\eea
It  satisfies $Q V=0$ in view of the pure spinor constraint on $\lambda$. The integrated vertex operator $U$ is a worldsheet $(1,0)$-form of ghost number 0 which is built out of the basic $(1,0)$-forms $\p \theta ^\a, \Pi^m, d_\a, N_{mn}$ times the corresponding linearized on-shell SYM field and is given by,
\bea
U = \p \theta ^\a A_\a (x, \theta) + \Pi ^m A_m(x,\theta) + d_\a W^\a(x, \theta) + \thalf N_{mn} F^{mn} (x, \theta)
\label{Uvertex}
\eea
Its BRST variation is a total derivative of the un-integrated vertex $V$,
\bea 
Q U = \p V
\eea
so that the integrals of  $U\bar U$ over a closed worldsheet and $U$ over a worldsheet boundary are BRST invariant.

\subsection{The $b$-ghost}

The RNS superstring naturally has a $(b,c)$ anti-commuting ghost system which results from gauge fixing worldsheet diffeomorphism symmetry, and a $(\beta, \gamma)$ commuting ghost system resulting from gauge fixing worldsheet local supersymmetry. The existence of an un-gauged-fixed formulation for the pure spinor superstring with a canonical $(b,c)$ ghost system is currently still under investigation \cite{Berkovits:2015yra, Berkovits:2016xnb}. The non-minimal formulation of the pure spinor  string was developed to produce a composite $b$-ghost \cite{Berkovits:2005bt}, without 
requiring a $c$-ghost companion. It is this formulation that we shall use here as a guide for the construction of the amplitude for five external states.

\sm

The key principle for the construction of the $b$-ghost is that it must be an anti-commuting Lorentz scalar, and a $(2,0)$-form on the worldsheet $\Sigma$ whose BRST transform is the chiral stress tensor $T_{{\rm tot}}$ which was given in (\ref{stress}),
\bea
\label{QbT}
Q \, b = T_{{\rm tot}}
\eea
Since $Q$ and $T_{{\rm tot}}$ have ghost number $1$ and $0$, respectively, $b$ must have ghost number~$-1$.  There is no canonical gauge-invariant field satisfying these conditions. However, there is a ghost number 0 composite spinor $G^\a $ given by, 
\bea
G^\a =\half \Pi _m (\gamma ^m d )^\a - { 1 \over 4} N_{mn} (\gamma ^{mn} \p \theta )^\a - { 1 \over 4} J_\lambda \p \theta ^\a - {1 \over 4} \p^2 \theta ^\a
\label{defgal}
\eea
whose BRST transform is proportional to the stress tensor,
\bea
Q \, G^\a  = \lambda ^\a T_{{\rm tot}}
\eea
The ghost field $\bar \lambda ^\a$ of the non-minimal pure spinor string allows one to formally solve (\ref{QbT})  for the $b$-ghost using the descent equations of BRST cohomology. The resulting $b$-ghost field is unique, up to BRST closed contributions,  and given by \cite{Berkovits:2005bt}, 
\bea
\label{bghost}
b & = & s^\a \p \bar \lambda _\a + { \bar \lambda _\a G^\a \over (\lambda \bar \lambda)} 
+ { (\bar \lambda \gamma ^{mnp } r) \over 192 (\lambda \bar \lambda )^2}  \Big ( (d \gamma _{mnp} d) 
+ 24 N_{mn} \Pi _p \Big )
\no \\ && \hskip 0.45in 
- { ( r \gamma _{mnp} r) \over 16 (\lambda \bar \lambda )^3 }  ( \bar \lambda \gamma ^m d) N^{np}
+ { (r \gamma _{mnp} r ) \over 128 (\lambda \bar \lambda)^4}  (\bar \lambda \gamma ^{pqr } r) N^{mn} N_{qr}
\eea
The solution is formal because the denominators in the holomorphic field
$(\lambda \bar \lambda )$ produce singularities.  A variety of regulators have
been proposed in \cite{Berkovits:2006vi, Aisaka:2009yp}. For the two-loop
amplitude with five external states, positive powers of $(\lambda \bar
\lambda)$ arise from the measure of the ghost fields, thereby regularizing
the singularities in the $b$-ghost (see section \ref{sec:33} below and
\cite{Berkovits:2005bt} for more details). The resulting expressions were used
to evaluate the two-loop four-point amplitude \cite{Gomez:2010ad} as well as
the leading low energy limits of the two-loop five-amplitude
\cite{Gomez:2015uha} and the three-loop four-point amplitude
\cite{Gomez:2013sla}.

\newpage

\section{Basics of genus-two amplitudes}
\setcounter{equation}{0}
\label{sec:3}

In this section, we shall review and further develop those computations in the
non-minimal pure spinor formalism on genus-two Riemann surfaces that are
needed for our construction of the genus-two chiral amplitude with five
external states. We re-iterate the strategy of our construction, as already
outlined in the Introduction: we shall combine ingredients from the BRST
cohomology of the pure spinor formulation and from the chiral splitting
procedure to conjecture the genus-two chiral amplitude for five external
states. We shall perform computations in the pure spinor formulation only to
the extent that their outcome guides us towards a compelling structure of the
amplitude, which will turn out to be unique. 

\sm

The final formula of the chiral amplitude will be derived in section
\ref{sec:4}, and different representations will be explored in section \ref{sec:4A}. 
The physical amplitudes for Type II and Heterotic strings,
obtained by assembling the contributions from the left and right moving chiral
parts and integrating over loop momenta, will be presented in section
\ref{sec:5}. Along the way, the amplitude for four external states will be
re-derived in subsection \ref{sec:34}.

\subsection{Genus-two correlators in the pure spinor formalism}

The ingredients needed to evaluate the correlators on genus-two Riemann
surfaces that arise in the non-minimal pure spinor formalism are the partition
functions, the zero mode counting, and the correlators of the non-zero mode
parts of the canonical worldsheet fields. A regulator of the ghost zero mode
integration is required to resolve indeterminacy issues in the pure spinor
formulation.  The discussion will be geared towards deriving the main target
of this work at the end of section \ref{sec:4}: the chiral genus-two amplitude
for five external massless states, formulated as an integral over pure spinor
superspace zero modes of a function of the external kinematics and the zero
modes of the spinor variables $\lambda^\alpha$ and $\theta^\alpha$. This
formulation economically contains the amplitudes with five external states
belonging to the gauge or supergravity multiplets which may be either bosons
or fermions.

\subsubsection{Partition functions}

All canonical chiral spinor fields in the non-minimal pure spinor formalism
occur in conjugate pairs of a $(1,0)$-form on $\Sigma$ and a $(0,0)$-form.
Since the central charges of the spinor fields along with that of the chiral
boson field $x_+$ add up to zero, the holomorphic anomaly cancels, and each
field contributes an effective chiral partition function. For the chiral
bosons $x_+^m$, as derived from chiral splitting, this contribution is
$Z (\Omega)^{-10}$ while for the pair of anti-commuting fields $(p_\a,
\theta ^\a)$ (or equivalently the pair $(d_\a, \theta ^\a)$) the contribution
is $Z (\Omega)^{32}$. 

\sm

The commuting pair of fields $(\lambda ^\a, w_\a)$ is subject to the pure
spinor constraint (\ref{pure}) and gauge-invariance (\ref{delw}) reducing
their effective number of spinor degrees of freedom from 16 to 11 for both
fields and producing a partition function $Z (\Omega)^{-22}$.
Therefore, in combination with the contribution $Z (\Omega)^{-10+32}$
from the matter variables, the combined partition function for the minimal
pure spinor string is 1 \cite{Berkovits:2004px}.

\sm

Finally, the pair of commuting fields $(\bar \lambda_\a , \bar w^\a)$ and
anti-commuting fields $(r_\a, s^\a)$ are  subject to the pure spinor
constraints (\ref{pure}) and gauge-invariances (\ref{delw}) reducing their
effective number of spinor degrees of freedom from 16 to 11 for each field.
Hence, the fields that are specific to the non-minimal pure spinor formalism
produce a combined partition function of~1, consistent with the interpretation
of this system as a topological field theory \cite{Berkovits:2005bt}\footnote{Due to the pure spinor
constraints, the ghost fields are actually not free fields on $\Sigma$.
However, decomposition of the $SO(10)$ spinors under the subgroup $U(5)$
allows one to change variables to a free field plus a $(\beta,\gamma)$ system
both of which may be handled with standard methods \cite{Berkovits:2000fe}.}.

\subsubsection{Zero modes of $(1,0)$-form spinor fields}

In this subsection, we shall discuss the zero modes of meromorphic $(1,0)$-form spinor fields on a compact worldsheet $\Sigma$ of genus $h$. It will be convenient to use the fields $d_\a, N_{mn}, J, \bar N_{mn}, \bar J$ instead of $p_\a, w_\a, \bar w^\a$ as discussed at the end of subsection \ref{sec:21}. These meromorphic $(1,0)$-form fields, on world-sheets of genus $h \geq 1$, have  zero modes which are linear combinations of the holomorphic $(1,0)$-forms $\om_I$ whose definition and properties are reviewed in appendix~\ref{sec:C}.  An explicit parametrization is obtained as follows, 
\bea
d_\a (z) = \hat d_\a (z) + d_\a ^I \, \om _I(z)
\hskip 1 in  
\oint _{\mA_I} \hat d_\a=0
\label{zvsnonz}
\eea
and similarly for the fields $N_{mn}, J, \bar N_{mn}, \bar J$, whose zero-mode coefficients will be denoted by $N_{mn}^I, J^I, \bar N_{mn}^I, \bar J^I$, respectively. The number of independent zero modes of these fields on a compact surface of genus $h$  is as follows, 
\bea
16 \times h \hbox{ zero modes } & \hskip 1in & d_\a
\no \\
10 \times h \hbox{ zero modes } & \hskip 1in & N_{mn}, \bar N_{mn}, S_{mn}
\label{16hd} \\
h \hbox{ zero modes } & \hskip 1in & J, \bar J, S
\no
\eea
The zero modes of $d_\a, S_{mn}, S$ are anti-commuting and those of $N_{mn}, \bar N_{mn}, J, \bar J$ commuting.

\subsubsection{Zero modes of $(0,0)$-form spinor fields and pure spinor superspace}

On a surface $\Sigma$ of arbitrary genus, the $(0,0)$-form fields $\theta^\alpha$, $\lambda^\alpha$, $\bar \lambda _\a$ and $r_\a$ have a single zero mode for each value of $\alpha$. Thus, the field $\theta ^\a$ may be decomposed as follows,
\bea
\theta ^\a (z) = \hat \theta ^\a (z) + \theta ^\a _0
\eea
where $\theta _0^\a$ is independent of $z$, and $\hat \theta^\a  (z)$ represents the non-zero mode contributions. The fields $\lambda^\alpha$, $\bar \lambda _\a$ and $r_\a$ admit analogous decompositions.  The integration over the zero modes of the fields will  guarantee that full correlators are independent of the prescription used to define $\hat \theta ^\a(z)$ from $\theta ^\a(z)$, for example by requiring that the integral of $\hat \theta ^\a(z)$ over $\Sigma$ vanish.

\sm

An ubiquitous ingredient in the pure spinor formulation is the following $\lambda$-dependent tensor with ghost number 3 (see section \ref{sec:32} for its further use), 
\bea
\label{mTdef}
\mT _{\a_1 \cdots \a_5} (\lambda) = 
( \lambda \gamma ^m )_{\a_1} ( \lambda \gamma ^n )_{\a_2} ( \lambda \gamma ^p )_{\a_3} 
(\gamma _{mnp}) _{\a_4 \a_5}
\eea
which is manifestly anti-symmetric in $\a_1, \a_2, \a_3$ as well as in $\a_4, \a_5$. Actually, $\mT$ is totally anti-symmetric in all five spinor indices as may be established by showing that the contractions of $\mT$ with $(\gamma _a)_{\a_1 \a_4}$ and $(\gamma _{abcde} )_{\a_1\a_4}$ vanish with the help of (\ref{gam2}), (\ref{gam3}), and (\ref{Fierz5}). The tensor $\mT$ projects the anti-symmetric tensor product of five spinors in the ${\bf 16}$ of $SO(10)$ onto the symmetric $\gamma$-traceless tensor product of three spinors $\lambda$ in the ${\bf 16}$ of $SO(10)$.

\sm

By spacetime supersymmetry and BRST-cohomology arguments, the zero-mode integrals of the fields $\theta ^\a$ and $\lambda ^\a$ only receive contributions from the cohomology at ghost number 3, specifically  from the combination $\mT  \theta \theta \theta \theta \theta $ \cite{Berkovits:2000fe}, or more explicitly,\footnote{Throughout, the integration over the zero mode part of the fields in the expectation value of an arbitrary operator $\cO$ will be denoted by $\< \cO\>_0$. It will be understood that the fields which enter into $\cO$ are to be evaluated on their zero-mode part only.}
\bea
\langle (\lambda \gamma^m \theta) (\lambda \gamma^n \theta) (\lambda \gamma^p \theta) 
(\theta \gamma_{mnp} \theta) \rangle_0 = 1
\label{pssbracket}
\eea
The above normalization (sometimes chosen to be 2880 in the literature)  affects the full chiral amplitude only by an overall multiplicative factor, which is not being sought after here, and may thus be chosen at will without loss of generality. The prescription (\ref{pssbracket}) annihilates BRST-exact superfields,
\bea
\< Q ( \cdots ) \>_0=0
\eea
a property which guarantees space-time gauge-invariance and supersymmetry of the expectation value of BRST-closed operators and allows us to carry out simplifications by adding $Q$-exact terms.

\sm

The goal of this paper is to derive the genus-two chiral  amplitude for five external massless states from the correlators of five BRST-closed vertex operators. More specifically, the amplitude will be presented as an integral over the zero modes of $\theta ^\a$ and $\lambda ^\a$ of a BRST-closed integrand in pure spinor superspace that contains all the external kinematic data of five arbitrary states in the supergravity multiplet \cite{Berkovits:2006ik}. BRST-exact contributions may be discarded to simplify the form of the amplitude. As we shall see in section \ref{sec:3.7}, the quest for BRST-closed integrands  will lead us to the unique construction of the genus-two five-point amplitude.

\subsubsection{The zero-mode regulator}

The above ingredients for the evaluation of higher-genus correlators in the non-minimal pure spinor formalism usually lead to an indeterminacy in the integrals over the ghost zero modes of the type 0/0. On the one hand, the singularities that arise  when $(\lambda \bar \lambda)$ vanishes in the expression (\ref{bghost}) for the $b$-ghost, or tends to $\infty$,  cause the functional integrals over bosonic ghosts to diverge. On the other hand, the fermionic zero modes would cause the functional integrals to vanish for sufficiently low genus and/or small number of external states, as is the case for instance in the two-loop five-point amplitude under investigation.

\sm
The vanishing of the fermion zero mode integrations may be resolved by the insertion of the following ``regulator" which was introduced in \cite{Berkovits:2005bt},
\bea
\label{regN}
\cN_h = \exp \Big \{ {-} (\lambda \bar \lambda) - (r \theta) + \sum_{I=1}^h (w^I \bar w^I + s^I d^I) \Big \}
\eea
where $\lambda, \bar \lambda, r, \theta$ are restricted to their zero mode contributions, as explained in footnote 8.
The argument of the exponential has been engineered to be BRST-exact, so that $\cN_h = 1 + Q(\cdots)$
does not have any effect in the cohomology as long as the functional integrals converge.\footnote{For the same reason, the usage of gauge-variant quantities in the exponential of \eqref{regN} instead of the original gauge-invariant
formulation in \cite{Berkovits:2005bt} has no effect in the amplitudes \cite{Mafra:2009wq}.} It has been argued in  \cite{Berkovits:2005bt} that for genus two no singularities arise when $(\lambda \bar \lambda) \to 0$ thanks to the $\lambda, \bar \lambda$-dependence of the measure, and the insertion of the regulator $\cN_2$ leads to convergent zero-mode integrals. Note that the summation symbol over the index $I$ has been kept explicitly because both factors in the summand have upper $I$-indices, for which no natural modular-invariant pairing exists.

\subsubsection{Wick contractions of non-zero-mode fields}

The Wick contractions for the vector field $x^m$ were already discussed in section \ref{sec:22} on
 the chiral splitting procedure.  The Wick contractions of the non zero-mode part of the 
 field $\theta^\a$ with itself vanishes, 
\bea
\label{thetacor}
\hat \theta ^\a (z) \hat \theta ^\beta (y) \sim 0
\eea
while the Wick contractions of the non-zero mode part  of the $(1,0)$-form spinor fields  generally produce meromorphic $(1,0)$ forms. For example, the Wick contractions of the fields 
$\hat p_\a(z)$, $\hat d_\alpha(z)$ and $\hat \Pi_m=\Pi_m-2 \pi p^I_m \om_I$
from  (\ref{zvsnonz}) and (\ref{rules2}) are given as follows,
\begin{align}
\hat p_\alpha(z) \, \theta^\beta(y) &\sim \partial_z \ln E(z,y) \, \delta_\alpha^{\beta} 
\notag \\
\hat d_\alpha(z) \, f \big ( x(y),\theta(y) \big ) &\sim  \partial_z \ln E(z,y) \, D_\alpha f \big  (x(y), \theta (y) \big )
\label{OPEwithE} \\
\hat \Pi_m(z) \, f \big ( x(y),\theta(y) \big ) &\sim -\partial_z \ln E(z,y)  \, \partial_m f \big (x(y), \theta (y)\big )
\notag
\end{align}
where $f(x,\theta)$ is an arbitrary function which depends on $x$ and $ \theta$, but not on the worldsheet derivatives of these fields. The meromorphic differential $\p_z \ln E(z,w)$ fails to be single-valued in its variables by itself, but the associated integrations over the  zero modes of these fields will render the full correlators, into which they are inserted, properly single-valued. This is familiar for the case of the correlators of the fields $\p x^m_+$ with $x^m_+$ thanks to momentum conservation, but also holds true for the Wick contractions of field $p_\a$ with $\theta^\a$. 

\sm

As should be expected, in the short distance limit $z \to y$, the Wick
contractions of (\ref{OPEwithE}) reproduce the OPE singularities of the
corresponding fields given in (\ref{D1}) and (\ref{D2}). While for genus zero,
the knowledge of the OPE suffices to evaluate any conformal correlator, this
is no longer true for higher genus. For the fields of the pure spinor string,
the missing information is provided by the contributions from the zero modes
of the $(1,0)$-form fields. One manifestation of this is that for genus two
and above, one has to distinguish the forms $\partial_z \ln E(z,y)
\omega_I(y)$ from $-\partial_y \ln E(z,y) \omega_I(z)$, whose short-distance
behaviors agree and which coincide for the sphere (genus zero)  and for the torus (genus one).
Fortunately, we shall not need the detailed evaluation of the full correlator
 for the genus-two five-point amplitude as  in
\cite{Gomez:2015uha}, since it will suffice to
extract all relevant information from the singularities at coinciding vertex points (see section~\ref{sec:multip}).

\subsubsection{The chiral correlator in pure spinor superspace}

The chiral amplitude for $N$ massless states at genus two is given by the correlator,
\bea
\label{amps}
\cF_{(N)}=   \left\langle \cN_2 \prod _{a=1}^3 (\mu _a, b)  \, 
\prod _{i=1}^N U_i (z_i)  \right\rangle  
\hskip 1in (\mu _a, b) = \int _\Sigma \mu _a b
\eea
provided this correlator is convergent. The Beltrami differentials are denoted by $\mu_a$ for $a=1,2,3$, and will be specified later with the help of  (\ref{Belt}). The bracket notation $\langle   \cdots  \rangle$ in (\ref{amps}) is used for the complete functional integral for the zero modes and non-zero modes of all the fields in the worldsheet action (\ref{action}). The subscript of $\langle \ldots\rangle_0$ in (\ref{pssbracket}), by contrast, refers to the zero-mode integrals for the $(0,0)$-form fields $\lambda^\alpha$ and $\theta^\alpha$.  The integrations over the positions $z_i$ and the loop momenta $p_m^I$ will be carried out after  the chiral blocks and their conjugates have been paired. 

\sm

The chiral correlator is evaluated by integrating over the chiral spinor fields and over the effective chiral 
scalar field $x_+^m$ of chiral splitting, considered at fixed loop momenta $p ^m_I$. Since each
of  the vertex operators include a plane wave factor, the correlator of the effective chiral scalar field $x_+^m$  
produces the chiral Koba-Nielsen factor $\cI_{(N)}$ given by (cf.\ (\ref{BandP})), 
\bea
\cI_{(N)} = \exp \left \{  i \pi  \Omega _{IJ} p ^I  \cdot p ^J 
+   \sum_{i=1}^N  2 \pi i p^I \cdot k_i  \int ^{z_i} _{z_0} \om _I  - \sum_{i<j}^N s_{ij} \ln E(z_i,z_j)\right \}
\label{KNN}
\eea
The dimensionless kinematic invariants $s_{ij}$ are given by,
\bea
s_{ij} = - { \alpha ' \over 4} (k_i+k_j)^2 = - k_i \cdot k_j
\eea
The second equality arises from our choice $\alpha ' =2$ and the mass-shell condition $k_i^2=k_j^2=0$. 

\sm

Since the Koba-Nielsen factor (\ref{KNN}) is an ubiquitous constituent of the chiral amplitude (\ref{amps}), the main goal of this work will be to evaluate the remaining factor $ {\cal K}_{(N)}$, 
\bea
\cF_{(N)} = \cI_{(N)}  \, \langle {\cal K}_{(N)} \rangle_0
\label{FisIK}
\eea
In order to obtain an amplitude representation in pure spinor superspace and keep any combination of external bosons and fermions accessible, the zero-mode integral (\ref{pssbracket}) is left to be performed. The desired superspace
expression ${\cal K}_{(N)}$ will be referred to as a chiral correlator and encodes the  dependence on the polarization vectors and spinors of bosons and fermions, respectively, in a supersymmetric manner.
Since the factor $ \cI_{(N)}$ already transforms according to \eqref{Bmon1} under homology shifts,
the reduced amplitude ${\cal K}_{(N)}$ must be strictly invariant under these shifts, without any phase factor.

\sm

In fact, chiral correlators $\cK_{(N)}$ fall into equivalence classes in two
respects: First, $Q$-exact terms do not contribute within the bracket $\langle \ldots \rangle_0$,
and second, total derivatives $\partial_{z_i} ( \cI_{(N)}  {\cal K}_{(N)} )$ integrate to zero 
after assembling  the overall amplitudes. Hence, it suffices to construct a particularly convenient
representative of ${\cal K}_{(N)}$ as we will do in the two-loop five-point case.

\subsection{Zero mode counting}
\label{sec:32}

The large number of zero modes of the spinor fields greatly simplifies the calculations and makes the evaluation of the correlator (\ref{amps})  with a small number $N$ of external states possible. We begin by observing that the vertex operators $U_i(z_i)$ do not involve the fields $\bar \lambda_\a, \bar w^\a, s^\a, r_\a$.  Since the $b$-ghost is also independent of the field $\bar w^\a$ the zero modes of $\bar w^\a$ must be paired with those of $w_\a$ via the regulator $\cN_2$ of (\ref{regN}). Equivalently, the zero modes of $\bar N_{mn}$ and $ J_{\bar \lambda}$ must be paired with the zero modes of $N_{mn}$ and $J_\lambda$. This leaves no room for zero modes of the fields $N_{mn}$ to occur either in the vertex operators or in the $b$-ghost insertions.

\sm

Next, we concentrate on the zero modes of the fields $s^\a$ and $d_\a$, which add up to 22 and 32 zero modes, respectively.
The vertex operators $U_i$ do not involve the field $s^\a$ and the $b$-ghost involves $s^\a$ only through its first term in (\ref{bghost}).  Let us denote by $\sigma$ the number of zero modes of the field $s^\a$ absorbed by the 3 $b$-ghosts. Each $b$-ghost may absorb at most 1  zero mode of~$s^\a$, so that $0 \leq \sigma \leq 3$. The regulator $\cN_2$ will absorb exactly as many $s^\a$ zero modes as it absorbs $d_\a$ zero modes. Therefore, the number of $d_\a$ zero modes absorbed by the integration over the $s^\a$ zero modes, the regulator, and the $s^\a$-dependent part of the $b$-ghosts equals $22 - \sigma$.

\sm

Further $d_\a$ zero modes may be absorbed by the remaining terms in the $b$-ghost, but this number is bounded from above by $6-2\sigma$. Tallying all contributions, we conclude that the maximal  number of $d_\a$ zero modes absorbed by the measure and the $b$-ghosts is $22-\sigma + 6 - 2 \sigma = 28-3\sigma$, leaving at least $4+3 \sigma$ zero modes to be absorbed by the vertex operators. Since each vertex operator is at most linear in $d_\a$, any amplitude whose number of  external states is 6 or fewer must have $\sigma =0$, leaving at least 4 zero modes of the $d_\a$ field to be absorbed by the vertex operators $U_i$. For amplitudes with 4 or 5 external massless states of interest in this paper, we thus have $\sigma =0$, and the integration over the zero modes of  $s^\a$ produces the following measure for the integration over the zero modes of the field 
$d_\a(z)=\hat d_\a(z) + d^I_\a\, \omega_I(z)$,
\bea
\prod _{I=1}^2 \int [d \, d^I] (\ep \cdot \mT \cdot d^I) 
\eea
Here the combination $(\ep \cdot \mT \cdot d^I)$ for each $I$  is given by,
\bea
(\ep \cdot \mT \cdot d^I) & = & \ep ^{\a_1 \cdots \a_{16}} \mT _{\a_1 \cdots \a_5} d^I _{\a_6}  \cdots d_{\a_{16}}^I
\eea 
where the $\lambda$-dependent tensor $\mT$ was introduced in (\ref{mTdef}), and $[d \, d^I]$ stands for the integration measure for the zero modes $d^I$.
Since $(\ep \cdot \mT \cdot d^I)$ involves 11 zero modes for each value of the
index $I$, a non-vanishing integral requires a further integrand with five
$d^I$ factors and, for a given value of $I$, we have,
\bea
\label{dT}
\int [d \, d^I] (\ep \cdot \mT \cdot d^I)  \, d^I_{\a_1} d^I _{\a_2} d^I _{\a_3} d^I _{\a_4} d^I_{\a_5} 
& = & c  \, \mT_{\a_1 \a_2 \a_3 \a_4 \a_5}
\no \\
\int [d \, d^I] (\ep \cdot \mT \cdot d^I)  \, d^I_{\a_1} d^I _{\a_2} d^I _{\a_3} (d^I \gamma ^{mnp}  d^I)
& = & 96 c  (\lambda \gamma ^{[m} )_{\a_1} (\lambda \gamma ^{n} )_{\a_2} (\lambda \gamma ^{p]} )_{\a_3} 
\eea
where on the right side of the second equation the indices $mnp$ are anti-symmetrized.  The normalization $c$ can be found in \cite{Gomez:2013sla} but is of no concern to us here, as the absolute normalization of the amplitude may be fixed by other methods such as unitarity.

\subsection{Zero modes absorbed by the $b$-ghosts}
\label{sec:33}

The non-vanishing of the genus-two  amplitude for $N$ massless states given in (\ref{amps}) requires that all the 32 zero modes $d^I_\a$ of the field $d_\a(z)$ be absorbed by a conspiracy of the $b$-ghost and the vertex operators.  As shown in the previous subsection, for $N \leq 5$, the $s \p \bar \lambda$ term of the $b$-ghost does not contribute and  the vertex operators can absorb at most 5 $d$-zero modes. As a result, the $b$-ghosts must contribute either 5 or 6 $d$-zero modes, which can arise only from the terms bilinear in $d$ or the term linear in $d_\a$ in the composite spinor $G^\a$ defined by (\ref{defgal}). (Note that the term linear in $d$ and linear in the field $N_{mn}$ in (\ref{bghost}) involves a zero mode of the field $N_{mn}$, but this cannot contribute as argued in the preceding subsection). In summary, the effective ghost field for $N\leq 5$ takes the form, 
\bea
\label{beff}
b =   { (\bar \lambda  \gamma ^m d ) \over 2 (\lambda \bar \lambda)} \Pi _m 
+ { (\bar \lambda \gamma ^{mnp } r) \over 192 (\lambda \bar \lambda )^2}  ( d \gamma _{mnp} d )
+ \cdots
\eea
where the ellipses stand for terms that do not contribute for $N\leq 5$. 

\sm

Parametrizing the insertion points of the $b$-ghosts by the variables $v_a$ for $a=1,2,3$, we use the fact that only the zero modes of the fields $\lambda, \bar \lambda, r, d$ contribute to the $b$-ghost insertions to render the $v_a$ dependence of the $b$-ghost explicit, 
\bea
b(v_a) =  \sum _I { (\bar \lambda  \gamma ^m d^I ) \over 2 (\lambda \bar \lambda)} \, \om_I (v_a)  \Pi _m (v_a)  
+ \sum_{I,J} { (\bar \lambda r d^I d^J)  \over 192 (\lambda \bar \lambda )^2} \, \om_I(v_a) \om _J (v_a)  + \cdots
\eea
where we have introduced the following convenient shorthand, 
\bea
(\bar \lambda r d^I d^J) = (\bar \lambda \gamma ^{mnp} r) (d ^I \gamma _{mnp} d^J)
\eea
We shall choose a system of local complex coordinates, $\tau_a$ with $a=1,2,3$, on moduli space and associated Beltrami differentials $\mu _a$ so that,
\bea
\label{Belt}
{ \p \Omega _{IJ} \over \p \tau_a} = \int _\Sigma \mu_a  \om_I \om_J
\eea
The chiral volume form on moduli space is given by,
\bea
d^3\Omega = d\Omega _{11} \wedge d\Omega _{12} \wedge d \Omega _{22} = \sum _{a,b,c} { \p \Omega _{11} \over \p \tau_a} 
{ \p \Omega _{12} \over \p \tau_b}  { \p \Omega _{22} \over \p \tau_c} \, d\tau_a \wedge d \tau _b \wedge d \tau_c
\eea
Non-vanishing contributions from the $b$-ghost insertions therefore require specific arrangements of the $d$-zero modes. Contributions from the $b$-ghosts with 6 and 5 $d$-zero modes, respectively, are given by the arrangements,
\bea
\label{pattern}
\hbox{ 6 zero modes } & \hskip 0.3in& (\bar \lambda r d^1d^1) (\bar \lambda r d^1d^2) (\bar \lambda r d^2d^2) 
\no \\
\hbox{ 5 zero modes } & &
(\bar \lambda r d^1d^1) \Big ( 2 (\bar \lambda \gamma ^m d^2)    (\bar \lambda r d^1 d^2) 
- (\bar \lambda \gamma^m d^1)  (\bar \lambda r d^2d^2)  \Big ) 
\no \\
& &
(\bar \lambda r d^2 d^2)  \Big ( 2 (\bar \lambda \gamma ^m d^1)  (\bar \lambda r d^1d^2)  
- (\bar \lambda \gamma^m d^2)  (\bar \lambda r d^1d^1)   \Big )
\eea
The contribution for 6 zero modes directly produces the measure on moduli space, as the coefficient of this term is a holomorphic quadratic  differential in each insertion point of the $b$-ghost. The contribution with 5 zero modes is contracted with the $(1,0)$-form field $\Pi_m(v_a)$ and, in view of the results of the chiral splitting procedure (\ref{rules2}), receives two different types of contributions. The term linear in loop momentum $p_m^I$ provides a holomorphic $(1,0)$ form, so that its contribution directly generates the measure on moduli space. The other two terms of $\Pi_m$ exhibited in (\ref{rules2}) are generally meromorphic rather than holomorphic; it is unclear at present how to evaluate their contribution directly, but we shall infer it by imposing various consistency conditions.

\subsection{The chiral amplitude for four external states}
\label{sec:34}

For four external states, the above counting shows that each $b$-ghost must
contribute exactly 2 $d$-zero modes, resulting in the pattern of the first
line of (\ref{pattern}), and each vertex must contribute exactly 1 $d$-zero
mode. Omitting the overall $\lambda \bar \lambda$-dependent normalization, the
structure of the remaining integration is as follows, 
\bea
\prod _{I=1}^2 \int [dd^I] ( \ep \cdot \mT \cdot d^I) (\bar \lambda r d^1d^1) (\bar \lambda r d^1d^2) (\bar \lambda r d^2d^2)
\prod_{i=1}^4 (d W_i) 
\eea
where only the zero modes of the field $d$ contribute in its pairing against the SYM fields $W_i$, 
\bea
(dW_i) \to \sum _{I=1}^2  (d^I W_i  )(z_i) \, \om_I(z_i)
\eea
By construction, the amplitude is Bose symmetric in the indices labeling the external states.

\sm

All dependence on the $d$-zero modes has now been made explicit, and its
integral may be carried out using (\ref{dT}). The contributions vanish unless
two of the four factors $(dW_i)$ carry the zero mode $d^1$ while the other two
carry the zero mode $d^2$. To evaluate these contributions we shall single out
one specific assignment and then sum over all permutations. Carrying out the
integral over $d$-zero modes, we find \cite{Gomez:2010ad}, 
\bea
&&
\prod _{I=1}^2 \int [dd^I] ( \ep \cdot \mT \cdot d^I) 
(\bar \lambda r d^1d^1) (\bar \lambda r d^1d^2) (\bar \lambda r d^2d^2)
(d^1 W_1) (d^1W_2) (d^2W_3) (d^2 W_4) 
\no \\ && \hskip 0.3in = (\lambda \gamma _{abcpt} \lambda)
(\bar \lambda \gamma ^{mnp} r) (\bar \lambda \gamma ^{qst} r) (\bar \lambda \gamma ^{abc} r) 
(\lambda \gamma _{m} W_1) (\lambda \gamma _{n} W_2) (\lambda \gamma _{q} W_3) (\lambda \gamma _{s} W_4) 
\quad
\label{fourds}
\eea
Carrying out the integration over the zero mode of the field $r$ converts each $r$ into a super derivative acting on the vertex operators, and we obtain,
\bea
(\lambda \gamma _{abcpt} \lambda)
(\bar \lambda \gamma ^{mnp} D) (\bar \lambda \gamma ^{qst} D) (\bar \lambda \gamma ^{abc} D) 
(\lambda \gamma _{m} W_1) (\lambda \gamma _{n} W_2) (\lambda \gamma _{q} W_3) (\lambda \gamma _{s} W_4) 
\eea
Given the choice of the zero mode assignments made here, this expression is
manifestly invariant under the permutations $1\leftrightarrow 2$ and
$3\leftrightarrow 4$ as well as under $(1,2) \leftrightarrow (3,4)$. 

\sm

Applying a single super derivative to a field $W_i$ produces the field
strength $F_i$, while applying more than one super derivative to the same field $W_i$
introduces bosonic derivatives $k_i W_i$ and $k_i F_i$.  Still, the latter
contributions are BRST equivalent to the terms of schematic form $W FFF$ from
applying each super derivative to a single one of the $W$ fields. See appendix
A of \cite{Gomez:2010ad} for further details. More specifically, carrying out
the integration over $\bar \lambda$ produces a sum of four distinct terms, 
\bea
\label{mT}
T _{1,2|3,4} = { 1 \over 4} \left ( \mt_{1,2|3;4} + \mt_{1,2|4;3} + \mt_{3,4|1;2}+ \mt_{3,4|2;1} \right )
\eea
where each term is given by,
\bea
\label{QM1}
\mt _{1,2|3;4} =  (\lambda \gamma _{mnpqr} \lambda) F_1^{mn} F_2^{pq} F_3^{rs} (\lambda \gamma _s W_4)
\eea
The manifest symmetry properties are  $\mt_{1,2|3;4} = \mt_{2,1|3;4}$ and   $T_{1,2|3,4}= T_{2,1|3,4}=T_{3,4|1,2}$ while, as a consequence of  (\ref{Fierz6}), we also have the following cyclic symmetries,
\bea
\label{symmT}
\mt_{1,2|3;4} + \mt_{2,3|1;4}+\mt_{3,1|2;4} & = & 0
\no \\
T_{1,2|3,4}+T_{1,3|4,2}+T_{1,4|2,3} & = & 0
\eea
To verify BRST closure of $\mt$, we use the results of (\ref{Q4}) that $(\lambda \gamma _s W_4)$ and $(\lambda _{mnpqr} \lambda) F_1^{mn} F_2^{pq}$ are BRST closed, so that it remains only to apply $Q$  to $F_3$ which gives,
\bea
\label{QM2}
Q \, \mt_{1,2|3;4} =
 (\lambda \gamma _{mnpqr} \lambda) F_1^{mn} F_2^{pq} 
 \Big ( (\lambda \gamma ^s \p^r W_3) - (\lambda \gamma ^r \p^s W_3) 
 \Big ) (\lambda \gamma_s W_4)=0
\eea
The contribution from the first and second terms in the parentheses vanishes in view of (\ref{Fierz2}) for pure spinors and (\ref{Fierz5}), respectively. As a result, $\mt_{1,2|3;4}$ and $T _{1,2|3,4}$ are BRST closed. 

\sm

The worldsheet dependence of the amplitude for four external states involves
the chiral Koba-Nielsen factor (\ref{KNN}), multiplied by a combination of 
holomorphic $(1,0)$-forms. We define the bi-holomorphic $(1,0)$-form,
\bea
\label{Del}
\Delta (z_1,z_2) = \om_1(z_1) \om_2(z_2) - \om _2(z_1) \om_1(z_2)
\eea
Recall that, following  our notations and conventions spelled out in footnote 2,  $\om_I(z)$ is the coefficient function of the $(1,0)$-form $\om_I(z) dz$ in local complex coordinates, and $\Delta(z_1,z_2)$ is similarly the coefficient function of the differential $\Delta (z_1,z_2) dz_1 \wedge dz_2$. With these conventions, $\Delta(z_1,z_2)$ is manifestly antisymmetric in $z_1,z_2$, and satisfies the following cyclic permutation sum identities,\footnote{Henceforth, when no
confusion is expected to arise, we shall denote the points $z_i$ as arguments of
functions and forms, simply  by their label $i$, and the derivative with respect to $z_i$ by
$\p_i$, so that for example $\om_I(i) = \om_I(z_i)$, $\Delta (i,j) = \Delta
(z_i,z_j)$, and $\p_i \ln E(i,j) = \p_{z_i} \ln E(z_i,z_j)$.}
\bea
\label{Delsym}
\om_I(1) \Delta (2,3) + \om_I(2) \Delta (3,1) + \om_I(3) \Delta (1,2) & = & 0
\no \\
\Delta(1,2)\Delta(3,4) + \Delta(1,3)\Delta(4,2) + \Delta(1,4)\Delta(2,3) & = & 0
\eea
The chiral amplitude is given by \cite{Berkovits:2005df}, 
\bea
\cK_{(4)} = T_{1,2|3,4} \, \Delta (1,3) \Delta (2,4) + T_{1,3|2,4} \, \Delta (1,2) \Delta (3,4)
\label{4ptcorel}
\eea
Symmetries under the permutations $(2 \leftrightarrow 3)$ and $(1
\leftrightarrow 4)$ are manifest from the above expression, while symmetry
under the permutation $(1 \leftrightarrow 2)$ may be established using both
the symmetries of $T$ in (\ref{symmT}) and of $\Delta$ in (\ref{Delsym}).
After performing the zero-mode integral (\ref{pssbracket}) for $\lambda$ and
$\theta$, the bosonic components of $\langle {\cal K}_{(4)} \rangle_0$ were
shown in \cite{Berkovits:2005ng} to reproduce the result of the RNS
computation \cite{D'Hoker:2005jc}. A proof of this equivalence using
pure spinor superspace cohomology techniques can be found in \cite{Mafra:2008ar}.

\newpage

\section{Genus-two amplitudes for five massless states}
\setcounter{equation}{0}
\label{sec:4}

In this section, we shall obtain the main result of this paper by carrying out
the construction of the genus-two chiral amplitude for five massless states.
To do so, we use chiral splitting, zero mode counting and BRST cohomology of
the pure spinor formulation.

\subsection{Structure of the chiral amplitude for five external states}

The starting point is the genus-two chiral amplitude for five external massless states, given by the correlator of (\ref{amps}) and (\ref{FisIK}) for the case $N=5$, 
\bea
\cF_{(5)} = \cI_{(5)}  \, \langle {\cal K}_{(5)} \rangle_0  
= \left \< \cN_2 \, \prod _{a=1}^3 (b, \mu_a) \, \prod _{i=1}^5 U_i (z_i)  \right \> 
\eea
The vertex operators $U_i$ are given by, 
\bea
\label{Ui}
U_i = \p \theta ^\a A_{i \a} (x, \theta) + \Pi _m A^m_i  (x, \theta) + d_\a W^\a_i  (x,\theta) +\half N_{mn} F^{mn}_i (x,\theta)
\eea
where each superfield multiplet $(A_{i \alpha}, A_i^m, W^\a_i, F_i ^{mn})$
encodes the polarization vector and spinor of the state $i$, as made explicit
in (\ref{thetexp}). Following the pattern for the distribution of $d$-zero
modes for five external states of (\ref{pattern}) derived in subsection
\ref{sec:33}, the $b$-ghosts can absorb either five or six $d$-zero modes,
leaving the vertex operators to absorb either five or four $d$-zero modes,
respectively. We shall now discuss each part in turn.

\subsubsection{Four $d$-zero modes and one loop momentum from vertex operators}

The contribution from the $b$-ghost that contains six $d$-zero modes is of the form,
\bea
\prod _{a=1}^3 (b, \mu_a) \to (\bar \lambda r d^1 d^1) (\bar \lambda r d^1 d^2) (\bar \lambda r d^2 d^2)
\eea
so that  the product of five vertex operators needs to supply four $d$ zero modes. The corresponding contribution to $\cK_{(5)}$ is given by,
\bea
\left \< ( \bar \lambda \gamma D)^3 \Big ( U_1 (d^1 W_2) (d^1 W_3)(d^2 W_4)(d^2 W_5) \Big ) \right \> + \hbox{ 14 permutations}
\label{wickterms}
\eea
where we recall that $D$ stands for the super derivative in (\ref{superD}) and
we have carried out the usual integration over $r$ which leads to $(\bar
\lambda \gamma r) \to (\bar \lambda \gamma D)$. The permutations consist of
all 120 permutations modulo those which swap $2\leftrightarrow 3$ as well as
those which swap $4\leftrightarrow 5$ and finally those which swap the pair
$(2,3) \leftrightarrow (4,5)$, in view of the symmetries of the distribution
of $d$ zero modes.

\sm

We start by considering the contributions to (\ref{wickterms}) that are linear in loop momentum:
Decomposing the operator $\Pi_m$ in $U_1$ according to the rules of chiral splitting in (\ref{rules2}),
we find a loop-momentum dependent term
\bea
2 \pi p^I_m \om_I(z_1)
\left \< ( \bar \lambda \gamma D)^3 \Big (A_1^m (d^1 W_2) (d^1 W_3)(d^2 W_4)(d^2 W_5) \Big ) \right \> + \hbox{ 14 permutations}
\label{linp}
\eea
and leave the leftover contributions $ \p x_+^m(z_1)  + \half (\theta \gamma ^m \p \theta )(z_1)$
from (\ref{rules2}) for the next section.

\sm

Applying the three super derivatives $D$ in (\ref{linp}) produces two types of
terms. Applying all three $D$ to $W_i$ vertex operators produces terms of the
form $A_1^m$ times the building block of the four-point amplitude
$T_{1,2|3,4}$ plus permutations thereof.  However, in addition to these
contributions, which are schematically of the form $A FFFW$, terms involving
$D A_1$ and terms in which several $D$ act on the same $W_i$ are also
produced. At four points, different partitions of the super derivatives to the
superfields $WWWW$ turn out to be BRST equivalent \cite{Gomez:2010ad}. We
expect that also at five points, the chiral correlator admits a cohomology
representative where the contributions of (\ref{linp}) are captured by
permutations of $A_1^mT_{2,3|4,5}$. They will produce a contribution
 to the ``vector block", as we will see in section \ref{sec:42}.
An explicit evaluation of (\ref{linp}) may be found in section 5 of \cite{Gomez:2015uha}.

\subsubsection{Four $d$-zero modes and one Wick contraction from vertex operators}

It remains to carry out the Wick contractions of $U_1$ with the fields $W_i$.
Using the vanishing of the Wick contractions of the non-zero modes of $\theta
^\a$ given in (\ref{thetacor}), we see that the contraction of the term
proportional to $\p \theta ^\a$ on the right side of $U_1$ in (\ref{Ui}) with
the remaining $W_i$ operators vanishes identically, so that this term in $U_1$
may be omitted. The contribution of the zero mode of $N_{mn} $ in $U_1$
similarly cancels as a factor of $\bar N_{mn}$ would be needed to give a non-zero
contribution. The Wick contractions of the non-zero mode of $N_{mn}$ with the
other fields similarly cancel. The remaining contribution is thus given by \cite{Gomez:2015uha},
\bea
\label{toAFFFW}
\left \< ( \bar \lambda \gamma D)^3 \Big (\hat \Pi_m  A_1^m(x, \theta) + \hat d_\a W_1^\a (x, \theta)  \Big )(z_1)
(d^1 W_2) (d^1 W_3)(d^2 W_4)(d^2 W_5) \right \>
\eea
where $\hat \Pi^m(z_1) = \p x_+^m(z_1)  + \half (\theta \gamma ^m \p \theta )(z_1)$ is obtained by
removing the loop momentum from the chiral-splitting prescription in (\ref{rules2}). Wick contractions
of $\hat \Pi^m$ give rise to contributions linear in external momenta which arise from four vertex operators of the form $(dW_i)$, two of which carry a $d^1$ zero mode with the other two carrying a $d^2$ zero mode.
\bea
\hat \Pi^m(z_1) W_i^\b (z_i)
\sim - i  \p_{z_1} \ln E(z_1,z_i) k_i^m W^\b _i
\label{toscal1}
\eea
Finally, the Wick contractions of $\hat d_\a (z_1)$ with $W_i^\beta $ for $i=2,3,4,5$ is given by the last formula of (\ref{OPEwithE}), and in this case simplifies as follows,
\bea
 \hat d_\a(z_1) W_i^\b (z_i)
\sim \p_{z_1} \ln E(z_1,z_i) D_\a W^\b _i = { 1 \over 4} (\gamma _{mn} ) _\a {}^\b F^{mn} _i \p_{z_1} \ln E(z_1,z_i)
\label{toscal2}
\eea
The two contributions (\ref{toscal1}) and (\ref{toscal2}) will produce terms in the
``scalar block", as we will see in section \ref{sec:2part}.

\subsubsection{Contributions with five $d$-zero modes from vertex operators}

The contribution from the $b$-ghost that contains five $d$-zero modes is of the form, 
\bea
\prod _{a=1}^3 (b, \mu_a) \to (\bar \lambda r d^2 d^2)  \Pi_m \Big ( 2 (\bar \lambda \gamma ^m d^1)  (\bar \lambda r d^1d^2)  
- (\bar \lambda \gamma^m d^2)  (\bar \lambda r d^1d^1)   \Big )
\eea
plus the same term with $d^1$ and $d^2$ zero modes swapped. As a result, the
product of the vertex operators needs to supply five $d$ zero modes, more
specifically three $d^1$ zero modes and two $d^2$ zero modes for the term
written down above. The corresponding contribution of the above term to
$\cK_{(5)}$ is given by \cite{Gomez:2015uha},
\bea
\label{fived}
&&
\Big \<  \Pi_m (d^2 \gamma d^2) \left ( 2 (\bar \lambda \gamma ^m d^1) (d^1 \gamma d^2) - (\bar \lambda \gamma ^m d^2) ( d^1 \gamma d^1) \right ) 
\no \\ && \hskip 0.5in \times 
(\bar\lambda \gamma D)^2 (d^1W_1) (d^1 W_2) (d^1 W_3) (d^2 W_4) (d^2 W_5) \Big \>
\eea
plus the same contribution with the zero modes $d^1$ and $d^2 $ swapped. Expanding  $\Pi_m$ as in (\ref{rules2}),  evaluated this time at one of the $b$-ghost insertions,  produces terms linear in loop momenta and terms which are linear in external momenta. The terms linear in loop momenta are accompanied by a holomorphic $(1,0)$-form at the $b$-ghost insertion point and will directly lead to the measure on moduli space. Terms linear in external momenta will not be computed directly but rather inferred by consistency.

\sm 

For the contributions linear in loop momenta we construct an expression of the
schematic form $FFWWW$ from cohomology arguments in the next section: Carrying
out the integration over $d$-zero modes and $r$-zero modes in (\ref{fived}),
we see that we now have two super derivatives acting on the vertex operators
(in contrast with the contribution with four $d$ zero modes from the vertex
operators, where we had three super derivatives).  When the super derivatives
act on two different vertex operators, the respective superfields $W_i$ will
be converted to $F_i$, leaving expressions of the schematic form $FFWWW$.
Contributions of the form $WWWWD^2W$ are expected to be BRST equivalent to
those of the form $FFWWW$ by analogy with the fate of the four-point
contributions $D^3(WWWW)$ \cite{Gomez:2010ad}.

\subsection{The vector block for the amplitude of five external states}
\label{sec:42}

Summarizing the structural information gathered in the previous subsection, we
have two distinct types of contributions to the chiral amplitude for five
external states. The first contribution is linear in the loop momenta and will
be referred to as the vector block, while the second contribution is
independent of loop momenta and will be referred to as the scalar block. Our
strategy will be to determine first the vector block, in part from information
obtained through its structural analysis in the previous section, and in part
from enforcing BRST invariance. The scalar block will not be computed
directly, but will be determined uniquely from the monodromy behavior of the
vector block (recall that, according to (\ref{Bmon}), loop momenta behave non-trivially under moving a
vertex operator point $z_i$ around a $\mB$-cycle on the surface) combined with
BRST invariance.

\sm

The vector block receives two different types of contributions, symbolically
of the form $AFFFW$ and $FFWWW$, as was derived in the previous section. It
will be convenient to label the contributions to the vector block with vertex
operator indices corresponding to the distribution of $d^1$ and $d^2$ zero
modes in the contribution with five $d$-zero modes on the vertex operators. 
Thus, a contribution with three $d^1$ zero modes on vertex operators $1,2,3$
and two $d^2$ zero modes on vertex operators $4,5$ will contribute to
$T^m_{1,2,3|4,5}$. We will also include in $T^m_{1,2,3|4,5}$ the contributions
with four $d$-zero modes on the vertex operators, specifically two  $d^2$-zero
modes on vertex operators $4,5$ with two $d^1$ zero modes and one
$A_m$ vertex distributed amongst the points $1,2,3$.  Thus, the vector block
$T^m_{1,2,3|4,5}$ takes the form \cite{Mafra:2015mja},
\bea
\label{TAW}
T^m _{1,2,3|4,5} = A_1^m \, T_{2,3|4,5} + A_2^m \, T_{3,1|4,5} + A^m _3 \, T_{1,2|4,5} + W^m _{1,2,3|4,5}
\eea
where $T_{2,3|4,5}$ and its permutations are the four-state blocks defined in
(\ref{mT}), and $W$ collects all the contributions of the structural form
$FFWWW$. The first three terms on the right side of (\ref{TAW}) are invariant
under all permutations of $1,2,3$ as well as under swapping $4,5$. Our goal
will be to construct $W^m_{1,2,3|4,5}$ and thus $T^m _{1,2,3|4,5}$ which are
invariant under these symmetries as well. 

\sm

A crucial ingredient in our construction will be the BRST transformation property of the vector block. Using the BRST invariance of  $T_{2,3|4,5}$ and its permutations, and the BRST transform of $A_m$ given in (\ref{Q3}), the  BRST transformation of the vector block is given by,
\bea
\label{QmT}
Q T^m _{1,2,3|4,5}  = i k^m_1 V_1 \, T_{2,3|4,5} + i k_2^m V_2 \, T_{3,1|4,5} + i k^m _3 V_3 \, T_{1,2|4,5} 
\eea
provided the BRST transform of $W^m_{1,2,3|4,5}$ satisfies,
\bea
\label{QWT}
Q W^m _{1,2,3|4,5} = - (\lambda \gamma ^m W_1) \, T_{2,3|4,5} 
- (\lambda \gamma ^m W_2) \, T_{3,1|4,5}  - (\lambda \gamma ^m W_3) \, T_{1,2|4,5} 
\eea
We shall now show that this equation may be solved for $W^m_{1,2,3|4,5}$, up to BRST exact contributions, by a sum of terms each of which is of the structural form $FFWWW$, as predicted in the previous subsection.  Three distinct types of contributions arise,
\bea
(\mw_1) _{3,4;5|1,2} ^m & = & 
- { 1 \over 8} (\lambda \gamma ^m W_5) \left \{ (\lambda \gamma _{pq} \gamma ^r W_1) F_2^{pq} + (1 \leftrightarrow 2) \right \} \left \{ (\lambda \gamma _{st} \gamma _r W_3) F_4^{st} + (3 \leftrightarrow 4) \right \} 
\no \\
(\mw_2) _{3,4;5|1,2} ^m & = & 
{1 \over 6}  (\lambda \gamma _t W_5) (\lambda  \gamma _{npqrs } \lambda ) F_1^{np} F_2^{qr} (W_3 \gamma ^{mst} W_4)
\no \\
(\mw_3) _{3,4;5|1,2} ^m & = & 
- { 1 \over 3} (\lambda \gamma _r W_5) \left \{ (\lambda \gamma _{pq} \gamma ^m W_1) F_2^{pq} + (1 \leftrightarrow 2) \right \}
\left \{ (\lambda \gamma _s W_3) F_4^{rs} + (3 \leftrightarrow 4) \right \} 
\eea
The overall coefficients have been chosen for later convenience. 
To make a connection with the structural analysis, the first term arises from four $d$-zero modes coming from the vertex operators, and one super derivative applied to $A_5^m$. The second and third terms arise from five $d$-zero modes coming from the vertex operators. Specifically, the second term arises from the first term in the large parentheses of (\ref{fived}) while the third term arises from the second term in the parentheses of (\ref{fived}).

\sm

The BRST transformations of these partial contributions are readily obtained using the results of (\ref{Q3}) and (\ref{Q4}), as well as the following identities,
\bea
\label{Q5}
 Q \{ (\lambda \gamma _{pq} \gamma _r W_i  ) F_j^ { pq} \} + (i \leftrightarrow j)  & = & 
- \half (\lambda  \gamma _{stpqr} \lambda ) F_i^{st} F_j^{pq}
\no \\
 Q (W_i \gamma ^{mst} W_j)  & = & { 1 \over 4} (\lambda \gamma_{pq} \gamma ^{mst} W_j) F_{i}^{pq} 
+ (i \leftrightarrow j)
\eea
The resulting BRST transformations are then given by, 
\bea
Q (\mw_1) _{3,4;5|1,2} ^m   & = & 
 {1 \over 4}  (\lambda \gamma ^m W_5) \left ( \mt_{1,2|3;4} + \mt_{1,2|4;3} - \mt_{3,4|1;2} - \mt_{3,4|2;1} \right )
 \no \\
Q (\mw_2) _{3,4;5|1,2} ^m  & = & 
- {1 \over 6}  (\lambda \gamma ^m W_5) \mt_{1,2|4;3}  - {1 \over 6}  (\lambda \gamma ^m W_5) \mt_{1,2|3;4} 
\no \\ &&
 - {1 \over 3}  (\lambda \gamma ^m   W_3  )  \mt_{1,2|4;5}  - {1 \over 3}  (\lambda \gamma ^m   W_4  )  \mt_{1,2|3;5}
  \no \\ &&
 + {1 \over 6}   (\lambda  \gamma ^m {}  _{npqr } \lambda )  F_1^{np} F_2^{qr} \Big [ F_4^{st} (\lambda  \gamma _s W_3  )  (\lambda \gamma _t W_5) 
 + ( 3 \leftrightarrow 4) \Big ]
 \no \\
 Q ( \mw _3)^m _{3,4;5|1,2} & = &  - { 1 \over 6}  ( \lambda ^m {} _{npqr} \lambda) F_1 ^{np} F_2^{qr}  F_4^{st} (\lambda \gamma _s W_3) (\lambda \gamma _t W_5) + (3 \leftrightarrow 4)
\eea 
where $\mt_{1,2|3;4}$ was defined in (\ref{QM1}).
An immediate simplification is obtained by adding $Q(\mw_2)$ and $Q (\mw_3)$. The sum of all three,
\bea
\mw^m _{3,4,5|1,2} = (\mw_1) _{3,4;5|1,2} ^m + (\mw_2) _{3,4;5|1,2} ^m + (\mw_3) _{3,4;5|1,2} ^m 
+ (5 \leftrightarrow 3,4)
\eea
has the following BRST transform, 
\bea
\label{QwT}
Q \mw_{3,4,5|1,2} =
- (\lambda \gamma ^m   W_3)  \, T_{1,2|4;5}  - (\lambda \gamma ^m   W_4)  \, T_{1,2|5,3}
 -  (\lambda \gamma ^m W_5) \, T_{1,2|3,4}
\eea
Thus, $\mw^m _{1,2,3|4,5}$ appears to provide a suitable candidate for $W^m _{1,2,3|4,5}$, except for the fact that it does not make the symmetries of $T^m_{1,2,3|4,5}$ manifest. 

\sm

Indeed, the symmetry of $T_{1,2|3,4}$ in (\ref{symmT}) implies that $QT^m _{1,2,3|4,5}$ satisfies the symmetry,
\bea
QT^m _{1,2,3|4,5} = QT^m_{3,4,5|1,2} +  QT^m_{2,4,5|1,3} +  QT^m_{1,4,5|2,3} 
\eea
The first three terms of $T^m _{1,2,3|4,5}$ in (\ref{TAW}) satisfy this same relation before applying  $Q$. Therefore, $T^m _{1,2,3|4,5}$ itself satisfies the following symmetry relation,
\bea
\label{mTsym}
T^m _{1,2,3|4,5} = T^m_{3,4,5|1,2} +  T^m_{2,4,5|1,3} +  T^m_{1,4,5|2,3} 
\eea
provided $W^m _{1,2,3|4,5}$ also satisfies this relation. The candidate $\mw^m_{1,2,3|4,5}$ we had obtained for $W^m_{1,2,3|4,5}$ satisfies the appropriate BRST relation (\ref{QwT}) but fails to satisfy (\ref{mTsym}).  The following symmetrization of $\mw^m_{1,2,3|4,5}$, 
\bea
W^m _{1,2,3|4,5} & = & \half \mw^m _{1,2,3|4,5} 
+ { 1 \over 6} (\mw^m _{3,4,5|1,2} + \mw^m _{2,4,5|1,3} + \mw^m _{1,4,5|3,2} )
\\ &&
- { 1 \over 6} (\mw^m _{1,2,4|3,5} + \mw^m_{1,2,5|3,4} + \mw^m_{1,3,4|2,5} + \mw^m _{1,3,5|2,4} + \mw^m_{3,2,4|1,5} +\mw^m _{3,2,5|1,4})
\no \eea
produces the  desired expression for $W^m_{1,2,3|4,5}$ which satisfies both the BRST condition (\ref{QWT}) and the cyclic symmetry (\ref{mTsym}).

\subsection{Worldsheet dependence of the vector block}

At fixed loop momenta the correlator of the field $x_+^m$ produces the chiral
Koba-Nielsen factor $\cI_{(N)}$ of (\ref{KNN}) for $N=5$, along with contributions
from the insertions of the operator~$\Pi^m$. In view of the substitution rule
(\ref{rules2}) of the chiral splitting procedure, the latter decomposes into
the operator $\p x_+^m + \frac{1}{2} \theta \gamma^m \partial \theta$ and
the part linear in loop momenta $p^I_m$ which is holomorphic in $z$. The
contributions to the chiral correlator ${\cal K}_{(5)}$ linear in $p^I_m$ is
captured by, 
\bea
\cK_{(5)}^{p} = 2\pi  p^I_m T_{1,2,3|4,5}^m   \omega_I(2) \Delta (3,4) \Delta (5,1)  + \hbox{ cycl}(1,2,3,4,5) 
\label{Fmin}
\eea
where the cyclic sum renders (\ref{Fmin}) invariant under all permutations of
the $z_i$ and external states\footnote{In the superfield formalism for the
external vertex operators used here, invariance of the amplitude for $N$
external states under all $N!$ permutations of the external states provides
the superfield implementation of Bose symmetry for external bosons and Fermi
symmetry for external fermions. Full permutation invariance may be verified by
repeatedly using the symmetries (\ref{Delsym}) and (\ref{mTsym}) of the forms
$ \omega_I(2) \Delta (3,4)$ and the kinematic factor $T_{1,2,3|4,5}^m$,
respectively.}. This combination has been chosen because it gives an
economical expression for a fully Bose symmetric amplitude contribution in
terms of cyclic permutations only, without the need to include all 120
permutations of five points. However, (\ref{Fmin}) fails to obey the homology invariance
properties (\ref{Bmon}) and (\ref{Bmon1}). 

\sm

To obtain homology invariance of (\ref{Fmin}), we shall now promote the dependence on the loop momenta to combinations which are homology invariant. As a first step, note that the insertion of a single operator 
$\p x^m$ multiplies the chiral Koba-Nielsen factor \eqref{KNN} by,
\bea
\label{veccor}
\cP^m(z_i) =  2 \pi i (p^I)^m \om_I(z_i) + \sum _{j\neq i} k_j ^m \p_i \ln E (z_i , z_j) 
\eea
Thanks to overall momentum conservation, the transformation law of the loop momenta  given in (\ref{Bmon1}), and the $\mA_I$ and $\mB_I$-cycle monodromies (\ref{mondE}) of the prime form, 
the one-form $\cP^m (z_i)$ is homology invariant.  
Hence, any loop momentum contracting the vector block $T^m_{1,2,3|4,5}$ in (\ref{Fmin})
will be promoted to the combination (\ref{veccor}). Since (\ref{Fmin})
additionally features bi-holomorphic $(1,0)$-forms $\Delta(i,j)$ defined in
(\ref{Del}), it is convenient to define the following vector-valued
meromorphic $(1,0)$-form in five variables $z_i$,
\bea
\cZ^m_{1|2,3|4,5} = \cP^m(1) \Delta (2,3) \Delta (4,5)
\label{vecZZ}
\eea
An immediate property which will be crucial soon is as follows,
\bea
\label{kZ}
k_1 ^m \cZ^m _{1|2,3|4,5} \cI_{(5)} = \p_1 \Big ( \cI_{(5)} \, \Delta (2,3) \Delta (4,5) \Big )
\eea
On these grounds, the homology-invariant completion of (\ref{Fmin}) is given by,
\bea
\cK_{(5)}^V = - i \, T_{1,2,3|4,5} ^m \, \cZ^m _{2|3,4|5,1}  + \hbox{ cycl}(1,2,3,4,5) 
\label{Fnonmin}
\eea
However, the terms proportional to $k_j^m T^m_{1,2,3|4,5} \partial_2 \ln
E(z_2,z_j) \Delta (3,4) \Delta (5,1)$, which are present in \eqref{Fnonmin}
in addition to the contributions of (\ref{Fmin}), do not preserve the Bose
permutation invariance of (\ref{Fmin}). At the same time, neither
(\ref{Fmin}) nor (\ref{Fnonmin}) are BRST closed.  In the next subsection, we
shall show that both shortcomings are cured by adding a loop-momentum
independent scalar block.

\subsubsection{BRST transformation of $\cK^V_{(5)}$}

In preparation for the construction of the scalar block in the next subsection, we begin by calculating and then simplifying the BRST transform of the vector block $\cK_{(5)}^V$. The BRST transform is obtained by using (\ref{QmT}) and is given by, 
\bea
Q \cK_{(5)}^V & = &  \left (  k^m_1 V_1 \, T_{2,3|4,5} +  k_2^m V_2 \, T_{3,1|4,5} +  k^m _3 V_3 \, T_{1,2|4,5} 
\right ) \cZ^m _{2|3,4|5,1}  
\no \\ &&
+ \hbox{ cycl}(1,2,3,4,5) 
\label{QFvec1}
\eea
Using the cyclic permutations to expose a single vertex operator $V_3$, we
have equivalently, 
\bea
Q \cK_{(5)}^V & = &  T_{1,2|4,5} \, V_3 \, k_3^m \, (\cZ^m_{2|3,4|5,1} +  \cZ^m_{4|2,3|5,1} )  
+ T_{2,4|5,1} \, V_3 \, k_3^m \,  \cZ^m_{3|4,5|1,2}  
\no \\ && 
  + \hbox{ cycl}(1,2,3,4,5) 
  \label{QFvec2}
\eea
Using the property (\ref{kZ}) and the fact that by now only zero mode
integrations remain for the vertex operator $V_3$ which depends only on
$\lambda$ and $\theta$, we see that the third term in $Q \cK_{(5)}^V
\cI_{(5)}$ is a total derivative in $z_3$ which vanishes upon integration over $z_3$.

\sm

The remaining terms may be simplified as follows. We begin by focussing on the loop momentum dependent part, which is given by,
\bea
k_3^m (\cZ^m_{2|3,4|5,1} +  \cZ^m_{4|2,3|5,1} ) \cI_{(5)} \Big |_p
& = &  2 \pi i k_3 \cdot p^I \Big ( \om_I(2) \Delta (3,4)+ \om_I(4)\Delta (2,3) \Big ) \Delta (5,1) \cI_{(5)}
\no \\ 
& = &
\p_3 \Big ( \Delta (2,4) \Delta (5,1) \cI_{(5)} \Big ) 
\no \\ && 
- \sum_{j\not=3} k_3 \cdot k_j \, \p_3 \ln E(3,j) \Delta (2,4) \Delta (5,1) \, \cI_{(5)}
\eea
where the second line has been obtained from the first by using the first identity in (\ref{Delsym}), and regrouping terms under the total derivative in $z_3$.  Upon including the terms without loop momenta in the $\cZ$-functions, and omitting the total derivative contributions, we find, 
\bea
k_3^m (\cZ^m_{2|3,4|5,1} +  \cZ^m_{4|2,3|5,1} ) \cI_{(5)} =   - \cL_3^0 \, \Delta (5,1) \, \cI_{(5)}
\eea
where $\cL_3^0$ is given by (recall that $s_{ij} = - k_i \cdot k_j$),
\bea
\label{L30}
\cL_3^0 & = & 
~ ~ s_{35} \big[ \partial_2 \ln E(2,5) \Delta(3,4) +  \partial_4 \ln E(4,5) \Delta(2,3) 
+ \partial_3 \ln E(3,5) \Delta(4,2) \big]  
\no \\ &&
+ s_{31} \big[ \partial_2 \ln E(2,1) \Delta(3,4)  +  \partial_4 \ln E(4,1) \Delta(2,3) 
+ \partial_3 \ln E(3,1) \Delta(4,2) \big] 
\no \\ &&
+ s_{32} \big[ \partial_4 \ln E(4,2) \Delta(2,3)  + \partial_3 \ln E(3,2) \Delta(4,2) \big] 
\no \\ &&
 +  s_{34} \big[ \partial_2 \ln E(2,4) \Delta(3,4) + \partial_3 \ln E(3,4) \Delta(4,2) \big]  
\eea
The form $\cL_3^0$ is invariant upon homology shifts of the points $z_i$
around $\mA$ and $\mB$ cycles, as may be shown using \eqref{mondE} and with the help of momentum
conservation, which implies the relation $s_{35}+s_{31}+s_{32}+s_{34}=0$.  To
render (\ref{L30}) manifestly invariant under homology shifts without the need
to invoke momentum conservation, it is convenient to add the following
combination which vanishes in view of momentum conservation,
\bea
\cL_3^1 &  = &
-\half (s_{35}+s_{31}+s_{32}+s_{34}) \Big [ \partial_4 \ln E(4,2) \Delta(2,3) + \partial_3 \ln E(3,2) \Delta(4,2) 
\no \\ && \hskip 1.7in 
+  \partial_2 \ln E(2,4) \Delta(3,4) + \partial_3 \ln E(3,4) \Delta(4,2) \Big ]
\eea
In summary, we have established that, up to total differentials in the vertex operator position points $z_i$, the contribution from the vector chiral block $\cK^V_{(5)}$ has  BRST transform, 
\bea
Q \cK_{(5)}^V & = & T_{1,2|4,5} \, V_3 \, \cL_3 \, \Delta (5,1) \,    + \hbox{ cycl}(1,2,3,4,5) 
\label{finalQFV}
\eea
where $\cL_3=\cL_3^0+\cL_3^1$. In particular, it  is  independent of loop momenta.

\subsection{Construction of the scalar block}
\label{sec:3.7}

By definition, the scalar block $\cK_{(5)} ^S$ is the part of the chiral
amplitude which is independent of loop momenta, and the full chiral amplitude
is the sum of both contributions,
\bea
\cK_{(5)} = \cK_{(5)} ^V + \cK_{(5)} ^S
\label{totKK}
\eea
BRST invariance of the full amplitude imposes the following constraint on the BRST variation of the scalar block,
\bea
Q\cK_{(5)} ^S = - T_{1,2|4,5} \, V_3 \, \cL_3 \, \Delta (5,1)  + \hbox{ cycl}(1,2,3,4,5) 
\label{QFscal}
\eea
To render $\cK_{(5)}$ BRST invariant, a solution must be found for $\cK_{(5)}^S$,
which is independent of the loop momenta, without discarding total derivative terms
(which would be allowed for the total chiral amplitude $\cK_{(5)} \cI_{(5)}$ but not for $\cK^S_{(5)}$).
In the next subsection, we shall construct the so-called
BRST ancestors, such as $S_{3;1|2|4,5}$, which satisfy, 
\bea
Q S_{3;1|2|4,5} & = & s_{31} V_3 T_{1,2|4,5}
\label{QvarS}
\eea
and obey symmetry properties analogous to $T_{1,2|4,5}$, see (\ref{symmT}),
\bea
S_{3;1|2|4,5} = S_{3;1|2|5,4} \hskip 1in S_{3;1|2|4,5} + S_{3;1|5|2,4} + S_{3;1|4|5,2} =0
\label{ssymm}
\eea
With these ancestors at hand, the BRST variation of $\cK_{(5)}^S$ may now be solved as follows,
\bea
 \cK_{(5)}^S = \mthree \, \Delta (5,1)  +  \hbox{ cycl}(1,2,3,4,5) 
 \label{Fscal}
\eea
where
\bea
\label{bigM}
\mthree& = & 
\half (S_{3;2|1|4,5}- S_{3;4|5|1,2}) \Big [ 
\partial_4 \ln E(4,2) \Delta(2,3) + \partial_3 \ln E(3,2) \Delta(4,2) 
\no \\ && \hskip 1.4in 
- \partial_2 \ln E(2,4) \Delta(3,4) - \partial_3 \ln E(3,4) \Delta(4,2) \Big ]
\no \\ && 
- \half S_{3;1|2|4,5} \Big [ \partial_4 \ln E(4,2) \Delta(2,3) + \partial_3 \ln E(3,2) \Delta(4,2) 
+ \partial_2 \ln E(2,4) \Delta(3,4)
\no \\ &&  \hskip 1.4in 
+ \partial_3 \ln E(3,4) \Delta(4,2) - 2 \partial_2 \ln E(2,1) \Delta(3,4) 
\no \\  && \hskip 1.4in 
-2  \partial_4 \ln E(4,1) \Delta(2,3) -2 \partial_3 \ln E(3,1) \Delta(4,2) \Big ]
\no \\ &&
-\half  S_{3;5|4|1,2} \Big [ \partial_4 \ln E(4,2) \Delta(2,3) + \partial_3 \ln E(3,2) \Delta(4,2) 
+ \partial_2 \ln E(2,4) \Delta(3,4)
\no \\ &&  \hskip 1.4in
+ \partial_3 \ln E(3,4) \Delta(4,2) - 2 \partial_2 \ln E(2,5) \Delta(3,4) 
\no \\ && \hskip 1.4in 
-2  \partial_4 \ln E(4,5) \Delta(2,3) -2 \partial_3 \ln E(3,5) \Delta(4,2) \Big ] 
\eea
Note that $\mthree$ is obtained from $V_3 T_{1,2|4,5}\cL_3$ by formally substituting $s_{31}V_3T_{1,2|4,5} \to S_{3;1|2|4,5}$ and permutations thereof, in keeping with the structure of (\ref{QvarS}).


\subsection{Scalar block in terms of two-particle superfields}
\label{sec:multip}

The construction of the scalar block $\cK_{(5)} ^S$ in the previous section
relies on the availability of a local scalar superfield $S_{3;1|2|4,5}$
subject to the BRST variation (\ref{QvarS}). To prove the existence of viable
solutions to the BRST condition and obtain their explicit construction, we
shall use the multi-particle superfield formalism, which was developed for
genus-zero applications in string theory in \cite{Mafra:2014oia} (see
\cite{Mafra:2011nv} for precursors) and tree-level applications in quantum
field theory in \cite{Lee:2015upy, Mafra:2015vca} (see \cite{Mafra:2010jq} for
precursors). Moreover, multi-particle superfields recursively capture 
the short-distance singularities of higher-genus correlators
\cite{Gomez:2013sla, Gomez:2015uha, Mafra:2018nla} and
tree-level subdiagrams of loop amplitudes in quantum field 
theory \cite{Mafra:2014gja, Mafra:2015mja}.

\subsubsection{Preamble} 

Chiral conformal field theory correlators of conformal primary operators of
dimension $(1,0)$ on a Riemann surface of genus zero are determined by the positions
and residues of their poles and their monodromy. In the absence of monodromy,
this statement is equivalent to the well-known result that a meromorphic
$(1,0)$ form on a sphere is completely determined by the positions and
residues of its poles. In particular, the positions of its zeros are
completely determined. In a conformal field theory, the singularity structure
is determined uniquely by the OPEs of the fields in the correlator, so that on
genus-zero surfaces the correlators may be recovered completely from the OPEs.
The chiral amplitudes $\cF_{(N)}$ of interest here are derived from the
insertion of chiral vertex operators $U_i$ of conformal dimension $(1,0)$ and
$b$-ghosts of conformal dimension $(2,0)$ whose monodromy is entirely
contained in the chiral Koba-Nielsen factor $\cI_{(N)}$. The reduced
amplitudes $\cK_{(N)}$ are monodromy-free.

\sm

By contrast, on a surface of higher genus, there exist holomorphic forms of
dimension $(1,0)$, so that specifying the positions and the residues of the
poles no longer suffices to determine the correlator, and additional
information on the contribution of the holomorphic forms is required. Thus,
the OPE is generally insufficient to reconstruct the correlators.

\subsubsection{Two-particle superfield formalism} 
\label{sec:2part}

The two-particle superfield formalism is based on exploiting the OPE structure
of chiral vertex operators $U_i$. Controlling the singularities in the OPE
(and its multi-particle generalization) allows for a complete determination of
the corresponding correlators at genus zero.  The operator product of two
chiral vertex operators enjoys the following structure  \cite{Mafra:2009bz,Mafra:2014oia},
\bea
U_1(z_1) U_2(z_2) \rightarrow
- z_{12}^{- s_{12}-1} \Big( \partial \theta^\a A_{12\a} + \Pi_m A_{12}^m + d_\a W^\a_{12} + \half N_{mn}F_{12}^{mn} \Big)  \label{multi.2}
\eea
up to total derivatives $\p_1$ and $\p_2$ of the product of $z_{12}^{-s_{12}}
$ times a single-valued function of $z_2$ plus non-singular terms.  Upon
integration of the vertex operators over their positions, the total derivative
contributions are expected to cancel. 

\sm

The prefactor $z_{12}^{-s_{12}}$ arises from the contractions of the
exponentials $e^{ik_1 \cdot x_+}$ with $e^{ik_2 \cdot x_+}$ and is contained
in the chiral Koba-Nielsen factor, where $k_1$ and $k_2$ are the momenta of
the external states. The extra factor of $z_{12}^{-1}$ arises from the Wick
contractions of the operator $\p x_+^m $ in $\Pi^m$ with the exponentials
$e^{ik_1 \cdot x_+}$ with $e^{ik_2 \cdot x_+}$ as well as from the pairwise
Wick contractions of the spinor fields. Double poles arise as well, but it was
shown \cite{Mafra:2009bz,Mafra:2014oia} that they may all be included in the total
derivatives which are being omitted. The composite fields $A_{12\a} , A_{12}^m
, W^\a_{12} , F_{12}^{mn} $ are referred to as {\sl two-particle superfields}.
Their expressions in terms of the one-particle superfields are given as
follows,
\begin{align}
(A_{12})_\a &=  \half\bigl[ A_{2\a}  (i k_2\cdot A_1) + A_2^m (\g_m W_1)_\a - (1\leftrightarrow 2)\bigr]
\notag \\
(A_{12})^m &=  \half\Bigl[ A_{1p} F_2^{pm} + A_2^m(i k_2\cdot A_1) + (W_1\g^m W_2) - (1\leftrightarrow 2)\Bigr] \notag \\
(W_{12})^\a &= {1\over 4}(\g_{mn}W_2)^\a F_1^{mn} + W_2^\a (i k_2\cdot A_1) - (1\leftrightarrow 2)
\label{multi.1}
\\
(F_{12})^{mn} &= F_2^{mn}(i k_2\cdot A_1) +  F_2^{[m}{}_{p}F^{n]p}_1 + i k_{12}^{[m}(W_1\g^{n]}W_2) - (1\leftrightarrow 2)
\notag
\end{align}
where $k_{12}^m = k_1^m + k_2^m$. The BRST transforms of the two-particle superfields which will be needed here are given as follows \cite{Mafra:2014oia}, 
\bea
Q W_{12}^{\alpha} & = &  {1\over 4}(\lambda \g_{mn})^\alpha F_{12}^{mn} + s_{12}(V_1 W_2^\alpha - V_2 W^\alpha_1) 
\no \\
Q F_{12} ^{mn} & = & i k_{12}^m (\lambda \gamma ^n W_{12}) - i k_{12}^n (\lambda \gamma ^m W_{12}) 
+s_{12} ( V_1 F_2^{mn} - V_2 F_1^{mn} ) 
\no \\ &&
+ s_{12}  \Big ( A_1^n (\lambda \gamma ^m W_2) - A_2^n (\lambda \gamma ^m W_1) 
- A_1^m (\lambda \gamma ^n W_2) + A_2^m (\lambda \gamma ^n W_1) \Big ) 
\eea 
Using the pure spinor constraint, we also deduce the following BRST transforms, which generalize the relations of (\ref{Q4}) to the case of two-particle superfields, 
\bea
\label{QllF}
Q (\lambda \gamma _s W_{12}) & = & s_{12} V_1 (\lambda \gamma _s W_2) - s_{12} V_2 (\lambda \gamma _s W_1)
\no \\
Q (\lambda \gamma _{mnpqr} \lambda ) F_{12}^{mn} & = & s_{12} (\lambda \gamma _{mnpqr} \lambda )
(V_1 F_2^{mn} - V_2 F_1^{mn}) 
\eea
Also at higher genus, the two-particle superfield formalism can be applied to determine
the singular parts of the correlators. However, since singularities of the OPE do not 
uniquely determine correlators beyond genus zero, the regular parts of the correlator
generically require additional input beyond the multi-particle superfield formalism. 
In the next subsection, the scalar block $S_{3;1|2|4,5}$ in the regular parts of the correlator
will be obtained by solving (\ref{QvarS}), i.e.\ taking BRST invariance and monodromies
into account. Our solution for $S_{3;1|2|4,5}$ turns out to be expressible in terms of the
vector (\ref{TAW}) and two-particle superfields, irrespectively of their OPE origin.

\subsubsection{Two-particle superfields for the five-point function}

To construct the scalar block $S_{3;1|2|4,5}$ solving (\ref{QvarS}), we begin by defining  the following composites of ghost number three,  built out of two-particle superfields in analogy with the construction of (\ref{QM1}) in the four-point case,
\begin{align}
\mt _{12,3|4;5} &=  (\lambda \gamma _{mnpqr} \lambda) F_{12}^{mn} F_3^{pq} F_4^{rs} (\lambda \gamma _s W_5) 
\no \\
 \mt _{4,5|3;12} &=  (\lambda \gamma _{mnpqr} \lambda) F_{4}^{mn} F_5^{pq} F_3^{rs} (\lambda \gamma _s W_{12})  
\label{multi.3}  \\
 \mt _{4,5|12;3} &=  (\lambda \gamma _{mnpqr} \lambda) F_{4}^{mn} F_5^{pq} F_{12}^{rs} (\lambda \gamma _s W_{3})   \no
\end{align}
The three composites are obtained from $\mt_{1,2|3;4}$ in (\ref{QM1}) by substituting the corresponding two-particle superfield for each single-particle field encountered in turn in (\ref{QM1}). Note that the substitution of $F_{12}$ for $F_1$ and $F_2$ in (\ref{QM1}) lead to the same expression $\mt_{12,3|4;5}$. Their BRST transforms are readily obtained from (\ref{QllF}) and (\ref{Q4}), and we find, 
\bea
Q \, \mt _{12,3|4;5} &= & s_{12} V_1 \, \mt_{2,3|4;5} - s_{12} V_2 \, \mt_{1,3|4;5} 
\no \\
Q \, \mt _{4,5|3;12} &=  & s_{12} V_1 \, \mt_{4,5|3;2} - s_{12} V_2 \, \mt_{4,5|3;1}
 \\
Q \, \mt _{4,5|12;3} &=  & s_{12} V_1 \, \mt_{4,5|2;3} - s_{12} V_2 \mt_{4,5|1;3}
 \no
\eea
Upon defining the following combination by analogy with (\ref{mT}),  
\bea
\label{Tij}
T _{12,3|4,5} = { 1 \over 4} \left ( \mt_{12,3|4;5} + \mt_{12,3|5;4} + \mt_{4,5|12;3}+ \mt_{4,5|3;12} \right )
\eea
we verify that its BRST transform is given by,
\bea
QT_{12,3|4,5} = s_{12} (V_1 T _{2,3|4,5} - V_2 T _{1,3|4,5})
\label{multi.4}
\eea
The composite $T_{12,3|4,5}$ by itself does not yet solve (\ref{QvarS}), but
it does exhibit a desired kinematic factor $s_{12}$, vertex operators $V_2$, and
the characteristic building block $T _{1,3|4,5}$, all of which are key
ingredients on the right side of (\ref{QvarS}).

\subsubsection{The scalar block in terms of two-particle superfields}
\label{sec:kinids}

The BRST variation of $T_{12,3|4,5}$ in (\ref{multi.4}), together with the expression (\ref{QmT}) for $QT^m_{1,2,3|4,5}$,
 imply the central result of this subsection, namely  that the combination,
\bea
S_{1;2|3|4,5} =
{1\over 2} \Big (  i (k_1^m{+}k_2^m{-} k_3^m) T^m_{1,2,3|4,5} + T_{12,3|4,5}+T_{13,2|4,5}+T_{23,1|4,5} \Big ) 
\label{multi.5}
\eea
yields the desired BRST variation (\ref{QvarS}). We note here, for later use
in section \ref{sec:5}, that the expressions for  $T^m_{1,2,3|4,5}$ in
(\ref{TAW}) and $T_{12,3|4,5}$ in \eqref{Tij} have been used in
\cite{Mafra:2015mja} to propose a BRST-invariant and manifestly local
representation for the integrands of two-loop five-point amplitudes in SYM and
maximal supergravity.

\sm

The steps in deriving the symmetries (\ref{symmT}) of the chiral blocks for four external states
carry over in identical form to the following relations \cite{Mafra:2015mja},
\bea
 T_{12,3|4,5}=T_{12,3|5,4} \hskip 1in  T_{12,3|4,5} + T_{12,4|5,3} +T_{12,5|3,4} =0 
\label{multi.6}
\eea
As a consequence, the symmetry,
\bea
S_{1;2|3|4,5}= S_{1;2|3|5,4}
\label{Sprops.1}
\eea 
is manifest from the definition (\ref{multi.5}), whereas the relation, 
\bea
S_{1;2|3|4,5}+ S_{1;2|4|5,3}+ S_{1;2|5|3,4} \cong 0
\label{Sprops.2}
\eea 
holds in the BRST cohomology, namely up to a $Q$-exact superfield (an equivalence which is denoted here and below by the symbol $\cong$). Similarly, the vector and scalar superfields are related via \cite{Mafra:2015mja}, 
\bea
ik_3^m (T^m_{1,2,3|4,5}+T^m_{3,4,5|1,2}) - T_{13,2|4,5} -  T_{23,1|4,5} +  T_{34,5|1,2} +  T_{35,4|1,2} \cong 0
\label{multi.7}
\eea
up to a $Q$-exact quantity, and it would be interesting to identify its BRST
ancestor. It is easy to show via momentum conservation
$s_{13}+s_{23}+s_{34}+s_{35}=0$ that the left-hand side of (\ref{multi.7}) is
BRST-closed, and exactness follows from an explicit check 
that its components $\langle \ldots \rangle_0$ vanish \cite{Mafra:2010pn}. 

\sm

More generally, any BRST-closed and \emph{local} combination of permutations of $k^m_jT^m_{1,2,3|4,5}$, and $T_{12,3|4,5} $ is checked to be BRST exact as well. Only non-local expressions such as
$s_{12}^{-1}  S_{1;2|3|4,5} - s_{13}^{-1}  S_{1;3|2|4,5}$ can be in the BRST cohomology.
The absence of local cohomology within our alphabet of kinematic building blocks $T^m_{1,2,3|4,5}$ and
$T_{12,3|4,5} $ is crucial for the viability of our approach.\footnote{For instance, for four external states it is possible to construct \emph{local} pure spinor superfield expressions in the cohomology of the BRST charge. This fact causes complications when applying the same ideas in an attempt to obtain the non-singular completion of the
three-loop four-point correlator from \cite{Gomez:2013sla}.} 

\sm

We will later on exploit that any contraction $k^m_jT^m_{1,2,3|4,5}$ of the vector
with external momenta is expressible via permutations of the scalar building block, 
\begin{align}
ik_1^m T^{m}_{1,2,3|4,5} &=  
S_{2;1|3|4,5}+S_{3;1|2|4,5}
\label{multi.9} \\
ik_5^m T^{m}_{1,2,3|4,5} &\cong  
S_{1;5|4|2,3}+S_{2;5|4|1,3}+S_{3;5|4|1,2}
\notag
\end{align}
The first identity is an immediate consequence of the definition (\ref{multi.5}) while
the second one is based on (\ref{multi.7}), i.e.\ only valid up to BRST-exact terms.
One can similarly show that
\bea
\label{SijT}
S_{1;2|3|4,5}-S_{2;1|3|4,5}=T_{12,3|4,5}
\eea
and, via momentum conservation and repeated application of (\ref{multi.7}), that,
\bea
S_{5;1|2|3,4}+S_{5;2|1|3,4}+S_{5;3|4|1,2}+S_{5;4|3|1,2}  \cong 0
\label{multi.10}
\eea
the last equality again holding up to BRST exact terms.

\newpage

\section{Structure of  the chiral amplitude}
\setcounter{equation}{0}
\label{sec:4A}
\label{sec:simp}

In this section, we shall simplify the expression for the genus-two chiral
amplitude for five external states and further explore its structure. Various
re-organizations between the vector block (\ref{Fnonmin}) and the scalar block
(\ref{Fscal}) lead to new representations that in turn expose manifest
homology invariance, BRST invariance, or locality.

\subsection{Theta functions and symmetry on the  Jacobian variety}
\label{sec:simp.1}

The chiral amplitude obtained in section \ref{sec:4} depends on the positions
of the vertex operators and the $b$-ghost entirely through the holomorphic
Abelian differentials $\om_I$, the prime form $E(z_i,z_j)$, and single
derivatives of its logarithm $\p_{i} \ln E(z_i, z_j)$. At genus zero and one,
the meromorphic form $\p_{i} \ln E(z_i, z_j)$ is odd under swapping the points
$z_i$ and $z_j$, but this property can no longer hold at higher genus since it
is a $(1,0)$ form in $z_i$ but a $(0,0)$ form in $z_j$. Under certain
conditions, which will turn out to be met for the 5-point amplitude, the
meromorphic form above can be recast directly in terms of $\om_I$ and
genus-two $\tet$-functions and their first order derivatives, and in this form
a higher-genus version of the swapping symmetry will be recovered. The present
subsection is devoted to exhibiting the associated simplifications of the
chiral amplitude.

\sm

To express the prime form in terms of genus-two $\tet$-functions we use the Abel-Jacobi map which sends a point $z_i$ in $ \Sigma$ to a point $\zeta _i$ in the Jacobian variety $J(\Sigma)$ (see appendix \ref{sec:C}), 
\bea
(\zeta_i)_I = \int_{z_0}^{z_i} \omega_I 
\label{simp.1}
\eea
Since only differences $\zeta _i - \zeta _j$ will be needed throughout, all dependence on the choice of the base point $z_0$ will cancel out. By the definition of the prime form in (\ref{Edef}), its logarithmic derivative may be decomposed as follows,
\begin{align}
\partial_{i} \ln E(z_i,z_j|\Omega)  =   \omega_I(z_i) g^I_{i,j} - \partial_{i} \ln h_\nu(z_i)
\label{simp.2}
\end{align}
where $\nu$ is an arbitrary odd spin structure, $h_\nu$ is the corresponding holomorphic $(\half, 0)$ form, and $g^I _{i,j}$ is given by the derivative of the logarithm of the $\tet$-function for spin structure $\nu$, 
\bea
g^I_{i,j} = \frac{ \partial }{\partial \zeta_I} \ln \tet [\nu ] (\zeta | \Omega) \bigg |_{\zeta = \zeta _i-\zeta_j} 
\label{simp.3}
\eea
While each term separately on the right side of (\ref{simp.2}) depends on $\nu$, their sum  is independent of the choice of $\nu$. The key advantage of the combination $g_{i,j}^I$
is the symmetry property,
\bea
g_{j,i}^I = - g_{i,j}^I
\eea
while the derivative of the prime form $\partial_{i} \ln E(z_i,z_j) $ exhibits no such symmetry.

\sm

Upon substituting the decomposition (\ref{simp.2}) of the derivative of the
prime form into the five-point amplitude, all dependence on the holomorphic
$(1/2,0)$-forms $h_\nu$ cancels between the vector and scalar blocks, provided
we choose one and the same odd spin structure for all substitutions. This
cancellation is guaranteed by the fact that the full chiral amplitude is a
well-defined $(1,0)$ form in each vertex point $z_i$ whose
monodromy is given solely by the monodromy of the chiral Koba-Nielsen factor.
It may also be verified directly on our final expressions for the vector and
scalar blocks. 

\sm

The contributions involving $h_\nu$ in the vector block are easy to track from (\ref{Fnonmin}), 
\bea
{\cal K}_{(5)}^V \, \Big|_{h_\nu} = - i \partial_{2} \ln h_\nu(z_2) \Delta(3,4) \Delta(5,1) k_2^m T^{m}_{1,2,3|4,5} + {\rm cycl}(1,2,3,4,5)
\label{simp.4}
\eea
where we have used momentum conservation to simplify. A slightly longer calculation is required to isolate the $h_\nu$-dependence of the quantity (\ref{Fscal}) in the scalar block,
\begin{align}
 \mthree \, \Big|_{h_\nu} &=  \frac{1}{2} (  S_{3;2|1|4,5} - S_{3;4|5|1,2})
 \big[\partial_{2} \ln h_\nu(z_2) \Delta(3,4) - \partial_{4} \ln h_\nu(z_4) \Delta(2,3) \big]
\no \\
&-\frac{1}{2} (S_{3;1|2|4,5}+S_{3;5|4|1,2}) 
 \big[\partial_{2} \ln h_\nu(z_2) \Delta(3,4) + \partial_{4} \ln h_\nu(z_4) \Delta(2,3) \big] \label{simp.5} 
 \no \\
 &\cong S_{3;2|1|4,5} \, \partial_{2} \ln h_\nu(z_2) \, \Delta(3,4) 
 + S_{3;4|5|1,2} \, \partial_{4} \ln h_\nu(z_4) \, \Delta(2,3)  
\end{align}
The last line has been obtained from the kinematic identity (\ref{multi.10}) in the BRST cohomology.
On these grounds, the sum  of all contributions $\partial_{i} \ln h_\nu(z_i)$ to the overall
amplitude can be obtained by combining (\ref{Fnonmin}) and (\ref{Fscal}), 
\bea
{\cal K}_{(5)} \, \Big|_{h_\nu} &=& \partial_{2} \ln h_\nu(z_2) \Delta(3,4) \Delta(5,1) \big( {-}ik_2^m T^m_{1,2,3|4,5}
+  S_{3;2|1|4,5}+  S_{1;2|3|4,5} \big) 
\no \\ && + {\rm cycl}(1,2,3,4,5) 
\label{simp.6}
\eea
The sum of the terms in the parentheses on the first line cancels in view of the first kinematic identity in (\ref{multi.9})
so that ${\cal K}_{(5)}|_{h_\nu} =0$, and all dependence on $h_\nu$ for all points $z_i$ cancels.

\subsection{Partition into sub-correlators}
\label{sec:simp.2}

In view of the results of the previous subsection, we may freely make the following substitutions of all partial derivatives of the logarithm of the prime form within ${\cal K}_5$, 
\bea
\label{gij}
\partial_{i} \ln E(z_i,z_j) \rightarrow \omega_I(z_i) \, g^I_{i,j}
\eea
It follows by inspection that both the contributions from the scalar and the
vector blocks may be expressed as linear combinations of holomorphic
differential forms of the type $\omega_I(i)\Delta(j,k) \Delta(\ell,m)$ with
coefficients given by the functions $g^I_{p,q}$, where $(i,j,k,\ell,m)$ is a
permutation of $(1,2,3,4,5)$. In view of the identities (\ref{Delsym}), 
the vector space spanned by  all such forms $\omega_I(i)\Delta(j,k) \Delta(\ell,m)$ is
five-dimensional and a basis is given by,\footnote{The number of independent such forms
follows from group theory. Each $\om_I(j)$ is an $SL(2)$ doublet and the
number of doublets occurring in the five-fold tensor product of doublets is
five. To see concretely that all the forms $\omega_I(i)\Delta(j,k)
\Delta(\ell,m)$ are linear combinations of the forms in (\ref{basisOM}), we
first use cyclic permutations to set $i=1$. There are three such forms,
$\omega_I(1)\Delta(2,3) \Delta(4,5)$, $\omega_I(1)\Delta(2,4) \Delta(3,5)$ and
$\omega_I(1)\Delta(2,5) \Delta(3,4)$. The second form is a linear combination
of the first and third by the second identity in (\ref{Delsym}) while the
third form may be decomposed using the first identity of (\ref{Delsym}),
$\omega_I(1)\Delta(2,5) \Delta(3,4)= - \omega_I(2) \Delta(3,4) \Delta(5,1) -
\om_I(5) \Delta (1,2) \Delta(3,4)$. This cyclic basis was already tacitly used
for the loop-momentum dependent part (\ref{Fmin}) in the opening line for the
vector correlator.}
\bea
\label{basisOM}
\om_I(1) \Delta (2,3) \Delta (4,5) \hskip 0,5in \hbox{and its 4 cyclic permutations of } (1,2,3,4,5)
\eea
Decomposing the correlator in the basis (\ref{basisOM}) we have,
\bea
{\cal K}_{(5)} = \omega_I(1) \Delta(2,3) \Delta(4,5) {\cal K}^I_{5,1,2|3,4} + {\rm cycl}(1,2,3,4,5)  
\label{simp.7}
\eea
We shall refer to the quantity ${\cal K}^I_{5,1,2|3,4}$ and its permutations  as {\sl sub-correlators}. 

\sm

The sub-correlators  comprise all the kinematic dependence, and the free index $I$ is carried by 
the  loop momentum $p^I_m$ or by a  function $g^I_{i,j}$ in (\ref{simp.3}). 
The explicit form of ${\cal K}^I_{5,1,2|3,4}$ resulting from (\ref{Fnonmin}), (\ref{Fscal}), (\ref{bigM}),
even after reduction to the basis of the five-forms, produces a large number of terms, but 
it drastically simplifies after use of  the kinematic identities in section \ref{sec:kinids}:
In terms of the scalar building block $S_{1;2|3|4,5}$ in (\ref{multi.5}) and their anti-symmetrized
combination $T_{12,3|4,5}$ in (\ref{SijT}), the coefficient
of each function $g^I_{i,j}$ reduces to just a single term,
\begin{align}
{\cal K}^I_{5,1,2|3,4} &= 2\pi p_m^I T^m_{5,1,2|3,4} - g^I_{1,2} T_{12,5|3,4} - g^I_{1,5} T_{15,2|3,4}  - g^I_{2,5} T_{25,1|3,4}  \notag \\
&- g_{1,3}^I S_{1;3|4|2,5}  - g_{2,3}^I S_{2;3|4|1,5}  - g_{5,3}^I S_{5;3|4|1,2}  \label{simp.8} 
\no \\
&- g_{1,4}^I S_{1;4|3|2,5}  - g_{2,4}^I S_{2;4|3|1,5}  - g_{5,4}^I S_{5;4|3|1,2} 
\end{align}
while the coefficient of $g^I_{3,4}$ vanishes.

\sm

As reflected by the notation for its subscripts, the sub-correlator
${\cal K}^I_{5,1,2|3,4} $ exhibits the same symmetries as the vector 
building block $T^m_{5,1,2|3,4}$ in (\ref{TAW}). It is manifest from (\ref{simp.8})
that ${\cal K}^I_{5,1,2|3,4}$ is symmetric with respect to labels that are separated by a comma,
\bea
{\cal K}^I_{5,1,2|3,4} = {\cal K}^I_{1,5,2|3,4}= {\cal K}^I_{5,2,1|3,4} 
\hskip 0.6in
{\cal K}^I_{5,1,2|3,4} = {\cal K}^I_{5,1,2|4,3} 
\label{simp.9}
\eea
Moreover, the symmetry relation  (\ref{mTsym}) of $T^m_{5,1,2|3,4}$ carries over to
\bea
{\cal K}^I_{5,1,2|3,4} + {\cal K}^I_{4,1,2|3,5} +{\cal K}^I_{3,1,2|4,5} \cong {\cal K}^I_{3,4,5|1,2}
\label{simp.10}
\eea
as can be verified from (\ref{SijT}) as well as the 
symmetries (\ref{Sprops.1}) and (\ref{Sprops.2}) of $S_{1;2|3|4,5}$. 

\sm

Based on (\ref{simp.9}) and (\ref{simp.10}), one can explain from a simple analogy why the correlator (\ref{simp.7}) is not only cyclically invariant but in fact Bose symmetric in the five external legs:  We have shown that $T^m_{5,1,2|3,4}$ and ${\cal K}^I_{5,1,2|3,4}$  have identical symmetry properties, and the correlator (\ref{simp.7}) is related to its 
loop-momentum dependent part ${\cal K}_{(5)}^p$ in (\ref{Fmin}) via $p^I_mT^m_{5,1,2|3,4} \leftrightarrow {\cal K}^I_{5,1,2|3,4}$. Hence, permutation invariance of ${\cal K}_{(5)}^p$ carries over to the full correlator in (\ref{simp.7}).

\sm

Note that (\ref{multi.5}) together with (\ref{simp.8}) reduce the superspace components $\langle {\cal K}^I_{5,1,2|3,4} \rangle_0$ to permutations of $\langle T^m_{5,1,2|3,4} \rangle_0$ and $\langle T_{51,2|3,4} \rangle_0$. The bosonic components of $\langle T^m_{5,1,2|3,4} \rangle_0$ and $\langle T_{51,2|3,4} \rangle_0$ can be
found in the files available for download from \cite{PSSsite}.

\subsection{Manifesting homology invariance}
\label{sec:simp.3}

We shall now verify that the sub-correlator ${\cal K}^I_{5,1,2|3,4}$ in (\ref{simp.8}) by itself is homology invariant as defined in (\ref{Bmon1}), so that the full amplitude is single-valued on $\Sigma$ after integration over the loop momenta. This statement is stronger than the statement that the sum $\cK_{(5)}$ of all sub-correlators  is homology invariant. The result will imply that, upon multiplication by the chiral Koba-Nielsen factor $\cI_{(5)}$, the contribution of each sub-correlator ${\cal K}^I_{5,1,2|3,4} \, \cI_{(5)}$ to the chiral amplitude  gives rise to the expected monodromies (\ref{Bmon}) all by itself.

\sm
 
The result is non-trivial because each function $g^I_{i,j}$   has non-trivial monodromy as
a point $z_\ell$ is shifted by a $\mB_L$-cycle (but is invariant under an $\mA_L$ shift),
\bea
\label{simp.11}
z_i \rightarrow z_i + \delta_{i\ell} \mB_L 
& \hskip 1in &
g^I_{i,j} \rightarrow g^I_{i,j} + 2\pi i \delta^I_L (\delta_{j\ell} - \delta_{i\ell})
\no \\
p^I \rightarrow p^I- \delta ^I_L k_\ell &&
\eea
which is readily established using the transformation laws of the prime form in (\ref{mondE}). Implementing the full homology transformations of (\ref{Bmon1}) on the loop momenta as well, we see that $\cK^I _{5,1,2|3,4}$ is invariant provided the following identities hold, 
\begin{align}
\label{simp.12}
 2\pi i \big( T_{12,5|3,4} + T_{15,2|3,4} + S_{1;3|4|2,5} + S_{1;4|3|2,5} \big) - 2\pi (k_1)_m T^m_{5,1,2|3,4} &\cong 0 \\
 - 2\pi i \big( S_{1;3|4|2,5} +S_{2;3|4|1,5} +S_{5;3|4|1,2}  \big) - 2\pi (k_3)_m T^m_{5,1,2|3,4} & \cong 0  \notag
\end{align}
The validity of these identities can be easily checked in the BRST cohomology by means of (\ref{multi.9}), (\ref{SijT}),  and (\ref{multi.10}). As a consequence, the integral over loop momenta of the chiral amplitude will be a single-valued function on $\Sigma$ (see section \ref{sec:5}).

\sm

Actually, an even stronger property may be obtained by decomposing $\cK^I_{5,1,2|3,4}$ into smaller blocks, each of which will by itself be homology invariant. The key to this re-organization of
$\cK^I_{5,1,2|3,4}$ is the following combination of $g_{i,j}^I$ functions, 
\bea
G^I_{i,j,k} = g^I_{i,j}+g^I_{j,k}+g^I_{k,i}
\label{simp.14}
\eea
for three distinct points $z_i, z_j, z_k$. The functions $G^I_{i,j,k}$ are
single-valued in view of the definition of $g^I_{i,j}$ and (\ref{simp.11}), but
they do depend on the spin structure $\nu$ involved in defining $g^I_{i,j}$. We
note that the combination $\om_I(z_i) G^I_{i,j,k}$ is the unique Abelian
differential of the third kind in $z_i$ having simple poles at $z_j$ and $z_k$
with residues $\pm 1$, whose $\mA_J$ period is $\p_J \ln \tet [\nu]
(\zeta_j{-}\zeta_k)$. 

\sm

The same kinematic identities (\ref{simp.12}) also allow us to decompose  $\cK^I _{5,1,2|3,4}$
into smaller blocks each of which is homology invariant. To see this we recast  $\cK^I _{5,1,2|3,4}$ as follows,
\bea
 \label{simp.13}  
{\cal K}^I_{5,1,2|3,4} &\cong & (2\pi p_m^I -ik_{2m} g^I_{1,2}-ik_{3m} g^I_{1,3}-ik_{4m} g^I_{1,4} -ik_{5m} g^I_{1,5}) T^m_{5,1,2|3,4} - G^I_{1,2,5} T_{25,1|3,4}
\no \\ &&
  - G^I_{1,2,3} S_{2;3|4|1,5}  - G^I_{1,5,3} S_{5;3|4|1,2}   - G^I_{1,2,4} S_{2;4|3|1,5}  - G^I_{1,5,4} S_{5;4|3|1,2}  
\eea
The expressions (\ref{simp.8}) and (\ref{simp.13}) agree in the BRST cohomology. To see this, we note that  the coefficients of $g^I_{2,5}$, $g^I_{2,3}$, $g^I_{5,3}$, $g^I_{2,4}$, $g^I_{5,4}$ and $p_m^I$  are manifestly the same, while the differences of the coefficients of $g_{1,2}^I,g_{1,3}^I,g_{1,4}^I$ and $g_{1,5}^I$
are BRST exact by permutations of (\ref{simp.12}). Inspection of (\ref{simp.13}) reveals that 
the combination of $p_m^I$ and $k_{jm} g^I_{1,j}$ in the first line is homology invariant by itself thanks to momentum conservation. Indeed,  it can be viewed as  the genus-two uplift of the generalized elliptic integrand $E^m_{1|2,3,4,5}$ in the genus-one five-point function \cite{Mafra:2017ioj, Mafra:2018qqe}.  Furthermore, each remaining term in (\ref{simp.13}) is single-valued by itself since its world-sheet dependence is through the  single-valued functions $G^I_{i,j,k}$.

\subsection{Manifesting BRST invariance}
\label{sec:simp.4}

Though the correlator ${\cal K}_{(5)}$ is BRST invariant by construction, it is instructive to see how this is realized in the decomposition (\ref{simp.7}) into sub-correlators. Combining the BRST transformations of the ingredients of $\cK^I _{5,1,2|3,4}$ from (\ref{QmT}), (\ref{QvarS}), and (\ref{multi.4}), we find, 
\bea
Q \cK^I_{5,1,2|3,4} & = &
 T_{1,2|3,4} V_5 \Big ( 2 \pi i p^I \cdot k_5 - \sum _{j \not= 5}  s_{5j} \, g^I_{5,j} \Big )
 + T_{2,5|3,4} V_1 \Big ( 2 \pi i p^I \cdot k_1 - \sum _{j \not= 1}  s_{1j} \, g^I_{1,j} \Big )
 \no \\ &&
 + T_{1,5|3,4} V_2 \Big ( 2 \pi i p^I \cdot k_2 - \sum _{j \not= 2}  s_{2j} \, g^I_{2,j} \Big )
\eea
Multiplying this result by $\om_I(1) \Delta (2,3) \Delta (4,5)$ and summing over all cyclic permutations gives the BRST variation of $\cK_{(5)}$ in the following form,
\bea
Q \cK_{(5)} & = & \Big (  \Delta (2,3) \Delta (4,5) T_{2,5|3,4} + \Delta (2,5) \Delta (4,3)  T_{2,3|4,5} \Big ) V_1 
\no \\ && \times 
 \om _I(1) \Big ( 2 \pi i p^I \cdot k_1 - \sum _{j \not= 1} g^I_{1,j} \, s_{1j} \Big ) + \hbox{ cycl}(1,2,3,4,5)
\eea
where we have used cyclic permutations and the first identity in (\ref{Delsym}) to regroup all terms in $\om_I(1)$.  The factor on the second line is readily recognized as the logarithmic derivative $\p_{z_1} \ln \cI_{(5)}$ of the chiral Koba-Nielsen factor (\ref{KNN}) 
\begin{equation}
\label{totz}
\p_1 \ln\cI_{(5)} = \om_I(1) \Big ( 2 \pi i p^I \cdot k_1 - s_{12} \, g^I_{1,2} - s_{13} \, g^I_{1,3}
-s_{14} \, g^I_{1,4} - s_{15} \, g^I_{1,5} \Big )
\end{equation}
so that we find,
\bea
Q (\cK_{(5)} \cI_{(5)}) & = & \Big (  \Delta (2,3) \Delta (4,5) T_{2,5|3,4} + \Delta (2,5) \Delta (4,3)  T_{2,3|4,5}  \Big ) 
V_1   \partial_1 \cI_{(5)}
\no \\ &&  
+ {\rm cycl}(1,2,3,4,5)  
\eea
Thus, the effect of acting by the BRST charge is to produce a total derivative in the vertex points (recall that only the $z_1$-independent zero mode parts of $V_1$ and $T_{i,j|k,\ell}$  remain). 

\sm

The above steps in checking BRST invariance serve as guidance to find a manifestly
BRST invariant representation of $\cK_{(5)} \cI_{(5)}$ by adding suitable total derivatives.
In the same way as the manifestly homology-invariant representation (\ref{simp.13}) was constructed 
by adding BRST exact terms to (\ref{simp.8}), we shall now add the following total derivatives, 
\bea
\widehat {\cal K}_{(5)} \cI_{(5)} &= & 
{\cal K}_{(5)} \cI_{(5)} 
- \frac{1}{4} \Big\{
\Big( {S_{1;2|3|4,5} \over s_{12}} +{S_{1;3|2|4,5} \over s_{13}} +{S_{1;4|5|2,3} \over s_{14}} +{S_{1;5|4|2,3} \over s_{15}}  \Big) \Delta(5,2) \Delta(3,4) \partial_{z_1}  \cI_{(5)} \label{simp.20} 
\no \\ &&  \hskip 0.8in 
+\Big( {S_{1;2|5|3,4} \over s_{12}} +{S_{1;5|2|3,4} \over s_{15}} +{S_{1;3|4|2,5} \over s_{13}} +{S_{1;4|3|2,5} \over s_{14}}  \Big) \Delta(2,3) \Delta(4,5) \partial_{z_1}   \cI_{(5)} 
\no \\ && \hskip 0.8in
+ {\rm cycl}(1,2,3,4,5) \Big\}
\eea
to express each sub-correlator in terms of BRST invariants superfield combinations. The factor of $\frac{1}{4}$ arises from averaging over the four possible ancestors $S_{1;2|3|4,5} / s_{12}$, $S_{1;3|2|4,5}/ s_{13}$, $S_{1;4|5|2,3} / s_{14}$ and $S_{1;5|4|2,3} / s_{15} $ of the BRST variation $V_1 T_{2,3|4,5}$. By expanding  the derivatives of the chiral Koba-Nielsen factor and expanding the five-forms in $\widehat {\cal K}_{(5)}$ in terms of  the five-element basis in (\ref{simp.7}), 
\bea
\widehat {\cal K}_{(5)} = \omega_I (1) \Delta(2,3) \Delta(4,5) \,
\widehat \cK^I_{5,1,2|3,4} + \hbox{ cycl}(1,2,3,4,5)
\eea
we find that the coefficients  of the sub-correlator associated with (\ref{simp.20}) are given by,
\begin{align}
\label{simp.21} 
\widehat {\cal K}^I_{5,1,2|3,4} &= 
2\pi p_m^I \, C^m_{5,1,2|3,4} - s_{12} \, g^I_{1,2} \, (C_{1;2|5|3,4} - C_{2;1|5|3,4})  \notag \\
&- s_{15} \, g^I_{1,5} \, (C_{1;5|2|3,4} - C_{5;1|2|3,4})  - s_{25} \, g^I_{2,5} \, (C_{2;5|1|3,4} - C_{5;2|1|3,4})  \notag \\
&- s_{13} \, g_{1,3}^I \, C_{1;3|4|2,5}  - s_{23} \, g_{2,3}^I \, C_{2;3|4|1,5}  - s_{35} \, g_{5,3}^I \, C_{5;3|4|1,2}   \no \\
&- s_{14} \, g_{1,4}^I \, C_{1;4|3|2,5}  - s_{24} \, g_{2,4}^I \, C_{2;4|3|1,5}  - s_{45} \, g_{5,4}^I \, C_{5;4|3|1,2}  
\end{align}
The superfields now enter through the following non-local combinations,
\bea
 C_{1;3|4|2,5} &= {1\over 4} \Big( \frac{ 3S_{1;3|4|2,5} }{s_{13}} -\frac{ S_{1;4|3|2,5}  }{s_{14}}- \frac{S_{1;2|5|3,4} }{s_{12}}  - \frac{ S_{1;5|2|3,4}  }{s_{15}}\Big) 
 \label{simp.23}
\eea
and
\bea
 \label{simp.22} 
C^m_{5,1,2|3,4} &= & T^m_{5,1,2|3,4} - {i\over 4} k_1^m \Big( \frac{ S_{1;2|5|3,4} }{s_{12}} + \frac{ S_{1;5|2|3,4}   }{s_{15}}+ \frac{ S_{1;3|4|2,5}   }{s_{13}}+ \frac{ S_{1;4|3|2,5}  }{s_{14}}\Big) 
\no \\ && \hskip 0.6in 
- {i\over 4} k_2^m \Big( \frac{S_{2;1|5|3,4} }{s_{12}}  +\frac{ S_{2;5|1|3,4}  }{s_{25}}+ \frac{S_{2;3|4|1,5} }{s_{23}} + \frac{S_{2;4|3|1,5}  }{s_{24}} \Big)
\no \\ && \hskip 0.6in
 - {i\over 4} k_5^m \Big( \frac{S_{5;1|2|3,4}  }{s_{15}}+ \frac{S_{5;2|1|3,4}  }{s_{25}}+ \frac{S_{5;3|4|1,2}  }{s_{35}}+ \frac{S_{5;4|3|1,2} }{s_{45}} \Big) 
\eea
Using (\ref{QmT}) and (\ref{QvarS}), it is straightforward to verify that both the scalar and the vector building block
are BRST invariant,
\bea
QC^m_{5,1,2|3,4} = 0 \hskip 1in  QC_{1;2|5|3,4} = 0
\label{simp.24}
\eea
The BRST invariants (\ref{simp.23}) and (\ref{simp.22}) can be viewed as the
analogues of the homology-invariant building blocks in (\ref{simp.13}) -- in both
cases, the respective invariance of the sub-correlator is made manifest term
by term. As another virtue of these BRST invariants, their superspace
components $\langle C_{1;2|5|3,4} \rangle_0$ and $\langle C^m_{5,1,2|3,4}
\rangle_0$ confirm the equivalence of the present approach in the {\it
minimal} pure spinor formalism with the {\it non-minimal} one: The bosonic
components are unchanged (up to identical normalization factors) when trading
the building blocks $T_{12,3|4,5}$ and $T^m_{1,2,3|4,5}$ \cite{Mafra:2015mja}
in the minimal pure spinor variables for their counterparts in the non-minimal
formalism (denoted by $T_{12,3|4,5}$ and $T^m_{1,2,3|4,5}$ in
\cite{Gomez:2015uha}).\footnote{
For the three-loop four-point amplitude, the building blocks in the minimal pure spinor formalism \cite{Mafra:2015gia} and the non-minimal one \cite{Gomez:2013sla} turn out to be inequivalent, due to the existence of non-trivial, local expressions in the BRST cohomology.}

\sm

The expansion of the two-loop BRST invariants \eqref{simp.23} and \eqref{simp.22} in terms of gluon polarizations
is related to the one-loop invariants $C^m_{1|2,3,4,5}$ and $C_{1|23,4,5}$ from \cite{Mafra:2014gsa} that completely determine the five-point correlator \cite{Mafra:2017ioj}. Using the files for
the bosonic components of $\langle C^m_{1|2,3,4,5} \rangle_0$ 
and $\langle C_{1|23,4,5} \rangle_0$
available to download from \cite{PSSsite} one can verify,
\begin{align}
C^m_{1,2,3|4,5} &\cong - {1\over 180}s_{45} C^m_{1|2,3,4,5}
+ {1\over360}(k_4^m - k_5^m) s_{45} C_{1|45,2,3}\label{Cmequiv} \nonumber \\
&+ {1\over720} k_2^m \big( s_{45} (C_{1|24,3,5} + C_{1|25,3,4}) + (
s_{13}+s_{23}) C_{1|23,4,5} \big)\no \\
&+ {1\over720} k_3^m\big( s_{45} (C_{1|34,2,5} + C_{1|35,2,4})  -
(s_{12}+s_{23}) C_{1|23,4,5}) \big)\no\\
&- {1\over720}(k_1^m + k_2^m + k_3^m)\big( s_{24} C_{1|24,3,5} + s_{25}
C_{1|25,3,4} + (2\leftrightarrow 3) \big)
\end{align}
and
\begin{align}
C_{1;3|4|2,5} &\cong {1\over720} \big( s_{35} C_{1|35,2,4} + s_{45} C_{1|45,2,3}
   - 2 s_{34} C_{1|34,2,5} - s_{23} C_{1|23,4,5} - s_{24} C_{1|24,3,5}
   \big)
   \label{Cequiv}
\end{align}
These identities reduce the components $\langle \widehat {\cal K}^I_{5,1,2|3,4} \rangle_0$ 
to one-loop building blocks and will play an important role in the discussion of
S-duality in a companion paper \cite{compone}. The identities \eqref{Cmequiv} and
\eqref{Cequiv} generalize the pure spinor superspace relation between the four-point kinematic
factors at one and two loops, and it would be similarly interesting to
find a superspace proof analogous to \cite{Mafra:2008ar}.

 \sm

We emphasize that the individual sub-correlators ${\cal K}^I_{5,1,2|3,4}
\cI_{(5)}$ and $\widehat {\cal K}^I_{5,1,2|3,4} \cI_{(5)}$ cannot be
identified since total derivatives only arise from the interplay between
different permutations.

\subsection{Simultaneous homology invariance and BRST invariance}
\label{sec:simp.5}

One can repeat the steps of subsection \ref{sec:simp.3} to obtain manifestly homology invariant
and manifestly BRST invariant sub-correlators (\ref{simp.21}). For this purpose, we rewrite the 
kinematic identities of section \ref{sec:kinids}  in terms of the BRST
invariants (\ref{simp.22}) and (\ref{simp.23}),
\begin{align}
\label{simp.27} 
ik_2^m C^m_{5,1,2|3,4} &= s_{12} C_{1;2|5|3,4}  +  s_{25} C_{5;2|1|3,4} 
\no \\
ik_3^m C^m_{5,1,2|3,4} &\cong s_{13} C_{1;3|4|2,5}  +
 s_{23} C_{2;3|4|1,5}  +   s_{35} C_{5;3|4|1,2}  
  \no \\
0&\cong s_{12} C_{2;1|5|3,4}  +  s_{25} C_{2;5|1|3,4}  +  s_{23} C_{2;3|4|1,5}  +  s_{24} C_{2;4|3|1,5} 
\no \\
0&\cong C_{2;1|5|3,4}+C_{2;1|4|5,3}+C_{2;1|3|4,5}
\no \\
0 &= C_{2;1|5|3,4} - C_{2;1|5|4,3}
\end{align}
These identities can be  obtained formally by promoting $T^m_{5,1,2|3,4} \rightarrow C_{5,1,2|3,4}^m $
and $S_{1;3|4|2,5} \rightarrow s_{13} C_{1;3|4|2,5}$ in the relations among local building blocks
in section \ref{sec:kinids}. Moreover, the same operations formally map the manifestly local
correlator representation (\ref{simp.8}) to the manifestly BRST invariant one in (\ref{simp.21}). 
There is an additional identity among BRST invariants,
\bea
0 = C_{2;1|5|3,4}  +  C_{2;5|1|3,4}  +   C_{2;3|4|1,5}  +  C_{2;4|3|1,5} 
\label{simp.30}
\eea
which directly follows from the definition (\ref{simp.23}) and does not seem to have any counterpart 
for the local superfields.

\sm

It is easy to show using the identities of (\ref{simp.27}) that the manifestly BRST-invariant 
sub-correlator (\ref{simp.21}) is cohomologically equivalent to,
\begin{align} 
\widehat {\cal K}^I_{5,1,2|3,4} &\cong  \Big(2\pi p_m^I -i \sum_{j=2}^5 (k_{j})_m g^I_{1,j}
 \Big) C^m_{5,1,2|3,4} - s_{25} G^I_{1,2,5} (C_{2;5|1|3,4} - C_{5;2|1|3,4})  \label{simp.31}
 \no \\
&  - s_{23} G_{1,2,3}^I C_{2;3|4|1,5}  - s_{35} G_{1,5,3}^I C_{5;3|4|1,2}  - s_{24} G_{1,2,4}^I C_{2;4|3|1,5}  - s_{45} G_{1,5,4}^I C_{5;4|3|1,2} 
\end{align}
This representation of the sub-correlator manifests both BRST invariance and homology invariance in each term, see (\ref{simp.14}) for the definition of the functions $G_{a,b,c}^I$. Moreover, one can verify that the 
symmetry property \eqref{mTsym} of  ${\cal K}^I_{5,1,2|3,4}$ carries over,
\bea
\widehat {\cal K}^I_{5,1,2|3,4} + \widehat {\cal K}^I_{4,1,2|3,5} + \widehat {\cal K}^I_{3,1,2|4,5} \cong \widehat {\cal K}^I_{3,4,5|1,2}
\label{simp.32}
\eea
This is most conveniently shown by repeating the steps that led to (\ref{simp.10}) with the above relations between
BRST invariants and using  (\ref{simp.27}).
Note that (\ref{simp.31}) also follows from the formal replacements
$T^m_{5,1,2|3,4} \rightarrow C_{5,1,2|3,4}^m $ and $S_{1;3|4|2,5} \rightarrow s_{13} C_{1;3|4|2,5}$ in
the manifestly local and homology-invariant correlator representation (\ref{simp.13}).

\sm

Similar representations with manifest homology invariance and BRST invariance
have been studied for multi-particle correlators at one loop. The one-loop analogues
of the representation (\ref{simp.27}) of ${\cal K}_{(5)}$ were the starting point to unravel 
double-copy structures in one-loop open-string amplitudes \cite{Mafra:2017ioj, Mafra:2018qqe}. 
The combinatorial structure of the one-loop correlators in the reference is identical to those of gravitational 
matrix elements with an insertion of the supersymmetrized curvature invariant $\cR^4$. Accordingly, it 
would be interesting if the two-loop five-point correlators based on (\ref{simp.31}) could be 
related to matrix elements of a similar gravitational counterterm of type $D^4 \cR^4$ and $D^2\cR^5$.

\subsection{The simplified correlator in terms of prime forms}
\label{sec:simp.6}

One can also rewrite the simplified representations of the five-point correlator in
terms of prime forms $\partial_i \ln E(z_i,z_j)$ 
instead of the function $g^I_{i,j}$ of the Abel maps. Given the
permutation symmetric contribution ${\cal K}^p_{(5)}$ in (\ref{Fmin}) linear in the loop momentum 
and the scalar quantity,
\bea
{\cal R}_{12} = \partial_1 \ln E(1,2) \big[ S_{1;2|3|4,5} \Delta(2,4) \Delta(3,5) 
+ S_{1;2|4|3,5} \Delta(2,3) \Delta(4,5) \big]+ (1\leftrightarrow 2)
\label{simp.34}
\eea
we claim that a BRST equivalent representation of the five-point correlator is given by,
\bea
{\cal K}_{(5)} = {\cal K}^p_{(5)} + \sum_{1\leq i<j}^5 {\cal R}_{ij}
\label{simp.35}
\eea
The expression (\ref{simp.34}) for ${\cal R}_{12}={\cal R}_{21}$ is permutation symmetric in $3,4,5$ up to
BRST-exact terms by the relations (\ref{Delsym}) and (\ref{ssymm}) of the forms and the superfields.
The $(\tfrac{1}{2},0)$-forms in the decomposition (\ref{simp.2}) of the prime form can be easily
checked to cancel from the permutation sum in (\ref{simp.35}) by repeated use of the
identity (\ref{multi.10}) in the BRST cohomology. Hence, one can effectively substitute
$\partial_{i} \ln E(z_i,z_j) \rightarrow \omega_I(z_i) g^I_{i,j}$ within
(\ref{simp.35}) and expand the correlator in terms of five-forms 
$ \omega_I(1)\Delta(2,3) \Delta(4,5)$. By matching the resulting expression with the basis
of five-forms in (\ref{simp.7}), we reproduce the sub-correlator in (\ref{simp.8}), validating 
(\ref{simp.35}) as an alternative representation of the five-point correlator.

\sm

The building blocks $\cR_{ij}$ in (\ref{simp.34}) conveniently track the short-distance singularities
of the correlator as pairs of punctures collide: the simple pole as $z_1 \rightarrow z_2$
stems solely  from setting $ \partial_1 \ln E(1,2) \rightarrow z_{12}^{-1}$ as well as 
$ \partial_2 \ln E(2,1) \rightarrow -z_{12}^{-1}$ and $\Delta(1,j) \rightarrow \Delta(2,j)$ 
in (\ref{simp.34}). This leads to a simple form of the residues
\begin{align}
{\rm Res}_{z_1 \rightarrow z_2} {\cal K}_{(5)}  
&= {\rm Res}_{z_1 \rightarrow z_2} {\cal R}_{12}
\notag \\
&= (S_{1;2|3|4,5} -S_{2;1|3|4,5})  \Delta(2,4) \Delta(3,5) 
\no \\ & \qquad
+ ( S_{1;2|4|3,5} - S_{2;1|4|3,5}) \Delta(2,3) \Delta(4,5) 
\notag \\
&=T_{12,3|4,5} \Delta(2,4) \Delta(3,5) + T_{12,4|3,5} \Delta(2,3) \Delta(4,5)
\label{res12}
\end{align}
where (\ref{SijT}) has been used in passing to the last line.
On the kinematic pole $(k_1{+}k_2)^{-2}$ resulting from integration over $z_1-z_2$, 
the two-particle
superfields factorize correctly on the single particle superfields of $T_{x,3|4,5}$
with a cubic vertex of the gauge-multiplet peeled off, see for instance appendix
A.4 of \cite{Gomez:2015uha}.

\subsubsection{Comparison with the OPE correlator from \cite{Gomez:2015uha}}

The non-minimal pure spinor prescription was used in \cite{Gomez:2015uha} to determine the genus-two five-point correlator up to holomorphic terms, namely terms with no worldsheet singularities.  These holomorphic terms are of course essential to obtain the full amplitude and for extracting the effective interactions in the low energy expansion beyond the lowest order \cite{compone}; indeed for four-point scattering  they are responsible for the entire correlator.

\sm

The result of the OPE analysis can be written as\footnote{In quoting equation (5.40) from \cite{Gomez:2015uha} we used the notation $\Pi_m^I\to 2\pi p_m^I$ and replaced $\eta_{12}\to\p_1\ln E(1,2)$. This last replacement rectifies the definition used in that reference in which $\eta_{ij}$ was
the derivative of the full Green function without stripping the zero modes.}
\begin{align}
{\cal K}^{\rm ope}_{(5)}
& =  \big[2\pi p_{m}^I T^m_{1,2,3|4,5} \Delta(5,1)\omega_I(z_2)\Delta(3,4)
+ {\rm cycl(1,2,3,4,5)}\big] \label{from1504} \\
&+ \Big[\p_1\ln E(1,2) (  T_{12,3|4,5} \Delta(2,4)\Delta(3,5) +
 T_{12,4|3,5} \Delta(2,3)\Delta(4,5))  + (1,2|1,2,3,4,5)\Big]
 \notag
\end{align}
where the notation $+(i,j|1,2,3,4,5)$ means a sum over all ordered choices of
$i$ and $j$ from the set $\{1,2,3,4,5\}$ for a total of ${5\choose 2}$ terms.

\sm

In order to relate \eqref{from1504} to the full correlator \eqref{simp.35} which includes regular terms  we first observe that the first line of \eqref{from1504} is equal to $\cK^p_{(5)}$ in \eqref{Fmin}. To relate the scalar terms we rewrite
$\cR_{12}$ using \eqref{SijT}
\begin{align}
\label{Rot}
\cR_{12} &= \p_1\ln E(1,2)\big( T_{12,3|4,5} \Delta(2,4)\Delta(3,5) +
T_{12,4|3,5}\Delta(2,3)\Delta(4,5)\big) \\
& \quad + S_{2;1|3|4,5}\big(\p_1\ln E(1,2) \Delta(2,4)\Delta(3,5) + \p_2\ln
E(2,1)\Delta(1,4)\Delta(3,5)\big)\notag \\
& \quad + S_{2;1|4|3,5}\big(\p_1\ln E(1,2) \Delta(2,3)\Delta(4,5) + \p_2\ln
E(2,1)\Delta(1,3)\Delta(4,5)\big)\notag
\end{align}
The first line of \eqref{Rot} contains singularities in the worldsheet and reproduces the corresponding terms in \eqref{from1504}. The second and third lines are non-singular on $\Sigma$ and therefore could not be
determined in the OPE analysis of \cite{Gomez:2015uha}.

\sm

Using \eqref{Rot},  the full five-point correlator at two loops \eqref{simp.35} can be written as,
\begin{equation}
\label{compCorr}
\cK_{(5)} = \cK_{(5)}^{\rm ope} + \big[\cK^{\rm reg}_{(12),3,4,5} + (1,2|1,2,3,4,5)\big]
\end{equation}
where $\cK_{(5)}^{\rm ope}$ is the result (\ref{from1504}) from \cite{Gomez:2015uha} while
\begin{align}
\label{deltaR}
\cK^{\rm reg}_{(12),3,4,5} &\equiv  S_{2;1|3|4,5}\big(\p_1\ln E(1,2)
\Delta(2,4)\Delta(3,5) + \p_2\ln E(2,1)\Delta(1,4)\Delta(3,5)\big)
\notag\\
& + S_{2;1|4|3,5}\big(\p_1\ln E(1,2) \Delta(2,3)\Delta(4,5) + \p_2\ln
E(2,1)\Delta(1,3)\Delta(4,5)\big)
\end{align}
is a non-singular function on the worldsheet.

\sm

It is interesting to observe that the regular functions in \eqref{deltaR} are natural from an OPE perspective as they correspond
to the difference in performing the OPEs as $z_1\to z_2$ or as $z_2\to z_1$, a distinction which is absent at genus zero or one.
Together with the existence of the building block $S_{1;2|3|4,5}$, this observation suggests a way to find the regular
completion of singular correlators such as \eqref{from1504}. The relative coefficient between the singular and regular
pieces can then be fixed by imposing overall BRST invariance. In hindsight, applied to the correlator \eqref{from1504},
this procedure yields the full five-point correlator derived in the previous sections.

\subsection{An alternative correlator in terms of prime forms}
\label{sec:moresimp}

A downside of the correlator representation (\ref{simp.35}) in terms of prime forms is that the loop
momentum dependence occurs via ${\cal K}_{(5)}^p$ in (\ref{Fmin}) instead of the homology-invariant
combinations ${\cal Z}^m_{1|2,3|4,5}$ in (\ref{vecZZ}). As an alternative to (\ref{simp.35}) with
more transparent monodromy properties, the correlator can be rewritten as,
\bea
\label{altrep}
\cK_{(5)} &=&  - i \eta_{mn} T^m_{5,1,2|3,4} {\cal Z}^n_{1|2,3|4,5}
\no \\ &&
 + \partial_1 \ln E(1,2) \big( S_{1;2|3|4,5} \Delta(2,5) \Delta(3,4) + S_{5;2|1|3,4} \Delta(2,3) \Delta(4,5)  \big) 
 \no \\ &&
+ \partial_1 \ln E(1,3) \big( S_{1;3|2|4,5} \Delta(2,5) \Delta(3,4) +S_{2;3|4|1,5} \Delta(2,3) \Delta(4,5)
+S_{5;3|4|1,2} \Delta(2,3) \Delta(4,5) \big)
\no \\ &&
+ \partial_1 \ln E(1,4) \big( S_{1;4|5|2,3} \Delta(2,5) \Delta(3,4) +S_{2;4|3|1,5} \Delta(2,3) \Delta(4,5)
+S_{5;4|3|1,2} \Delta(2,3) \Delta(4,5) \big) 
\no \\ && 
+ \partial_1 \ln E(1,5) \big( S_{1;5|4|2,3} \Delta(2,5) \Delta(3,4) + S_{2;5|1|3,4} \Delta(2,3) \Delta(4,5)  \big)  
\no \\ &&
+ {\rm cycl}(1,2,\ldots,5) \!\!
\eea
Once again, the dependence on the half-differentials cancels\footnote{This cancellation is based on the kinematic identities (\ref{multi.9}),  (\ref{multi.10}) and occurs separately for all  five terms in the cyclic orbit.} between the contributions $(k_1)_m T^m_{5,1,2|3,4} \partial_1 \ln h_\nu(1)$ from ${\cal Z}^n_{1|2,3|4,5}$ and the remaining terms in (\ref{altrep}), so one can again replace $\partial_i \ln E(i,j) \rightarrow \omega_I(i) g^I_{i,j}$.
Under this rule, ${\cal Z}^n_{1|2,3|4,5}$ directly reproduces the coefficient of $T^m_{5,1,2|3,4}$ 
in the manifestly homology-invariant representation (\ref{simp.13}) of the sub-correlator.  The contributions proportional to $ G^I_{i,j,k}$ to (\ref{simp.13}) in turn can be recovered from the  explicit prime forms in (\ref{altrep}). For the latter class of terms, the symmetries (\ref{Delsym}) of the forms and kinematic identities including (\ref{multi.10}) need to be used, and different terms in the cyclic orbit of (\ref{altrep}) contribute to the sub-correlator
${\cal K}^I_{5,1,2|3,4}$ multiplying the basis form $\omega_I(1)\Delta(2,3) \Delta(4,5)$.

\bigskip

\newpage

\section{Type II and Heterotic 5-point amplitudes}
\setcounter{equation}{0}
\label{sec:5}

In this section, we shall use the chiral amplitude $\cF_{(5)}$, derived in the previous section, to construct the genus-two amplitude for five external states for the Type II and Heterotic strings. We begin by recalling the structure of the chiral amplitude,
\bea
\label{FK}
\cF_{(5)} & = & \left \<  \cK_{(5)} \right \> _0 \cI_{(5)} 
\eea
where $\cI_{(5)}$ is the chiral Koba-Nielsen factor (\ref{KNN}) and $ \left \<  \cK_{(5)} \right \> _0 $
is the integral (\ref{pssbracket}) of the  chiral correlator $\cK_{(5)}$ over  the zero modes of $\lambda$ and $\theta$. The chiral correlator $\cK_{(5)} = \cK_{(5)} ^V + \cK_{(5)}^S $ was initially constructed in section \ref{sec:4} from two terms $\cK_{(5)} ^V$ and $ \cK_{(5)}^S$ each of which individually is a single-valued function of the vertex points $z_i$ upon integration over loop momenta, and whose sum is BRST closed even though neither term individually  is BRST closed. Section \ref{sec:4A} then presents various simplified forms of $ \cK_{(5)}$ where different subsets of its properties are made manifest. For the purpose of integrating over loop momenta, it is the forms (\ref{simp.35}) and (\ref{altrep})  that will be particularly convenient.

\subsection{Assembling both chiralities for closed string amplitudes}

Scattering amplitudes of closed strings are obtained by pairing left-moving and right-moving chiral blocks and integrating over loop momenta $p_I$ in $\RR^{10}$, over vertex operator positions $z_i$ in $\Sigma$, and over the moduli space $\cM_2$ of compact genus-two Riemann surfaces, which we parametrize locally by the period matrix $\Omega _{IJ}$ in the Siegel upper half-plane \cite{DHoker:1988pdl,DHoker:2002hof,DHoker:1989cxq}.  As a result, the amplitude takes the following form, up to an overall numerical normalization factor that remains to be determined by
unitarity, 
\bea
\label{pairing}
\cA_{(5)} = \delta \Big(\sum_{i=1}^5 k_i \Big) 
\int _{\cM_2} |d^3 \Omega|^2   \int _{\Sigma^5} 
\int _{\RR^{20}} dp  \, 
\cF_{(5)} (z_i,k_i,p^I) \, \overline{ \tilde \cF_{(5)} (z_i, -k_i^*, -p^I) } 
\eea
where $d^3 \Omega = d \Omega _{11} d \Omega _{12} d \Omega _{22}$ produces the holomorphic top  form on $\cM_2$. For each of the closed superstring theories, $\cF_{(5)}$ is the supersymmetric chiral amplitude given in (\ref{FK}), while  the second chiral amplitude $\tilde \cF_{(5)}$ depends on the type of superstrings under consideration.
In either case, the combined integrals will be absolutely convergent for purely imaginary values of the kinematic variables $s_{ij}$. The amplitude obtained this way may be analytically continued to values of $s_{ij}$ throughout the complex plane thereby producing the expected physical poles and branch cuts, as was shown explicitly for the  genus-one amplitude in \cite{DHoker:1994gnm}.

\sm

The dependence on the polarization vectors, polarization spinors, or internal degrees of freedom for the Heterotic string  of both $\cF_{(5)}$ and $\tilde \cF_{(5)}$ will be suppressed throughout. In all cases, the product 
$\cF_{(5)} \tilde \cF_{(5)}$ includes  the absolute value of the chiral Koba-Nielsen factor $\cI_{(5)}$ as a universal factor. This factor  is conveniently rearranged as follows, 
\bea
\label{hatp}
\big |\cI_{(5)} \big |^2 
& = &
\exp \left  \{ - 2 \pi  Y_{IJ} \, \hat p^I \cdot \hat  p^J  + \sum _{i<j} s_{ij} \, \cG(z_i,z_j) \right \}
\no \\
\hat p^I & = & p^I + Y^{IJ} \sum _i k_i  \, \Im \int ^{z_i} _{z_0} \om_J 
\eea
where $\cG$ is the Arakelov Green function of (\ref{GreenA}), which may be replaced by the  string Green function \eqref{Green} since the total momentum is conserved. In addition to the exponential factor, both $\cF_{(5)}$ and $\tilde \cF_{(5)}$ generically also have explicit dependence on the momenta $p^I$ through a polynomial prefactor, which it will be convenient to trade for a dependence on the shifted momentum $\hat p^I$. Note that the measure $dp$ is unaffected by this shift.

\sm

In preparation for integrating over the  loop momenta, we shall recast the dependence of the supersymmetric
chiral correlator (\ref{FK}) on the loop momentum in a form that exhibits the single-valued  Arakelov Green function $\cG$.  To do so, we eliminate $\p_i \ln E(i,j)$ from $\cZ^m_{1|2,3|4,5}$ in favor of  $-\p_i \cG(i,j)$ plus Abelian differentials, Abel-Jacobi integrals and the shifts $\gamma(z_i)$ in (\ref{gashift}). The Abelian differentials and integrals precisely combine with the loop momenta into their shifted versions $\hat p$ in (\ref{hatp}), and we obtain,
\begin{align}
\label{vecZG}
\cZ^m_{1|2,3|4,5} 
&= \Big ( 2 \pi i  (\hat p^I)^m \om_I(1) 
- \sum _{j= 2}^5 k_j ^m \, \p_1 \cG(1 , j) + k_1^m  \p_1 \gamma(z_1)  \Big ) \, \Delta (2,3) \Delta (4,5)
\end{align}
The remaining terms in the correlator representation (\ref{altrep})  are independent of loop momenta and cancel all instances of $  \p_i \gamma(z_i) $.  We now rearrange $\cK_{(5)}$ as follows,
\bea
\cK_{(5)} = \cW + 2 \pi i  \hat p^I _m \cV^m _I 
\eea
where the combinations $\cV^m_I$ are similar to (\ref{vecZG}) and $\cW$ collects the scalar leftover terms,
\begin{align}
\cV^m_I & =  T^m_{1,2,3|4,5} \, \om_I(2) \Delta (3,4)\Delta (5,1)  + \hbox{ cycl}(1,2,3,4,5)
\label{defcv} 
 \\
\cW & =i  T^m_{5,1,2|3,4} \, \sum _{j=2}^5 k_j^m \, \p_1 \cG(1,j) \, \Delta (2,3)\Delta (4,5) 
\label{defcw} \\
&\ \ \ \; {-}\partial_1 \cG(1,2) \big( S_{1;2|3|4,5} \Delta(2,5) \Delta(3,4) + S_{5;2|1|3,4} \Delta(2,3) \Delta(4,5)  \big)
 \notag\\
&\ \ \ \; {-} \partial_1 \cG(1,3) \big( S_{1;3|2|4,5} \Delta(2,5) \Delta(3,4)
+S_{2;3|4|1,5} \Delta(2,3) \Delta(4,5)
+S_{5;3|4|1,2} \Delta(2,3) \Delta(4,5) \big)
\notag\\
&\ \ \ \; {-}\partial_1 \cG(1,4) \big( S_{1;4|5|2,3} \Delta(2,5) \Delta(3,4)
+S_{2;4|3|1,5} \Delta(2,3) \Delta(4,5)
+S_{5;4|3|1,2} \Delta(2,3) \Delta(4,5) \big) \notag\\
&\ \ \ \; {-} \partial_1 \cG(1,5) \big( S_{1;5|4|2,3} \Delta(2,5) \Delta(3,4) + S_{2;5|1|3,4} \Delta(2,3) \Delta(4,5)  \big) 
\no \\ & \ \ \
 + \hbox{ cycl}(1,2,3,4,5)  \notag
\end{align}
and the cyclic sum in the expression for $\cW$ is to be applied to all five lines. 
To obtain the expression (\ref{defcw}) for $\cW$, we have substituted (\ref{vecZG}) into
(\ref{altrep}) and replaced everywhere $\partial_i \ln E(i,j)$ by $- \partial_i G (i,j) - 2\pi i \omega_I(i) Y^{IJ} \Im \int^{z_i}_{z_{j}} \omega_J$. One can then observe that all such terms proportional to $Y^{IJ}$ 
cancel in the cyclic sum between ${\cal W}$ and $2\pi i \hat p^I_m {\cal V}^m_I$. This cancellation follows from the same manipulations that were described in section \ref{sec:moresimp} to relate (\ref{altrep}) to (\ref{simp.13}). 
Finally,  we have replaced all derivatives 
$\partial_i G(i,j)$ of the string Green function \eqref{Green} by derivatives $\partial_i \cG(i,j)$ 
of the Arakelov Green function \eqref{GreenA}, since the difference $\partial_i \gamma(z_i)$ between
the two cancels in the complete
chiral correlator, by the same mechanism which ensures the cancellation of the derivatives of the half-forms $\partial_{i} \ln h_\nu(z_i)$ in section \ref{sec:simp.1}.  In the new representation \eqref{defcw}, 
both $\cV^m_I$ and $\cW$ are  now manifestly single-valued in $z_i$.

\sm

While the expression (\ref{defcw}) for the scalar correlator is adapted to  the representation (\ref{altrep}) of ${\cal K}_{(5)}$, we can bring the loop-momentum-independent part $\cW$ into an alternative form that is more reminiscent of representation (\ref{simp.35}). For this purpose, the manipulations of the forms and kinematic factors that relate (\ref{simp.35}) to (\ref{altrep}) can be readily repeated with $\hat p^I$ and $- \partial_i G(i,j)$ in place 
of $p^I$ and $\partial_i \ln E(i,j)$. Hence, we can immediately rewrite (\ref{defcw}) by performing 
the appropriate replacements in (\ref{simp.35}),
\bea
{\cal W} =  \sum_{1\leq i<j}^5 {\cal Q}_{ij}
\label{defcw.1} 
\eea
where ${\cal Q}_{ij}$ is given by the following simple combinations,
\bea
{\cal Q}_{12} &= & - \partial_1 \cG(1,2) \big[ S_{1;2|3|4,5} \Delta(2,4) \Delta(3,5) 
+ S_{1;2|4|3,5} \Delta(2,3) \Delta(4,5) \big]
\no \\ && -  \partial_2 \cG(2,1) \big[ S_{2;1|3|4,5} \Delta(1,4) \Delta(3,5) 
+ S_{2;1|4|3,5} \Delta(1,3) \Delta(4,5) \big] 
\label{defcw.2}
\eea
To proceed further, we distinguish between the different string theories.

\subsection{Type II amplitudes}

The complete amplitudes are simplest to organize for the Type II superstrings, since the massless sectors of these theories consist only of the unique Type IIA or Type IIB supergravity multiplet. Type IIA and Type IIB amplitudes involve the chiral amplitude
$\tilde \cF_{(5)}  = \langle  \tilde \cK_{(5)} \rangle_0  \overline{\cI_{(5)} }$, where
$\tilde \cK_{(5)}$ is obtained from $\cK_{(5)}$ by substituting the left-moving vector and spinor polarizations by the right-moving vector and spinor polarizations of opposite (Type IIA) or same space-time chirality (Type IIB), respectively. In either case, the structure of $\tilde \cK_{(5)}$ is as follows,
\bea
\label{K5tW}
\tilde \cK_{(5)} = \tilde \cW + 2 \pi i \, \hat p^I_m \tilde \cV_I^m
\eea
With the help of this expression, the loop momentum integrations may now be carried out,
\bea
\int _{\RR^{20}} dp  \, \cK_{(5)}  \, \overline{ \tilde \cK_{(5)} } \, \big |\cI_{(5)} \big |^2
=
{ 1 \over \det (2 Y)^5}  \left ( \cW \,  \overline{ \tilde \cW}  - \pi 
Y^{IJ} \, \cV^m _I  \, \overline{\tilde \cV^m _J}\right ) \,
\prod_{i<j} e^{s_{ij} \cG(i,j)}
\label{intR20}
\eea
The full amplitude therefore becomes, 
\bea
\cA_{(5)} = \delta \Big ( \sum _{i=1}^5 k_i \Big ) \int _{\cM_2} d \mu \, 
{ 1 \over \det (2 Y)^2}   \int _{\Sigma^5}  \left \langle \cW \, \overline{ \tilde \cW} - \pi 
Y^{IJ} \, \cV^m _I 
\, \overline{\tilde \cV^m _J} \right \rangle_0 \,
\prod_{i<j} e^{s_{ij} \cG(i,j)} 
\quad
\label{LRcontract}
\eea
where $\langle \ldots \rangle_0$ collects the zero-mode integrals (\ref{pssbracket}) of
the $\theta^\alpha$ and $\lambda^\alpha$ in both chiral halves.
Three of the powers of $\det (2 Y)$ have been regrouped to produce the modular invariant measure on $\cM_2$, given by, 
\bea
d \mu = { |d^3 \Omega | ^2 \over \det ( 2Y) ^3}
\eea
The remaining two factors of $\det (2 Y)$ combine with the products of bi-holomorphic forms $\Delta$  of (\ref{Del}) and their complex conjugates so that the combinations, 
\bea
{ \Delta (i,j) \, \overline{\Delta (k, \ell)} \over \det (2 Y) }
\eea
are modular invariant. In summary, after integration over loop momenta, the resulting integrand for the scattering amplitude is  invariant under the full modular group $Sp(4,\ZZ)$.

\sm

Scattering amplitudes for Type II strings compactified on a torus $T^d$  are obtained as usual  by restricting the polarizations of the external particles and  inserting a sum over solitonic configurations of the compact coordinates \cite{AlvarezGaume:1986es}, namely the Siegel-Narain theta series 
\bea
\label{defGamma}
\Gamma_{d,d,2}(g,B| \Omega) = \det(2Y)^{d/2}
\sum_{m^I_{\alpha}\in \mathbb{Z}^{2d} \atop n^{I,\alpha}\in \mathbb{Z}^{2d}}
e^{-\pi \cL^{IJ} Y_{IJ} + 2\pi i\, m^I_{\alpha} n^{J,\alpha} X_{IJ}}\ ,\quad
\eea
where $X=\Re \Omega$ and 
$m^I_{\alpha}, n^{I,\alpha}$ are the momenta and windings along the $\alpha$-th direction 
of the torus, and 
\bea
\cL^{IJ} =  
 (m^I_{\alpha} + B_{\alpha\gamma} n^{I,\gamma}) 
 g^{\alpha\beta}  (m^J_{\beta} + B_{\beta\delta} n^{J,\delta}) + n^{I,\alpha} g_{\alpha\beta}  n^{J,\beta}
\eea
where $g_{\alpha\beta}$ and $B_{\alpha\beta}$ are the constant metric and B-field along the 
torus, and  $g^{\alpha\beta}$ is the inverse metric, measured in units of $\alpha'$. The Siegel-Narain theta series 
 \eqref{defGamma} is 
invariant under modular transformations in $Sp(4,\mathbb{Z})$ and T-duality transformations 
in $O(d,d,\mathbb{Z})$ acting on the usual way on $(g,B)$. The prefactor $\det(2 Y)^{d/2}$ 
cancels the part of factor $\det( 2Y)^{5}$ in \eqref{intR20} which would have come from integrating over the loop momenta $p^{\alpha}_I$.

\subsection{Heterotic string amplitudes}

We shall now construct the five-point genus-two amplitude for Heterotic strings. In this case, 
the massless sector in ten dimensions consists of two types of multiplets, namely
the  $\cN=1$ supergravity (SG) multiplet and the  $\cN=1$ super Yang-Mills (SYM) multiplet
with gauge group $E_8 \times E_8$ (for the HE string) or $Spin(32)/\ZZ_2$  (for the HO string) \cite{Gross:1985fr,Gross:1985rr}.

\sm

Similar to the Type II superstring, the five-point  amplitude for Heterotic strings is given as an integral (\ref{pairing}) of the product of the chiral amplitude $\cF_{(5)}$ in (\ref{FK}) for the superstring, and the (conjugate of)  the chiral amplitude $\tilde \cF_{(5)}$ for the bosonic string, compactified on the  tori associated with the root lattice of $E_8 \times E_8$ or $Spin(32)/\ZZ_2$, respectively.  The latter is given by the product of the chiral measure for the bosonic string at genus two, given by the inverse of the Igusa cusp form\footnote{Recall that $\Psi_{10}=\prod_\kappa 
\tet^2[\kappa](0)$ where the product runs over all even spin structures.} $\Psi_{10}$  \cite{Moore:1986rh,Belavin:1986tv}, times the correlator  of the right-moving vertex operators, given by either,
\bea
\cV ^{{\rm SYM}}_i (z_i) & = & \sum _a t_i ^a j^a(z_i) \, e^{i k_i \cdot x _+(z_i)}
\no \\
\cV^{{\rm SG}} _i(z_i) & = & \tilde \ep _i ^* \cdot \Big ( \p x_+(z_i) + 2 \pi p^I \om_I(z_i) \Big ) \, e^{i  k_i \cdot x_+(z_i)}
\label{vertexop}
\eea
where $t^a_i$ is the gauge field polarization, $j^a(z_i)$ is the corresponding holomorphic current, and $\tilde \ep _i ^* $ is the polarization vector for the right movers. For the five-point amplitude, each external state may belong either to the SYM or the SG multiplet, thereby giving rise to six different types of amplitudes.  Schematically representing the states in the SYM multiplet by $F$ (the field strength), and the states in the SG multiplet by $R$ (including the Riemann tensor, the anti-symmetric tensor field, and the dilaton), the six possible structures  correspond to $R^5, R^4 F, R^3 F^2, R^2 F^3, RF^4,$ and~$F^5$. Since the gauge groups for both Heterotic theories are simple, it is immediate that the amplitude corresponding to $R^4 F$ vanishes.

\sm

Correlators of the chiral vertex operators $\cV^{{\rm SG}}_i$ for the supergravity multiplet may be computed straightforwardly using the Wick contractions \eqref{wick}. Although gauge invariance under $\tilde \ep _i ^m \to \tilde \ep _i ^m + \alpha k^m_i$ is not immediately manifest, it is possible to recast the result in terms of the gauge invariant combinations $f_i^{mn} = \tilde \ep_i ^m k_i ^n - \tilde \ep_i ^n k_i ^m$ by discarding exact differentials which do not contribute to the integrated amplitude. This process was carried out for the four-point amplitude in sections 12.4 and 12.5 of  \cite{D'Hoker:2005jc}  and may be generalized to the five-point amplitude in a straightforward, if tedious, manner which is beyond the scope of this paper. Decomposing the resulting chiral correlator in the same way as in \eqref{K5tW},
in terms of the shifted loop momenta $\hat p^I_m$ in (\ref{hatp}), the integral over loop momenta \eqref{LRcontract} produces a term  proportional to $Y^{IJ}  (\tilde \ep _i ^* \cdot \cV_J)$, which has no analogue for the four-point amplitude.

\sm 

For scattering amplitudes of SYM multiplets, it is convenient to fermionize the 16 chiral  compact  bosons into 32 
chiral worldsheet fermions $\lambda ^I (z)$ for $I=1,\cdots, 32$ (not to be confused with the pure spinor ghost field $\lambda ^\a$). For the case of HO, all 32 fermions transform in the defining representation of $SO(32)$ and have the same spin structure $\kappa$ (independent, and to be distinguished from the spin structure on the supersymmetric side).  For the case of HE, the 32 fermions are split into two groups of 16 transforming under the defining representation of $SO(16)_1 \times SO(16)_2$, the maximal orthogonal subgroup of $E_8 \times E_8$, and  $\kappa = (\kappa_1, \kappa_2)$ labels the corresponding independent spin structures $\kappa_1$ and $\kappa_2$. In
absence of fermionic insertions, the partition functions for the internal fermions are given by
\bea
\label{ZHOHE}
\cZ_{HO} = \sum _\kappa  \tet [\kappa](0)^{16} 
\hskip 1in \cZ_{HE} =\sum _{\kappa_1,\kappa_2}  \tet [\kappa _1](0)^8 \tet [\kappa _2](0)^8
\eea
where the sum runs over all even spin structures.

\sm

The current $j^a(z)$ appearing in the vertex operator \eqref{vertexop} for either of the two Heterotic strings 
is given in terms of $\lambda ^I(z)$ by,
 \bea
 j^a (z) = \half \sum_{I,J=1}^{32} T^a _{IJ} \, \lambda ^I(z) \lambda ^J (z)
 \eea
Here,  $T^a _{IJ}$ are the anti-symmetric generators in the defining representations of the Lie algebras of $SO(32)$ and $SO(16)_1 \times SO(16)_2$, respectively. The remaining generators of $E_8 \times E_8$ are accounted for by spin fields, which will not be needed here. The correlators of the holomorphic fields $\lambda ^I(z)$ are given by,
\bea
\< \lambda ^I (z) \lambda ^J(w) \>_{\kappa} = - \delta ^{IJ} \, S_\kappa (z,w)
\eea
where $S_\kappa$ is the Szeg\"o kernel for the spin structure $\kappa$ for the HO theory, and $\kappa$ equals $\kappa_1$ or $\kappa _2$ for the HE theory, depending on whether both $I,J$ belong to $SO(16)_1$ or $SO(16)_2$. Self-contractions on the current are absent so that $\< j^a (z)\>_{\kappa} =0$. The current correlators required for the case of the four-point amplitude \cite{D'Hoker:2005jc}  are,\footnote{Note that $\tr (T^{a_1} \cdots T^{a_n})=0$ whenever generators of both $SO(16)_1$ and $SO(16)_2$ occur under the trace.} 
\bea
\label{jfour}
\< j^{a_1} (z_1) j^{a_2} (z_2) \>_\kappa  & = & \tfrac{1}{2} \tr (T^{a_1} T^{a_2} ) \, S_\kappa (z_1,z_2)^2
\\
\< j^{a_1} (z_1) j^{a_2} (z_2) j^{a_3} (z_3) \>_\kappa  & = & 
 \tr (T^{a_1} T^{a_2} T^{a_3} ) \, S_\kappa (1,2) \, S_\kappa (2,3) \, S_\kappa (3,1)
\no \\
\< j^{a_1} (z_1) j^{a_2} (z_2) j^{a_3} (z_3) j^{a_4} (z_4)\>_\kappa  & = & 
- \tr (T^{a_1}T^{a_2} T^{a_3}T^{a_4}) S_\kappa (1,2) S_\kappa (2,3)S_\kappa (3,4) S_\kappa (4,1)
\no \\ &&
+ \tfrac{1}{4} \tr (T^{a_1}T^{a_2}) \tr (T^{a_3}T^{a_4}) S_\kappa (1,2)^2 S_\kappa (3,4)^2
+ ( 2 \leftrightarrow  3,4)
\no
\eea
where we denote as usual  $S_\kappa (i,j) =S_\kappa (z_i, z_j)$. For the five-point amplitude, we require the correlators of (\ref{jfour}) as well as the following five-point correlators,
\bea
\Big \< \prod _{i=1}^5 j^{a_i} (z_i) \Big \>_\kappa & = & 
\tfrac{1}{2}  \sum _{(i,j|k,\ell,m)} \tr (T^{a_i} T^{a_j}) \tr (T^{a_k} T^{a_\ell} T^{a_m})  
S_\kappa (i,j)^2 S_\kappa (k, \ell) S_\kappa (\ell, m) S_\kappa (m, k)
\no \\ && 
+ \sum _{(i,j,k,\ell)}  \tr (T^{a_1} T^{a_i} T^{a_j} T^{a_k} T^{a_\ell}) 
S_\kappa (1,i) S_\kappa (i,j) S_\kappa (j,k) S_\kappa (k, \ell) S_\kappa (\ell, 1)
\qquad \quad
\label{j5}
\eea
where the first sum is over all 10 inequivalent partitions of five into 2+3, and the second sum is over all 12 permutations of 2,3,4,5 modulo reversal $(i,j,k,\ell) \to (\ell, k, j, i)$. 

\sm

The spin structure sums required for amplitudes with up to five SYM states can be expressed in terms of the Siegel modular forms $\Psi_{4k}$ of weight $4k$, 
\bea
\Psi_{4k} & = & \sum _\kappa \tet [\kappa](0)^{8k} 
\eea
and the following correlators, 
\bea
\label{defPsi4k}
F_{4k}^{(2)} (z_1,z_2) & = & \sum _\kappa \tet [\kappa](0)^{8k} S_\kappa (1,2)^2
\\
F_{4k}^{(3)} (z_1,z_2,z_3) & = & \sum _\kappa \tet [\kappa](0)^{8k} 
S_\kappa (1,2) S_\kappa (2,3) S_\kappa (3,1) 
\no \\
F_{4k}^{(2,2)} (z_1,z_2;z_3,z_4) & = & \sum _\kappa \tet [\kappa](0)^{8k} S_\kappa (1,2)^2 S_\kappa (3,4)^2
\no \\
F_{4k}^{(4)} (z_1,z_2,z_3,z_4) & = & \sum _\kappa \tet [\kappa](0)^{8k} 
S_\kappa (1,2) S_\kappa (2,3) S_\kappa (3,4) S_\kappa (4,1) 
\no \\
F_{4k}^{(2,3)} (z_1,z_2;z_3,z_4,z_5) & = & \sum _\kappa \tet [\kappa](0)^{8k} S_\kappa (1,2)^2 S_\kappa (3,4) S_\kappa (4,5) S_\kappa (5,3) 
\no \\
F_{4k}^{(5)} (z_1,z_2,z_3,z_4,z_5) & = & \sum _\kappa \tet [\kappa](0)^{8k} S_\kappa (1,2) S_\kappa (2,3) S_\kappa (3,4) S_\kappa (4,5) S_\kappa (5,1) 
\no
\eea
The first sum $F^{(2)} _{4k}$ can be computed in terms of $\Psi_{4k}$ through \cite[Eq. 12.7]{D'Hoker:2005jc},
\bea
F_{4k}^{(2)} (z,w)  = \Psi_{4k} \p_z \p_w \ln E(z,w) + { \pi i \over 2} \om_I(z) \om_J(w) \p^{IJ} \Psi_{4k}
\eea
where $\p_{IJ}$ is the derivative with respect to $\Omega _{IJ}$ for $I\leq J$. The product of three Szeg\"o kernels may 
be decomposed onto a sum of squares of Szeg\"o kernels times functions that are independent of spin structures \cite{comptwo}, so that $F_{4k} ^{(3)}$ may be similarly decomposed onto a sum of $F^{(2)}_{4k}$ functions. Similarly, it will be shown in \cite{comptwo} that the products of four and five Szeg\"o kernels may all be decomposed onto sums of the product of two squares of Szeg\"o kernels, so that $F_{4k}^{(4)} $, $F_{4k}^{(2,3)}$, and $F_{4k}^{(5)} $ may all be decomposed onto sums of $F^{(2,2)}_{4k}$ with known coefficients. 

\sm

We end with perhaps the simplest example of a Heterotic amplitude for five external SYM states, two belonging to the first $E_8$, and three belonging to the second $E_8$. The corresponding chiral amplitude may be read off from the ingredients presented above, and is given by, 
\bea
\tilde {\cal F}_{(5)} = {1 \over 4 \Psi_{10}(\Omega)} 
\tr (T^{a_1} T^{a_2}) \tr ( T^{a_3} T^{a_4} T^{a_5}) F_4 ^{(2)} (z_1, z_2) 
F^{(3)} _4 (z_3,z_4,z_5)
\eea
where $a_1, a_2$ refer to $SO(16)_1$ while $a_3,a_4,a_5$ refer to $SO(16)_2$.

\sm 

As usual,  the HE and HO Heterotic strings become indistinguishable after compactifying on a torus $T_d$. The chiral integrand ${\cal F}_{(5)} $ is obtained by replacing  the partition function $\cZ_{HO}$ or $\cZ_{HE}$ in \eqref{ZHOHE} by the Siegel-Narain theta series  $\Gamma_{d+16,d,2}$, with suitable insertions of  lattice momenta for each current as in the four-point amplitude discussed in \cite{Bossard:2018rlt}.

\newpage

\section{The supergravity limit}
\setcounter{equation}{0}
\label{sec:6}

In this section we shall study the field theory limit of the string amplitudes for five external massless states derived in the
earlier sections of this paper. In the limit $\alpha' \rightarrow 0$, keeping the external momenta $k_i$ fixed, the Type~II
superstring amplitudes are expected to reduce to the two-loop field theory amplitudes of $\cN=2$ supergravity, while in the
Heterotic strings the amplitudes are expected to reduce to those of $\cN=1$ supergravity plus super-Yang-Mills
\cite{Green:1982sw}.  For four-dimensional external states, the loop integrand for two-loop supergravity was determined in
\cite{Carrasco:2011mn} using the spinor-helicity formalism and color-kinematics duality \cite{Bern:2008qj, Bern:2010ue} (see
\cite{Bern:2019prr} for a review). This result was later extended to external states in ten dimensions in
\cite{Mafra:2015mja} by making use of pure spinor superspace.

\sm

\begin{figure}
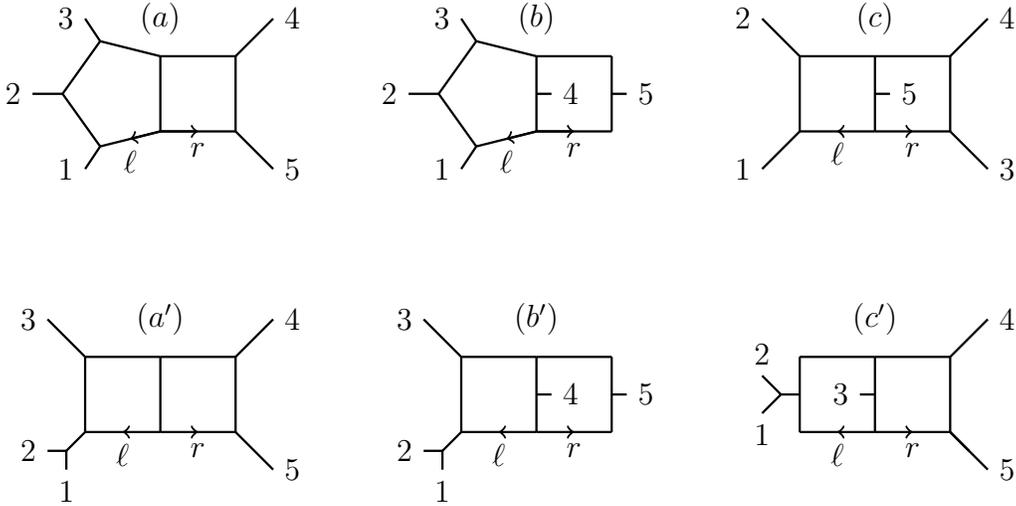

\begin{center}
  \tikzpicture [scale=1, line width=0.30mm]
   \scope[xshift=0cm]
   \draw(0,1) node{$(a)$}; 
  \draw (0,-0.5) -- (0,0.5);
   \draw (1,-0.5) -- (1,0.5);
  \draw (0,-0.5) -- (1,-0.5);
    \draw (0,0.5) -- (1,0.5);
    \draw[->] (0,-0.5) -- (0.5,-0.5)node[below]{$r$};
     \draw (0,0.5) -- (-0.8,0.7);  
     \draw (0,-0.5) -- (-0.8,-0.7);  
     %
     \draw[->] (0,-0.5) -- (-0.4,-0.6);  
     \draw (-0.4,-0.9)node{$\ell$};
     \draw (-1.3,0) -- (-0.8,0.7);  
     \draw (-1.3,0) -- (-0.8,-0.7);      
    \draw (1,0.5) -- (1.5,1) node[right]{4};
    \draw (1,-0.5) -- (1.5,-1) node[right]{5};
    \draw(-1.3,0) -- (-1.7,0) node[left]{2}; 
   \draw(-0.8,0.7) -- (-1,1) node[left]{3}; 
     \draw(-0.8,-0.7) -- (-1,-1) node[left]{1};
     \endscope
   \scope[xshift=5cm]
   \draw(0,1) node{$(b)$}; 
  \draw (0,-0.5) -- (0,0.5);
   \draw (1,-0.5) -- (1,0.5);
   \draw (0,-0.5) -- (1,-0.5);
    \draw (0,0.5) -- (1,0.5);
    \draw[->] (0,-0.5) -- (0.5,-0.5)node[below]{$r$};
     \draw (0,0.5) -- (-0.8,0.7);  
     \draw (0,-0.5) -- (-0.8,-0.7);  
     %
     \draw[->] (0,-0.5) -- (-0.4,-0.6);  
     \draw (-0.4,-0.9)node{$\ell$};
     \draw (-1.3,0) -- (-0.8,0.7);  
     \draw (-1.3,0) -- (-0.8,-0.7);      
    \draw (0,0) -- (0.2,0) node[right]{4};
    \draw (1,0) -- (1.2,0) node[right]{5};
    \draw(-1.3,0) -- (-1.7,0) node[left]{2}; 
   \draw(-0.8,0.7) -- (-1,1) node[left]{3}; 
     \draw(-0.8,-0.7) -- (-1,-1) node[left]{1};
     \endscope
  \scope[xshift=9.5cm]
  \draw(0,1) node{$(c)$}; 
  \draw (0,-0.5) -- (0,0.5);
   \draw (1,-0.5) -- (1,0.5);
   \draw (0,-0.5) -- (1,-0.5);
    \draw (0,0.5) -- (1,0.5);
      \draw (-1,-0.5) -- (-1,0.5);
   \draw (0,-0.5) -- (-1,-0.5);
    \draw (0,0.5) -- (-1,0.5);    
    \draw (1,0.5) -- (1.5,1) node[right]{4};
     \draw (1,-0.5) -- (1.5,-1) node[right]{3};
    \draw (-1,0.5) -- (-1.5,1) node[left]{2};
     \draw (-1,-0.5) -- (-1.5,-1) node[left]{1};
      \draw (0,0) -- (0.2,0) node[right]{5};
      \draw[->] (-0.49,-0.5)--(-0.5,-0.5);
      \draw[->] (0.49,-0.5)--(0.5,-0.5);
      \draw (-0.5,-0.5)node[below]{$\ell$};
      \draw (0.5,-0.5)node[below]{$r$};
      \endscope
 \scope[xshift=0cm, yshift=-4cm]
 \draw(0,1) node{$(a')$}; 
 \draw (0,-0.5) -- (0,0.5);
   \draw (1,-0.5) -- (1,0.5);
   \draw (0,-0.5) -- (1,-0.5);
    \draw (0,0.5) -- (1,0.5);
      \draw (-1,-0.5) -- (-1,0.5);
   \draw (0,-0.5) -- (-1,-0.5);
    \draw (0,0.5) -- (-1,0.5);    
    \draw (1,0.5) -- (1.5,1) node[right]{4};
     \draw (1,-0.5) -- (1.5,-1) node[right]{5};
    \draw (-1,0.5) -- (-1.5,1) node[left]{3};
     \draw (-1,-0.5) -- (-1.25,-0.75) ;
     \draw (-1.25,-0.75) --(-1.25,-1)node[below]{1};
     \draw (-1.25,-0.75) --(-1.5,-0.75)node[left]{2};
      \draw[->] (-0.49,-0.5)--(-0.5,-0.5);
      \draw[->] (0.49,-0.5)--(0.5,-0.5);
      \draw (-0.5,-0.5)node[below]{$\ell$};
      \draw (0.5,-0.5)node[below]{$r$};
     \endscope
   \scope[xshift=5cm, yshift=-4cm]
   \draw(0,1) node{$(b')$}; 
 \draw (0,-0.5) -- (0,0.5);
   \draw (1,-0.5) -- (1,0.5);
   \draw (0,-0.5) -- (1,-0.5);
    \draw (0,0.5) -- (1,0.5);
      \draw (-1,-0.5) -- (-1,0.5);
   \draw (0,-0.5) -- (-1,-0.5);
    \draw (0,0.5) -- (-1,0.5);    
  \draw (0,0) -- (0.2,0) node[right]{4};
    \draw (1,0) -- (1.2,0) node[right]{5};
    \draw (-1,0.5) -- (-1.5,1) node[left]{3};
     \draw (-1,-0.5) -- (-1.25,-0.75) ;
     \draw (-1.25,-0.75) --(-1.25,-1)node[below]{1};
     \draw (-1.25,-0.75) --(-1.5,-0.75)node[left]{2};
      \draw[->] (-0.49,-0.5)--(-0.5,-0.5);
      \draw[->] (0.49,-0.5)--(0.5,-0.5);
      \draw (-0.5,-0.5)node[below]{$\ell$};
      \draw (0.5,-0.5)node[below]{$r$};
 \endscope
  \scope[xshift=9.5cm, yshift=-4cm]
 \draw(0,1) node{$(c')$}; 
 \draw (0,-0.5) -- (0,0.5);
   \draw (-1,-0.5) -- (-1,0.5);
   \draw (0,-0.5) -- (-1,-0.5);
    \draw (0,0.5) -- (-1,0.5);
      \draw (1,-0.5) -- (1,0.5);
   \draw (0,-0.5) -- (1,-0.5);
    \draw (0,0.5) -- (1,0.5);    
   \draw (-1,0) -- (-1.25,0) ;
    \draw (-1.25,0) --(-1.5,0.25)node[above]{2};
    \draw (-1.25,0) --(-1.5,-0.25)node[below]{1};
  \draw (0,0) -- (-0.2,0) node[left]{3};
    \draw (1,0.5) -- (1.5,1) node[right]{4};
      \draw (1,-0.5) -- (1.5,-1) node[right]{5};
     \draw (1,-0.5) -- (1.25,-0.75) ;
      \draw[->] (0.49,-0.5)--(0.5,-0.5);
      \draw[->] (-0.49,-0.5)--(-0.5,-0.5);
      \draw (0.5,-0.5)node[below]{$r$};
      \draw (-0.5,-0.5)node[below]{$\ell$};
      \endscope
   \endtikzpicture
  \end{center}
\caption{The six graphs contributing to two-loop five-point amplitudes in maximally supersymmetric Yang-Mills and supergravity \cite{Carrasco:2011mn}. The reducible diagrams $a', b', c'$ were denoted $d,e,f$ respectively  in \cite{Mafra:2015mja}.   \label{fig:1}}
\end{figure}

Whether the external states of the superstring amplitude are in a supergravity or super-Yang-Mills multiplet, the corresponding field theory amplitudes involve a sum over the six Feynman graph topologies depicted in Figure \ref{fig:1}.  As we shall demonstrate below for Type~II superstrings (and sketch for the Heterotic and Type~I cases), the field theory limit of the integrand over loop momenta, moduli, and vertex points of the superstring amplitude for five external massless states, derived  in earlier sections,  reduces, at leading order in $\alpha'$, to the integrand over loop momenta and Feynman parameters of the corresponding supergravity amplitude \cite{Carrasco:2011mn,Mafra:2015mja}. The precise matching of these integrands provides a  strong consistency check on the validity of our construction. Higher-order terms in the $\alpha'$ expansion of the  integrated amplitude produce  higher-derivative effective interactions  to the  supergravity and/or super-Yang-Mills Lagrangian which will be investigated in a companion paper \cite{compone}.

\sm

To leading order in the $\alpha'$ expansion, the amplitude is dominated by the contribution from maximally degenerate Riemann surfaces. In order to study these degenerations systematically, it will be useful to interpret the vertex operator positions as punctures on the Riemann surface, and use the Deligne-Mumford compactification of the moduli space of punctured Riemann surfaces, in the present case of genus two with five punctures. All degenerations are then obtained by a finite sequence of the following two elementary degenerations,
\begin{enumerate}
\itemsep=-0.05in
\item the separating degeneration, in which a trivial homology cycle shrinks, thereby degenerating the surface into two disconnected surfaces;
\item the non-separating degeneration, in which a non-trivial homology cycle shrinks, thereby degenerating the dual cycle into a long and thin funnel. 
\end{enumerate}
The degeneration by which two or more  punctures collide is equivalent, in the Deligne-Mumford compactification, to a separating degeneration in which a sphere with three or more punctures separates from the remaining surface. The maximal degeneration of the Riemann surface is obtained by a maximal sequence of separating and non-separating degenerations in which for example all  the $\mA$-cycles of the surface shrink, and the $\mB$-cycles become long thin funnels. These funnels are effectively connected by internal interaction vertices, just as in field theory Feynman diagrams. A maximal degeneration may be described by a trivalent graph $\Gamma$, sometimes known as a tropical Riemann surface (see e.g. \cite{Itenberg:2011,Tourkine:2013rda}), which reproduces the on-shell Feynman graphs  of quantum field theory. The vertices of the graph correspond to genus zero components with three punctures, while the edges $e_a$ correspond to the long thin funnels.  The lengths $L_a\in \mathbb{R}^+$ and twists  $\sigma_a\in [0,2\pi[$ of the funnels provide an appropriate set of coordinates on the moduli space near the maximal degeneration locus.

 In the limit where all $L_a$ are scaled to infinity at the same rate, the string integrand is expected to reduce to the field theory integrand in the world-line formalism \cite{Schmidt:1994zj,Roland:1996np,Dai:2006vj},  where $L_a$ is the  Schwinger parameter for the  propagator on edge $e_a$. Upon using the chiral splitting procedure in string theory, the momentum  $p^I$ is identified with the loop  momentum in field theory \cite{Tourkine:2019ukp}. For the pure spinor superstring, the string integrand is expected  to reduce to the field theory integrand in pure spinor world-line formalism \cite{Bjornsson:2010wm,Bjornsson:2010wu} and the double-copy structure of the loop integrand in supergravity
should be manifest  \cite{Mafra:2011kj, Mafra:2014gja, He:2015wgf}.

\subsection{Maximal degeneration of a genus-two Riemann surface}

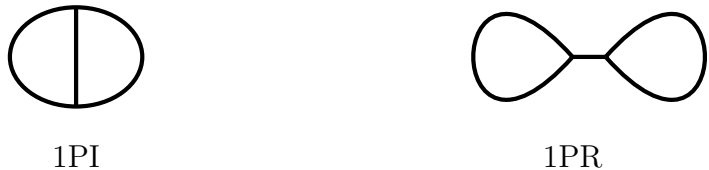
\begin{figure}[h]
\begin{center}
\begin{tikzpicture}[scale=.22]
\begin{scope}[shift={(-15,0)}]
\draw [ultra thick] (0,0) ellipse (4 and 3);
\draw [ultra thick] (0,3) -- (0,-3);
\draw (0, -6) node{1PI};
\end{scope}
\begin{scope}[shift={(15,0)}]
\begin{scope}[rotate=180,shift={(-2,0)}]   
\draw [ultra thick] (0,0) .. 
controls (-8,9) and (-8,-9) .. 
(0,0);
\end{scope}
\draw [ultra thick] (0,0) .. 
controls (-8,9) and (-8,-9) .. 
(0,0);
\draw [ultra thick] (0,0) -- (2,0);
\draw (0, -6) node{1PR};
\end{scope}

\end{tikzpicture}
\caption{1PI versus 1PR two-loop skeletons. \label{pivspr}}
\end{center}
\end{figure}

For a compact genus-two Riemann surface without punctures, there are two possible maximal  degenerations, corresponding to the one-particle irreducible (1PI) or  one-particle reducible (1PR) two-loop skeletons depicted in Figure \ref{pivspr}. In principle, there can also be contact  terms supported on ``figure-eight" diagrams where the length of the middle edge in either of the two skeletons shrinks to zero.\footnote{Such contact terms are known to arise in the field theory limit of  Heterotic amplitudes \cite{Bossard:2018rlt} and Type I partition functions in a magnetic background 
\cite{Magnea:2015fsa}.}
 
 \vskip -0.4in
 
{\color{red} {}}
{\color{red} {}}

 \vskip -0.2in

For a genus-two Riemann surface with punctures, the various different maximal degenerations correspond to the various different ways of attaching external legs to either skeleton of the case without punctures, possibly by forming trees, such that the resulting graph is still connected. For five punctures, many different connected graphs may be drawn. It will be convenient to arrange the graphs into two classes (1) graphs which contain no triangle or bubble subgraphs; and (2) all other graphs. All graphs obtained from the 1PR vacuum graph fall in class (2).

 \vskip 0.1in
{\color{red} {}}
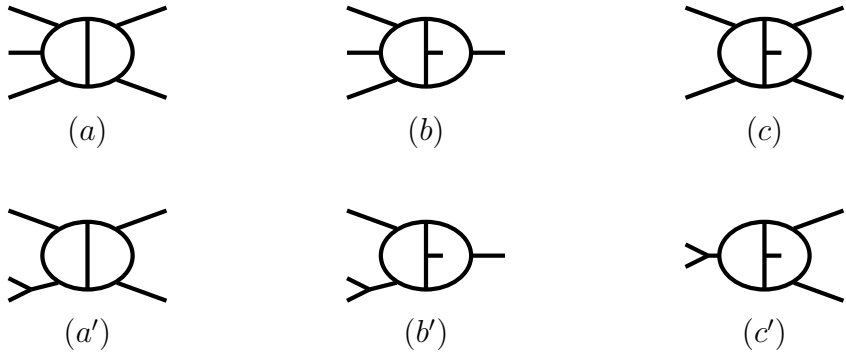
\begin{figure}[h]
\begin{center}
\begin{tikzpicture}[scale=.15]
\begin{scope}[shift={(0,0)}]
\draw [ultra thick] (-2.6,-2.4) -- (-7,-4);
\draw [ultra thick] (-2.6,2.4) -- (-7,4);
\draw [ultra thick] (2.6,-2.4) -- (7,-4);
\draw [ultra thick] (2.6,2.4) -- (7,4);
\draw [ultra thick] (0,0) ellipse (4 and 3);
\draw [ultra thick] (0,3) -- (0,-3);
\draw [ultra thick] (-4,0) -- (-7,0);
\draw (0, -7.0) node{$(a)$};
\end{scope}

\begin{scope}[shift={(30,0)}]
\draw [ultra thick] (-2.6,-2.4) -- (-7,-4);
\draw [ultra thick] (-2.6,2.4) -- (-7,4);
\draw [ultra thick] (0,0) ellipse (4 and 3);
\draw [ultra thick] (0,3) -- (0,-3);
\draw [ultra thick] (-4,0) -- (-7,0);
\draw [ultra thick] (4,0) -- (7,0);
\draw [ultra thick] (0,0) -- (1.5,0);
\draw (0, -7.0) node{$(b)$};
\end{scope}

\begin{scope}[shift={(60,0)}]
\draw [ultra thick] (-2.6,-2.4) -- (-7,-4);
\draw [ultra thick] (-2.6,2.4) -- (-7,4);
\draw [ultra thick] (2.6,-2.4) -- (7,-4);
\draw [ultra thick] (2.6,2.4) -- (7,4);
\draw [ultra thick] (0,0) ellipse (4 and 3);
\draw [ultra thick] (0,3) -- (0,-3);
\draw [ultra thick] (0,0) -- (1.5,0);
\draw (0, -7.0) node{$(c)$};
\end{scope}

\begin{scope}[shift={(0,-18)}]
\begin{scope}[rotate=0]   
\end{scope}
\draw [ultra thick] (0,0) ellipse (4 and 3);
\draw [ultra thick] (2.6,-2.4) -- (7,-4);
\draw [ultra thick] (2.6,2.4) -- (7,4);
\draw [ultra thick] (-2.6,2.4) -- (-7,4);
\draw [ultra thick] (-2.6,-2.4) -- (-5,-3);
\draw [ultra thick] (-5,-3) -- (-7,-4);
\draw [ultra thick] (-5,-3) -- (-7,-2);
\draw [ultra thick] (0,3) -- (0,-3);
\draw (0, -7.0) node{$(a')$};
\end{scope}

\begin{scope}[shift={(30,-18)}]
\begin{scope}[rotate=0]   
\end{scope}
\draw [ultra thick] (0,0) ellipse (4 and 3);
\draw [ultra thick] (-2.6,2.4) -- (-7,4);
\draw [ultra thick] (-2.6,-2.4) -- (-5,-3);
\draw [ultra thick] (4,0) -- (7,0);
\draw [ultra thick] (0,0) -- (1.5,0);
\draw [ultra thick] (-5,-3) -- (-7,-4);
\draw [ultra thick] (-5,-3) -- (-7,-2);
\draw [ultra thick] (0,3) -- (0,-3);
\draw (0, -7.0) node{$(b')$};
\end{scope}

\begin{scope}[shift={(60,-18)}]
\begin{scope}[rotate=0]   
\end{scope}
\draw [ultra thick] (0,0) ellipse (4 and 3);
\draw [ultra thick] (2.6,-2.4) -- (7,-4);
\draw [ultra thick] (2.6,2.4) -- (7,4);
\draw [ultra thick] (-5,0) -- (-4,0);
\draw [ultra thick] (0,0) -- (1.5,0);
\draw [ultra thick] (-5,0) -- (-7,-1);
\draw [ultra thick] (-5,0) -- (-7,1);
\draw [ultra thick] (0,3) -- (0,-3);
\draw (0, -7.0) node{$(c')$};
\end{scope}
\end{tikzpicture}
\end{center}
\caption{All maximal degeneration graphs of class (1), namely containing no subgraphs with one, two, or three external edges. \label{fig:2}}
\label{classone}
\end{figure}

All the graphs in class (1) are  represented in Figure \ref{fig:2} and, by inspection, are seen to be in one-to-one correspondence with the field theory graphs of Figure \ref{fig:1}. The graphs in class (2) correspond to field theory graphs that vanish in view of the extended supersymmetry of the corresponding supergravity or super-Yang-Mills theory, a property that is sometimes referred to as ``no bubble or triangles" \cite{Bern:1994zx}. In both Type II and Heterotic superstring theories, on-shell amplitudes with one, two, or three external massless states are expected to vanish. General arguments to this effect have been given in \cite{Martinec:1986wa,Witten:2013cia} while the result was proven by explicit calculation in the genus-two case in \cite{D'Hoker:2005jc} for both Type II and Heterotic strings. Our proof here that the genus-two five-point amplitude reduces to the corresponding supergravity amplitude in the $\alpha ' \to 0$ limit, will be based on showing that the diagrams of class (1) precisely match those of field theory and that those of class (2) vanish.

\begin{figure}[h]
\begin{center}
\begin{tikzpicture}[scale=.15]

\begin{scope}[shift={(0,0)}]
\draw [ultra thick] (-2.6,-2.4) -- (-7,-4);
\draw [ultra thick] (-2.6,2.4) -- (-7,4);
\draw [ultra thick] (-3.6,-1.5) -- (-7,-2);
\draw [ultra thick] (-3.6,1.5) -- (-7,2);
\draw [ultra thick] (0,0) ellipse (4 and 3);
\draw [ultra thick] (0,3) -- (0,-3);
\draw [ultra thick] (-4,0) -- (-7,0);
\end{scope}

\begin{scope}[shift={(25,0)}]
\draw [ultra thick] (-2.6,-2.4) -- (-7,-4.5);
\draw [ultra thick] (-2.6,2.4) -- (-7,4.5);
\draw [ultra thick] (-3.8,-1) -- (-7,-1.5);
\draw [ultra thick] (-3.8, 1) -- (-7,1.5);
\draw [ultra thick] (0,0) ellipse (4 and 3);
\draw [ultra thick] (0,3) -- (0,-3);
\draw [ultra thick] (4,0) -- (6,0);
\end{scope}

\begin{scope}[shift={(50,0)}]
\begin{scope}[rotate=180,shift={(-2,0)}]   
\draw [ultra thick] (0,0) .. 
controls (-8,9) and (-8,-9) .. 
(0,0);
\end{scope}
\draw [ultra thick] (0,0) .. 
controls (-8,9) and (-8,-9) .. 
(0,0);
\draw [ultra thick] (0,0) -- (2,0);
\draw [ultra thick] (-8,0) -- (-6,0);
\draw [ultra thick] (-5,-2.4) -- (-7,-4);
\draw [ultra thick] (-5,2.4) -- (-7,4);
\draw [ultra thick] (7,-2.4) -- (9,-4);
\draw [ultra thick] (7,2.4) -- (9,4);
\end{scope}

\begin{scope}[shift={(75,0)}]
\begin{scope}[rotate=180,shift={(-2,0)}]   
\draw [ultra thick] (0,0) .. 
controls (-8,9) and (-8,-9) .. 
(0,0);
\end{scope}
\draw [ultra thick] (0,0) .. 
controls (-8,9) and (-8,-9) .. 
(0,0);
\draw [ultra thick] (0,0) -- (2,0);
\draw [ultra thick] (-8,0) -- (-6,0);
\draw [ultra thick] (-5,-2.4) -- (-7,-4);
\draw [ultra thick] (-5,2.4) -- (-7,4);
\draw [ultra thick] (8,0) -- (10,0);
\draw [ultra thick] (1,0) -- (1,4);
\end{scope}
\end{tikzpicture}
\end{center}
\caption{Some of the maximal degeneration graphs of class (2), namely containing one or several subgraphs with one, two,  or three external edges, whose contributions to the genus-two amplitude with five massless external states vanish.}
\label{classtwo}
\end{figure}
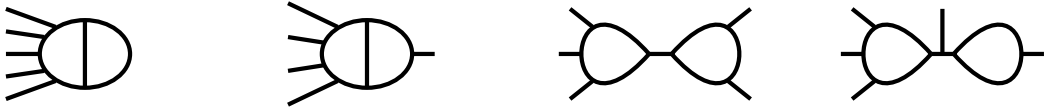

The Schwinger parameters $L_1,L_2,L_3$ for the  two-loop 1PI skeleton may be identified with the 
imaginary part $Y= \Im \Omega$ of the period matrix $\Omega$ via the relation \cite{Green:2005ba,Green:2008bf},  
\be
\label{YL123}
Y = \frac{1}{\alpha'} \, \begin{pmatrix} L_1+L_3 &-L_3 \\ -L_3 & L_2+L_3 \end{pmatrix}
\ee
in the limit $\alpha'\to 0$ holding the $L_i$'s fixed. The location of the external legs along the two loops gives five additional parameters $t_1,\dots, t_5$ lying in one of the intervals $[0,L_a]$, depending on the topology of the diagram. 
The topologies $a',b',c'$ where two external legs form a tree before attaching to the skeleton are included by allowing two of these parameters to coincide.

\subsection{Tropical limit of the Abelian differentials and prime form}

\begin{figure}[h]
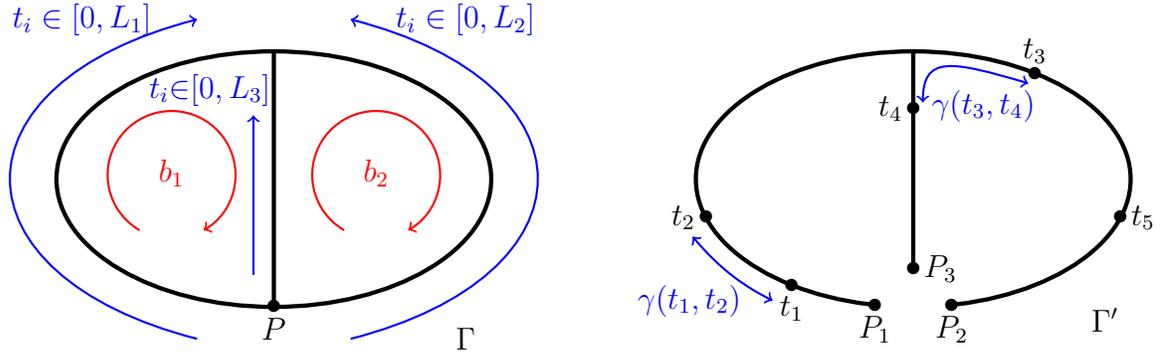

\begin{center}
\tikzpicture[scale=0.85]
\draw[ultra thick] (0,0) -- (0,4);
\draw [ultra thick] (0,2) ellipse (3.4 and 2);
\draw[thick,<-,red] (-1.1,1.2) arc(-60:240:1);
\draw[thick,<-,red] (2.1,1.2) arc(-60:240:1);
\draw[red](-1.6,2.1)node{$b_1$};
\draw[red](1.6,2.1)node{$b_2$};
\draw [thick,blue,->] (-1.2,-0.5) .. controls (-5.1,0.5)  and (-5.1,3.5) .. (-1.2,4.5);
\draw [thick,blue,->] (1.2,-0.5) .. controls (5.1,0.5)  and (5.1,3.5) .. (1.2,4.5);
\draw [thick,blue,->] (-0.32,0.5) -- (-0.32,3.0);
\draw[blue](-3,4.5)node{$t_i \in [0,L_1]$};
\draw[blue](3,4.5)node{$t_i \in [0,L_2]$};
\draw[blue](-1,3.4)node{$t_i {\in} [0,L_3]$};
\draw(0,0)node{$\bullet$} node[below]{$P$};
\draw(3,-0.5)node{$\Gamma$};
\begin{scope}[xshift=10cm]
\draw[ultra thick] (0,0) -- (0,4);
\draw [ultra thick] (0,2) ellipse (3.4 and 2);
%
\draw(-3.24,1.4)node{$\bullet$}node[left]{$t_2$};
\draw(-1.9,0.33)node{$\bullet$}node[below]{$t_1$};
\draw [thick,blue,<->] (-3.45,1.1) .. controls (-3,0.56) and (-2.6,0.3) .. (-2.2,0.1);
\draw(-3.5,0.1)node[blue]{$\g(t_1,t_2)$};
\draw(1.9,3.65)node{$\bullet$}node[above]{$t_3$};
\draw(-0,3.1)node{$\bullet$}node[left]{$t_4$};
\draw [thick,blue,<->] (1.8,3.5) .. controls (0.6,3.85) and (0.15,3.95) .. (0.15,3.2);
\draw(1.1,3.15)node[blue]{$\g(t_3,t_4)$};
\draw(3.24,1.4)node{$\bullet$}node[right]{$t_5$};
\draw[white, fill=white](-0.52,0.16) rectangle (0.52,-0.10);
\draw[white, fill=white](-0.1,0) rectangle (0.1,0.6);
\draw(0,0.6)node{$\bullet$} node[right]{$P_3$};
\draw(0.6,0.025)node{$\bullet$} node[below]{$P_2$};
\draw(-0.6,0.025)node{$\bullet$} node[below]{$P_1$};
\draw(3,-0.2)node{$\Gamma'$};
\end{scope}
   \endtikzpicture
  \end{center}
\caption{The left panel exhibits the two-loop 1PI skeleton graph $\Gamma$ with a choice of homology basis and parametrization. The right panel exhibits the simply connected graph $\Gamma'= \Gamma \setminus P$  obtained by removing one vertex~$P$ from $\Gamma$, and labeling $P_a$ the endpoint of the edge  $e_a$. On $\Gamma'$ each
pair of points $t_i,t_j$  is connected by a unique path $\g(t_i,t_j)$. When $t_i,t_j$ are on the same edge we have $\partial_{i} L(t_i,t_j)= \sgn(t_i-t_j)\, dt_i$, 
while when $t_i,t_j$ are on different  edges we have $\partial_{i} L(t_i,t_j)= - dt_i$. For the purpose of illustration, we have displayed vertices corresponding 
to (a permutation of) graph ($c$) in Figure \ref{fig:1}, the other graphs being analogous.}
\label{conven}
\end{figure}

Before analyzing the tropical limit of the string integrand, we review some basic results about the tropical limit of Abelian differentials and Green functions \cite{Tourkine:2013rda,DHoker:2017pvk,DHoker:2018mys}.  We choose a canonical homology basis of cycles $\mA_I$ and $\mB_I$ and conjugate normalized holomorphic Abelian differentials $\om_I$  on the Riemann surface $\Sigma$ (see appendix \ref{sec:C} for a summary). First, let  $b_I$ be a homology basis on the skeleton graph $\Gamma$ arising by degenerating the  homology basis $(\mA_I, \mB_I)\to (0,b_I)$ on $\Sigma$ (see  Figure \ref{conven}).  In the tropical limit, the Abelian differentials  scale as follows,
\be
\omega_I(z_j) \rightarrow  \frac{ i \, \omega\trop_I(t_j)}{\alpha'}
\label{tropomega}
\ee
where $\omega\trop_I$ is equal to 
 $\pm d t_j$ on the  edge $e_a$ if $e_a$ belongs to the cycle $b_I$, and 0 otherwise. The sign is fixed by the  orientation of $e_a$ with respect to the cycle $b_I$. For the choice of parametrization and  homology basis for the skeleton graph in Figure \ref{conven}, we have,
\bea
\omega\trop_1(z_j) =\left\{ \begin{array}{rl} {+}dt_j &: \  {\rm on} \ {\rm left} \ {\rm edge} \\
{-}dt_j &: \  {\rm on} \ {\rm middle} \ {\rm edge} \, ,\\
0 &: \  {\rm on} \ {\rm right} \ {\rm edge}  \end{array} \right.   \ \ \ 
\omega\trop_2(z_j) = \left\{ \begin{array}{rl} 0&: \  {\rm on} \ {\rm left} \ {\rm edge} \\
{+}dt_j &: \  {\rm on} \ {\rm middle} \ {\rm edge} \\
{-}dt_j &: \  {\rm on} \ {\rm right} \ {\rm edge}  \end{array} \right.
\label{detsgn}
\eea
The imaginary part of the period matrix $Y_{IJ}\sim \int_{b_I} \omega\trop_J/\alpha'$ reproduces \eqref{YL123} above.

In order to discuss the tropical limit of the prime form, careful account must be taken of the fact that the prime form is  a multi-valued form on $\Sigma \times \Sigma$. A single-valued representation may be obtained 
by considering the prime form on the simply connected domain obtained by fixing a base point $P$ on $\Sigma$ and then cutting $\Sigma$ along four canonical homology basis cycles $\mA_I$, $ \mB_I$ chosen to pass through $P$ (see e.g.\ Figure 12 in \cite{DHoker:1988pdl}).  In the tropical limit of a genus-two Riemann surface,  the point $P$ will lie at one of the vertices of the skeleton $\Gamma$ such that the graph  $\Gamma'=\Gamma \setminus P$ becomes simply connected \cite{Tourkine:2019ukp}, as shown in the right panel of Figure~\ref{conven} where the vertex $P$ has been replaced by endpoints $P_a$ for the open edges $e_a$. Between any two points $t_i, t_j \in \Gamma'$, corresponding to the  tropical limit of $z_i,z_j$ on $\Sigma$,  there is now a single path $\g(t_i,t_j)$ lying inside $\Gamma'$ \footnote{The path $\gamma(t_i,t_j)$ is not to be confused with the functions $\gamma(z|\Omega)$ which relate the string to the Arakelov Green functions in (\ref{GreenA}).}, such that the Abel-Jacobi map  scales like,
\be
(\zeta_i-\zeta_j)_I\rightarrow \frac{i}{\alpha'}  \zeta_{i,j,I}\trop
\hskip 1in
\zeta_{i,j,I}\trop= -\int_{\gamma(t_i,t_j)} \omega\trop_I
\ee
 in the tropical limit. As explained in \cite{Tourkine:2013rda}, the logarithm of the  prime form  then scales  as the length of the path,  
\bea
\label{tropE}
\ln E(z_i,z_j|\Omega) \rightarrow \frac{\pi}{\alpha'} L(t_i,t_j)
\eea
To establish this\footnote{We are grateful to Piotr Tourkine for helpful discussions on this matter.}  
one shows that, for an adapted choice of the odd spin structure $\nu=[\kappa',\kappa'']$, the theta series in  \eqref{hnudef} and \eqref{Edef} are dominated by a single vector $n$ in the sum \eqref{deftheta}, such that, 
\be
\label{tropEratio}
\ln E(z_i,z_j|\Omega) \rightarrow   \frac{2\pi}{\alpha'} |\zeta_{i,j}\trop \cdot \kappa'|
- \frac{1}{2} \ln | \omega\trop(t_i) \cdot \kappa' | - \frac{1}{2} \ln |  \omega\trop(t_j) \cdot \kappa'|
\ee
Here, ``adapted"  means that the two arguments of the logarithms, coming from the tropical limit of the
half-differentials, are non-zero. Whether a given spin structure is adapted  or not strongly depend on the positions $t_i,t_j$: e.g for the two paths in the right panel of Figure \ref{conven}, we have (omitting a factor $dt_i$ 
in the first three columns),
\bea
\hspace*{-5mm}
\begin{array}{|c||c|c|c||c|c|c|} 
\hline
\kappa' & \sigma_{1}, \sigma_2  & \sigma_3 & \sigma_4
& 2\zeta\trop_{1,2} \cdot \kappa' & 2\zeta\trop_{2,3} \cdot \kappa' & 2\zeta_{3,4}\trop \cdot \kappa' \\
\hline
(\frac12,0) & 1 & 0 & -1 & t_1-t_2 & t_2-L_1 & L_2-t_4\\
(0,\frac12) & 0 & -1 & 1 & 0 & t_3-L_3 & t_3+t_4-L_2-L_3\\
(\frac12,\frac12 ) & -1 & 1 & 0 &  t_1-t_2 & t_2+t_3-L_1-L_3& t_3-L_3 \\ 
\hline
\end{array}
\nonumber\\
\label{kaomtable}
\eea
where we have used the following abbreviations for $i=1,2,3,4$ in the table, 
\bea
\sigma_i = 2 \omega\trop(t_i)  \cdot \kappa' 
\eea
For the path $\g(t_1,t_2)$, the spin structures $(\frac12,0)$ and $(\frac12,\frac12)$ are both adapted,
and the first term in \eqref{tropEratio} is proportional to the length $L(t_i,t_j)$. For the path 
$\g(t_2,t_3)$, only the spin structure $(\frac12,\frac12)$ is adapted, and the same 
conclusion holds. 

\sm

For other spin structures, deemed ``not adapted", one of the combinations $\omega\trop(t_i)\cdot \kappa'$ or $\omega\trop(t_j)\cdot \kappa'$ or both in the arguments of the logarithms of \eqref{tropEratio} may vanish in taking the tropical limit naively. Instead, one must retain sub-leading corrections near the tropical limit. Since the prime form $E(z_i,z_j|\Omega)$ is independent of the choice of odd spin structure $\nu$, these sub-leading corrections must conspire to reproduce the behavior \eqref{tropE}. 

\sm

It follows from \eqref{tropE} that the one-form $\partial_i \ln E(z_i,z_j)$ reduces to $\pm \pi dt_i/\alpha '$ in the tropical limit, where the sign depends whether the variation $dt_i$ increases or decreases the length $L(t_i,t_j)$. With the conventions of Figure \ref{conven}, the sign is always negative  if the two points are on different edges (e.g.\ for the path $\g(t_3,t_4)$), while it depends on the sign of $t_i{-}t_j$ if the two points are on the same edge (e.g.\ for the path $\g(t_1,t_2)$). 

\sm

As a first application,  the tropical limit of the homology-invariant one-form \eqref{veccor} is given by, 
\bea
\label{veccorbis}
 \cP^m(z_i) ~ \longrightarrow ~
\frac{2 \pi}{\alpha'}   \Big(  {-}\ell^m  + \frac12 \sum_{j\in J} \sgn(t_i-t_j) k_j^m
 - \frac12 \sum _{j\notin J} k_j ^m  \Big) \, d t_i 
\eea
where $J$ is the set of external legs on the same edge as $i$ (we include $i$ in the set $J$, but set $\sgn(0)=0$), and $\ell^m$ is the loop momentum flowing through the point $i$ on the skeleton diagram (in absence of other external vertices). By momentum conservation, this can be rewritten as, 
\be
 \frac{2 \pi}{\alpha'}  \Big(  {-}\ell^m + \frac12 \sum_{j\in J} (1+\sgn(t_i-t_j))  k_j^m  \Big) \, d t_i 
\ee
which is recognized as the average of the  momenta flowing into and out of the vertex point $t_i$ along the graph $\Gamma'$.

\sm

As a second application, we consider the tropical limit of the function $g_{i,j}^I$ defined in \eqref{simp.3}, 
\bea
g^I_{i,j} = \frac{ \partial }{\partial \zeta_I}  \ln \tet [\nu ] (\zeta | \Omega) \Big |_{\zeta = \zeta _{i}-\zeta_{j}}
\hskip 1in 
(\zeta _{i}-\zeta_j)_{I} = \int _{z_j} ^{z_i} \om _I
\eea
Unlike the derivative of the prime form it has the antisymmetry property $g^I_{j,i} = - g^I_{i,j}$. For a choice of odd spin structure $\nu=[\kappa',\kappa'']$ such that $\zeta\trop_{i,j} \cdot \kappa'\neq 0$, the tropical  limit of the theta series $\ln \tet [\nu ] (\zeta | \Omega)$ is given by the first term in \eqref{tropEratio}, whose derivative with respect to $\zeta_{i,j}^I$  gives,
\be
g_{i,j}^I \rightarrow  -2i\pi \, \sgn(\zeta\trop_{i,j} \cdot \kappa')\, \kappa'_I
\ee  
One may check that this result is consistent with the relation \eqref{simp.2} in the tropical limit. 
For the specific choice of spin structure $(\frac12,\frac12)$ and any pair of points in the right panel of Figure  \ref{conven}, we conclude that the tropical limit of $g_{i,j}^I$ is independent on $I$ and given by, 
\bea
g_{i,j}^I \rightarrow \left\{ \begin{array}{cl}  
 + i\pi &: \ t_i,t_j \  \mbox{on distinct edges} \ (L_1, L_2), \ (L_1, L_3)  \ \mbox{or}\  (L_3, L_2)
\\
 - i\pi &: \ t_i,t_j \  \mbox{on distinct edges} \ (L_2, L_1), \ (L_3, L_1)  \ \mbox{or}\  (L_2, L_3)
\\
i  \pi \, \sgn(t_j-t_i)&: \ t_i,t_j  \ {\rm both} \ {\rm on} \ {\rm edge} \ L_1
\\
i  \pi \, \sgn(t_i-t_j)&: \ t_i,t_j  \ {\rm both} \ {\rm on} \ {\rm edge} \ L_2  \end{array} \right.
   \label{tropgij}
\eea
This conclusion would not hold for pairs of points on the middle edge of Figure \ref{conven}, as the contraction $\zeta\trop_{i,j}\cdot \kappa'$ would vanish in that case. The fact that \eqref{tropgij} is  independent on $I$ makes the spin structure $(\frac12,\frac12)$ particularly convenient, although one could in principle use any other odd spin structure.

\subsection{Tropical limit of the chiral integrand: pentaboxes}
\label{sec:62}

We shall now analyze the behavior of  the chiral integrand in the regime where the Abel-Jacobi map between the vertex points scales to infinity at the same rate   $\zeta_i-\zeta_j \sim i\zeta\trop_{i,j}/\alpha'$  as the period matrix $\Omega\sim Y/\alpha'$. This degeneration will turn out to reproduce precisely
the pentabox diagrams $(a,b,c)$ which occur both in supergravity and SYM theory.  Contact
terms responsible for the double-box diagrams $(a',b',c')$ require a discussion of  the full
integrand, which is deferred to the  next subsection. 

\sm

Recall that the chiral integrand is given by  \eqref{simp.7}, 
which we copy for convenience after cyclically permuting the legs,
\bea
{\cal K}_{(5)} = \omega_I(2) \Delta(3,4) \Delta(5,1) \,{\cal K}^I_{1,2,3|4,5} + {\rm cycl}(1,2,3,4,5)  
\label{simp.7bis}
\eea
where 
${\cal K}^I_{1,2,3|4,5}$ is the sub-correlator \eqref{simp.8}, cyclically permuted,
\begin{align}
{\cal K}^I_{1,2,3|4,5} &= 2\pi p_m^I T^m_{1,2,3|4,5} - g^I_{2,3} T_{23,1|4,5} - g^I_{2,1} T_{21,3|4,5}  - g^I_{3,1} T_{31,2|4,5}  \notag \\
&- g_{2,4}^I S_{2;4|5|1,3}  - g_{3,4}^I S_{3;4|5|2,1}  - g_{1,4}^I S_{1;4|5|2,3}  \label{simp.8bis} 
\no \\
&- g_{2,5}^I S_{2;5|4|3,1}  - g_{3,5}^I S_{3;5|4|2,1}  - g_{1,5}^I S_{1;5|4|2,3} 
\end{align}
where we recall that $\Delta(i,j)$ is the bi-holomorphic $(1,0)$ form \eqref{Del}. 

\sm

In the tropical limit, $\Delta(i,j)$ vanishes by antisymmetry if the vertices $t_i$, $t_j$ lie on the same edge of the skeleton diagram, and reduces to $\pm dt_i\, dt_j$ otherwise with the sign determined by (\ref{detsgn}). 
This implies that the three edges of the graph can carry (3,2,0), (3,1,1) or $(2,2,1)$ external legs
and therefore rules out the first two graphs in Figure \ref{classtwo} with bubble and triangle
subdiagrams. The third and fourth graph of Figure \ref{classtwo} in turn involve bubble and triangle
subdiagrams within a 1PR skeleton and drop out from the field theory limit for a different reason: 
Graphs obtained from the 1PR vacuum graph in
the right panel of figure \ref{pivspr} cannot contribute by unitarity as a consequence of the
non-renormalization theorems for three-point functions of on-shell massless states
at one loop \cite{Green:1982sw} and two loops \cite{D'Hoker:2005jc}.

\sm

We shall assign the external legs such that, for the  odd spin structure $\kappa'=(\frac12,\frac12)$, the inner product 
$\zeta\trop_{i,j}\cdot \kappa'$ in (\ref{tropEratio}) is non-zero for all pairs of points, so that \eqref{tropgij} applies. This is for convenience only, since  the result cannot depend on the choice of $\kappa'$ since the correlator (\ref{simp.7bis}) is expressible in terms of prime forms, see (\ref{simp.35}) or (\ref{altrep}), which are independent of the spin structure.
At the same time, the tropical limit of (\ref{simp.7bis}) is unaffected by the vanishing of certain $\omega(t_j)\cdot \kappa'$ in (\ref{kaomtable}) since they descend from the $(\frac{1}{2},0)$-forms $h_\nu(z_j)$ that were shown to 
cancel from ${\cal K}_{(5)}$ in section \ref{sec:simp.1}. 

\sm

Consider first the case where  the external legs are distributed as in the planar pentabox ($a$) of Figure \ref{fig:1}. By (\ref{detsgn}), the Abelian differentials $\omega_I(z_j)$ reduce to 
\bea
\tikzpicture
\draw(-2,0.5)node{$(a) \, :$};
\draw [thick] (0,-0.1) -- (0,1.1);
\draw [thick] (0,0.5) ellipse (0.7 and 0.6);
\draw (-0.5,0.07)node{$\bullet$}node[left]{$t_1$};
\draw (-0.7,0.5)node{$\bullet$}node[left]{$t_2$};
\draw (-0.5,0.93)node{$\bullet$}node[left]{$t_3$};
\draw (0.6,0.8)node{$\bullet$}node[right]{$t_4$};
\draw (0.6,0.2)node{$\bullet$}node[right]{$t_5$};
\draw(2.1,0.5)node{$\Longrightarrow$};
\draw(6.5,0.5)node{$\displaystyle \begin{pmatrix} \omega_1(z_j)  \\ \omega_2(z_j)   \end{pmatrix} 
 \stackrel{(a)} {\longrightarrow}  \begin{pmatrix} 
 1 & 1 & 1 & 0 & 0 \\
0 & 0 & 0& -1 & -1 \end{pmatrix} \times \frac{i\, dt_j}{\alpha'} $};
\endtikzpicture 
%
\eea
Thus the only non-vanishing term in the sum over cyclic permutations in \eqref{simp.7bis}
is the first one proportional to $\omega_I(2)  \Delta(3,4) \Delta(5,1)$ with $\omega_I(2) \rightarrow 
i\delta_{I,1}dt_2/\alpha'$, namely
\be
{\cal K}_{(5)}  \stackrel{(a)} {\rightarrow} -\frac{ i }{(\alpha')^5} {\cal K}_{1,2,3|4,5}^1 \, dt_1\dots d t_5
 \stackrel{(a)} {\rightarrow} - \frac{2\pi }{(\alpha')^5} {\cal N}^{(a)}_{1,2,3|4,5}(\ell)  \, dt_1\dots d t_5
 \label{plusnum}
\ee
with
\begin{align}
{\cal N}^{(a)}_{1,2,3|4,5}(\ell)  &=  i p^1_m T^m_{1,2,3|4,5} + \frac12 \left( T_{23,1|4,5}  + T_{12,3|4,5}  + T_{13,2|4,5} \right)  \notag \\
&\ \ \ \ + \frac12 \left(  S_{2;4|5|1,3}  + S_{3;4|5|2,1}  + S_{1;4|5|2,3}
+ S_{2;5|4|3,1}  + S_{3;5|4|2,1}  + S_{1;5|4|2,1} \right) 
\label{K5sugraa}  \\
&= i \Big( p^1_m - \frac{1}{2} (k_1{+}k_2{+}k_3)_m \Big) T^m_{1,2,3|4,5}  + \frac12 \left( T_{23,1|4,5}  + T_{12,3|4,5}  + T_{13,2|4,5} \right)  \notag
\end{align}
One can identify $p^1_m$ with the loop momentum $\ell$ in Figure \ref{fig:1} ($a$) which
is in the lower end of the edge supporting the external particles $1,2,3$. 
The combination $(k_1{+}k_2{+}k_3)_mT^m_{1,2,3|4,5}$ is obtained
from the six permutations of $S_{2;4|5|1,3} $ via (\ref{multi.9}). Up to a global rescaling
of internal and external momenta by a factor of $i$ which was left implicit in \cite{Mafra:2015mja}, this is in precise agreement with the numerator for the diagram ($a$) computed in that reference.

\sm

Next, consider the case where  the external legs are distributed as in the non-planar pentabox ($b$) of Figure \ref{fig:1}. By (\ref{detsgn}), the Abelian differentials $\omega_I(z_j)$ now reduce to 
\be
\tikzpicture
\draw(-2,0.5)node{$(b) \, :$};
\draw [thick] (0,-0.1) -- (0,1.1);
\draw [thick] (0,0.5) ellipse (0.7 and 0.6);
\draw (-0.5,0.07)node{$\bullet$}node[left]{$t_1$};
\draw (-0.7,0.5)node{$\bullet$}node[left]{$t_2$};
\draw (-0.5,0.93)node{$\bullet$}node[left]{$t_3$};
\draw (0.7,0.5)node{$\bullet$}node[right]{$t_5$};
\draw (0,0.5)node{$\bullet$}node[right]{$t_4$};
\draw(2.1,0.5)node{$\Longrightarrow$};
\draw(6.5,0.5)node{$\displaystyle \begin{pmatrix} \omega_1(z_j)   \\ \omega_2(z_j)    \end{pmatrix} 
 \stackrel{(b)}{\rightarrow}
 \begin{pmatrix} 1 & 1 & 1 & -1 & 0 \\
0 & 0 & 0& 1 &- 1\end{pmatrix} \times \frac{i\, dt_j}{\alpha'} $};
\endtikzpicture 
\ee
The only non-vanishing term in the sum over cyclic permutations in \eqref{simp.7bis}
is again the first one proportional to $\omega_I(2) \Delta(3,4) \Delta(5,1)$ with 
$\omega_I(2)\rightarrow i\delta_{I,1}dt_2/\alpha'$, leading to the same 
integrand as in \eqref{K5sugraa} up to an overall sign from the fourth column,
\be
\label{Nba}
{\cal K}_{(5)}  \stackrel{(b)} {\rightarrow} \frac{i}{(\alpha')^5} {\cal K}_{1,2,3|4,5}^1 \, dt_1\dots d t_5
 \stackrel{(b)} {\rightarrow} - \frac{2\pi }{(\alpha')^5} {\cal N}^{(b)}_{1,2,3|4,5}(\ell)  \, dt_1\dots d t_5
\ee
with 
\bea
{\cal N}^{(b)}_{1,2,3|4,5}(\ell)  = - {\cal N}^{(a)}_{1,2,3|4,5}(\ell) 
\label{pentarel}
\eea
The tropical limit of $ {\cal K}_{1,2,3|4,5}^1$ is identical in the cases of ($a$) and ($b$) since $g^I_{4,5}$ does
not occur in (\ref{simp.8bis}).
The non-planar pentabox numerator (\ref{pentarel}) is again in precise agreement with the numerator for the diagram ($b$) computed in \cite{Mafra:2015mja}.

\sm

Finally, let consider the case where  the external legs are distributed as in the non-planar pentabox ($c$) 
of Figure \ref{fig:1} (also see the right panel of Figure \ref{conven}). The Abelian differentials $\omega_I(z_j)$ 
now reduce to,
\be
\tikzpicture
\draw(-2,0.5)node{$(c) \, :$};
\draw [thick] (0,-0.1) -- (0,1.1);
\draw [thick] (0,0.5) ellipse (0.7 and 0.6);
\draw (-0.6,0.8)node{$\bullet$}node[left]{$t_2$};
\draw (-0.6,0.2)node{$\bullet$}node[left]{$t_1$};
\draw (0.6,0.8)node{$\bullet$}node[right]{$t_4$};
\draw (0.6,0.2)node{$\bullet$}node[right]{$t_3$};
\draw (0,0.5)node{$\bullet$}node[right]{$t_5$};
\draw(2.1,0.5)node{$\Longrightarrow$};
\draw(6.5,0.5)node{$\displaystyle \begin{pmatrix} \omega_1(z_j) \\ \omega_2(z_j)   \end{pmatrix} 
 \stackrel{(c)}{\rightarrow}
\begin{pmatrix} 1 & 1 & 0& 0 & -1 \\
0 & 0 & -1 & -1 & 1\end{pmatrix} \times \frac{i\, dt_j}{\alpha'}  $};
\endtikzpicture 
\ee
There are now two non-vanishing terms in the sum over cyclic permutations in \eqref{simp.7bis}, namely
$ \omega_I(1) \Delta(2,3) \, \Delta (4,5)$ and $ \omega_I(4) \Delta(5,1) \, \Delta (2,3)$,
\bea 
{\cal K}_{(5)}  \stackrel{(c)} {\rightarrow} \frac{i}{(\alpha')^5} ({\cal K}_{5,1,2|3,4}^1 -{\cal K}_{3,4,5|1,2}^2)\, dt_1\dots d t_5
 \stackrel{(c)} {\rightarrow}  \frac{2\pi}{(\alpha')^5} {\cal N}^{(c)}_{1,2|4,3|5}(\ell,r)  \, dt_1\dots d t_5
 \label{signissues}
\eea
with loop momenta $p^1= \ell$ as well as $p^2= - r$
in Figure \ref{fig:1} ($c$) and
\be
\label{Nca}
 {\cal N}^{(c)}_{1,2|4,3|5}(\ell,r) = \cN^{(a)}_{1,2,5|3,4}(p^1) + \cN^{(a)}_{3,4,5|1,2}(-p^2)
 = \cN^{(a)}_{1,2,5|3,4}(\ell) + \cN^{(a)}_{3,4,5|1,2}(r)\
\ee
again in precise agreement with the numerator for the diagram ($c$) computed in \cite{Mafra:2015mja}.
The degenerations ${\cal K}_{5,1,2|3,4}^1 \rightarrow  -2\pi i \cN^{(a)}_{1,2,5|3,4}(\ell)$ and 
${\cal K}_{3,4,5|1,2}^2 \rightarrow 2\pi i\cN^{(a)}_{3,4,5|1,2}(r)$ are obtained by repeating the steps of (\ref{K5sugraa})
which are now sensitive to all the five cases of $g^I_{i,j}$ covered in (\ref{tropgij}). The change of orientation
in $p^2= - r$ stems from the fact that the definition (\ref{defloopmom}) of loop momenta via $\mA_I$-cycle integrals 
leads to both of $p^1$ and $p^2$ pointing to the left in Figure \ref{fig:1} ($c$), whereas $r$ is drawn to point to the right. Moreover, note the relative sign between the right-hand sides of (\ref{signissues}) and (\ref{plusnum}), 
(\ref{Nba}) in identifying the numerators: This sign reflects the orientation of leg 5 in Figure \ref{fig:1} ($c$) 
whether its external edge points to the left or right and drops out from the gravity numerator 
$ {\cal N}_{1,2|4,3|5}^{(c)}(\ell,r) \tilde {\cal N}_{1,2|4,3|5}^{(c)}(\ell,r)$ that we are deriving from 
the tropical limit.

\sm

Note that the relations (\ref{pentarel}) and \eqref{Nca} among pentabox numerators are the kinematic Jacobi identities which are consequences of color-kinematics  duality \cite{Carrasco:2011mn}.  In our setup, the kinematic Jacobi identities among $\cN^{(a)}, \, \cN^{(b)}, \, \cN^{(c)}$ follow from the degenerations of the five-forms in the correlator (\ref{simp.7bis}) and the tropical limit (\ref{tropgij}) of $g^I_{i,j}$.

\subsection{Tropical limit of the Type II string integrand: double boxes}
\label{sec:63}

Scattering amplitudes in Type II strings involve an integral  \eqref{pairing} of the product 
${\cal K}_{(5)}\tilde {\cal K}_{(5)} |\cI_{(5)}|^2$ over the loop momentum, vertex points $z_i$ and complex
structure moduli parametrized by $\Omega$. As we review in subsection \ref{sec_assemble} below, 
the tropical limit of the chiral integrand discussed in the previous subsection reproduces exactly the 
contribution of the pentabox diagrams $(a,b,c)$ in Figure \ref{fig:1}. However, there are additional  
contributions from maximal degenerations of the genus-two Riemann surface where two punctures 
collide, which are responsible for the double-box diagrams $(a',b',c')$, as we now show.

\sm

Due to short-distance singularities in the chiral integrand arising from derivatives of the prime form,
\be
\partial_{z_i} \ln E(z_i,z_j) = \frac{1}{z_i{-}z_j}+{\cal O}(z_i{-}z_j)
\ee
the integral  of the product ${\cal K}_{(5)}\tilde {\cal K}_{(5)} |\cI_{(5)}|^2$ 
over vertex points $z_i$ is not finite in the low energy expansion, 
but rather has kinematical poles of the form
\bea
 \int_{|z|<R} d^2 z\, |z|^{-2s-2} \, f(z) = - \pi \frac{f(0)}{s}+ {\cal O}(s^0)
\label{dbox.1}
\eea
where we assume that the function $f(z)$ is continuous at the origin. 
The ${\cal O}(s^0)$ term depends on the radius $R>0$ used to excise the singularity at $z=0$, but does not contribute to the field theory limit  at leading order and can  be ignored. 

\sm
The coefficients of the kinematic poles can be computed by collecting the four possible
sources of poles of the form $1/ |z_i {-}z_j|^2$, and performing the replacement, 
\bea
\left. \begin{array}{r}
|\partial_{z_i} \ln E(z_i,z_j) |^2  \\ 
\bigg. -\partial_{z_i} \ln E(z_i,z_j) \overline{\partial_{z_j} \ln E(z_j,z_i)}
\end{array} \right\} \sim \frac{1}{|z_i {-}z_j|^2} 
\rightarrow -\frac{ \pi \delta^2(z_i,z_j) }{s_{ij}}
\label{dbox.2}
\eea
Note that products of prime forms with different arguments $\partial_{z_i} \ln E(z_i,z_j)\overline{\partial_{z_i} \ln E(z_i,z_k)}$ with $k\neq j$ do not  lead to any kinematical pole since the resulting singularity $(z_i-z_j)^{-1}(\bar z_i-\bar z_k)^{-1}$ integrates to zero after  integration
over the phase of $z_{i}{-}z_j$. Moreover, maximal degenerations with three of more punctures colliding do not
contribute to the field theory limit at five points since they would require more than one prime form in the
chiral correlators such that the integration rule (\ref{dbox.1}) can be used multiple times.

\sm

The singularities $(z_{1}{-}z_2)^{-1}$ of the chiral correlator were already extracted in
(\ref{res12}) based on the representation (\ref{simp.35}). The residue  at $s_{12}=0$ of the 
relevant chiral contributions is given by, 
\bea
{\cal K}_{(5)}^{12} = {\rm Res}_{z_1 \rightarrow z_2} {\cal K}_{(5)}  = T_{12,3|4,5} \Delta(2,4) \Delta(3,5) + T_{12,4|3,5} \Delta(2,3) \Delta(4,5)
\label{dbox.3}
\eea
which is permutation symmetric in $3,4,5$, by virtue of the symmetries (\ref{symmT}) 
and (\ref{Delsym}). Hence, the graphs where the  vertices $1$ and $2$ collide are 
captured by applying the replacement (\ref{dbox.2}) to,
\begin{align}
{\cal K}_{(5)}\tilde {\cal K}_{(5)} &\rightarrow \frac{ 
{\cal K}_{(5)}^{12}  \tilde {\cal K}_{(5)}^{12}  }{ |z_1 {-}z_2|^2}
\rightarrow -\frac{ \pi \delta^2(z_1,z_2) }{s_{12}}{\cal K}_{(5)}^{12}  \tilde {\cal K}_{(5)}^{12}
\label{dbox.4}
\end{align}

\sm

We will now extract the chiral contributions to double-box numerators for diagrams ($a'$), ($b'$), ($c'$) in 
Figure \ref{fig:1}. Given that the chiral contribution (\ref{dbox.3}) shares the structure of the four-point
correlator (\ref{4ptcorel}), the computations below closely follow the tropical limit of the two-loop four-point amplitude in \cite{Tourkine:2013rda}.

\sm

In the planar case ($a'$), the abelian differentials $\omega_I(z_j)$
with $j=2,3,4,5$ reduce to (see (\ref{detsgn})) 
\bea
\tikzpicture
\draw(-2.3,0.5)node{$(a') \, :$};
\draw [thick] (0,-0.1) -- (0,1.1);
\draw [thick] (0,0.5) ellipse (0.7 and 0.6);
\draw (-0.6,0.8)node{$\bullet$}node[left]{$t_3$};
\draw (-0.6,0.2)node{$\bullet$}node[left]{$t_1{,}t_2$};
\draw (0.6,0.8)node{$\bullet$}node[right]{$t_4$};
\draw (0.6,0.2)node{$\bullet$}node[right]{$t_5$};
\draw(2.1,0.5)node{$\Longrightarrow$};
\draw(6.5,0.5)node{$\displaystyle \begin{pmatrix} \omega_1(z_j)   \\ \omega_2(z_j)  \end{pmatrix} 
 \stackrel{(a')} {\rightarrow}  \begin{pmatrix}  1& 1 & 0 & 0 \\
 0 & 0&-1& -1 \end{pmatrix} \times \frac{i\, dt_j}{\alpha'}  $};
\endtikzpicture 
 \label{dbox.5}
\eea
and (\ref{dbox.3}) reduces to the first term $ \Delta(2,4) \Delta(3,5) \rightarrow dt_2\ldots dt_5/(\alpha')^4$.
The resulting numerator agrees with the result of \cite{Mafra:2015mja} (denoted by 
${\cal N}^{(d)}_{12,3|4,5}(\ell) $ in the reference)
\be
(\alpha')^4{\cal K}^{12}_{(5)}  \stackrel{(a')} {\rightarrow} T_{12,3|4,5} \, dt_2\dots d t_5
 \stackrel{(a')} {\rightarrow} {\cal N}^{(a')}_{12,3|4,5}(\ell)  \, dt_2\dots d t_5
 \label{dbox.6}
\ee
Moreover, this expression for planar double-box numerators matches antisymmetric 
combinations of planar pentabox numerators $ {\cal N}^{(a)}$ in (\ref{K5sugraa})
\be
{\cal N}^{(a')}_{12,3|4,5}(\ell)= T_{12,3|4,5} = {\cal N}^{(a)}_{1,2,3|4,5}(\ell) - {\cal N}^{(a)}_{2,1,3|4,5}(\ell)
 \label{dbox.7}
\ee
and therefore realizes another kinematic Jacobi identity required by the color-kinematics duality~\cite{Carrasco:2011mn}.

\sm

The above steps can be repeated to determine the non-planar double-box numerators for diagrams ($b'$) and ($c'$) in Figure \ref{fig:1}. The degeneration of the Abelian differentials, 
\begin{align}
\tikzpicture
\draw(-2.3,0.5)node{$(b') \, :$};
\draw [thick] (0,-0.1) -- (0,1.1);
\draw [thick] (0,0.5) ellipse (0.7 and 0.6);
\draw (-0.6,0.8)node{$\bullet$}node[left]{$t_3$};
\draw (-0.6,0.2)node{$\bullet$}node[left]{$t_1{,}t_2$};
\draw (0.7,0.5)node{$\bullet$}node[right]{$t_5$};
\draw (0,0.5)node{$\bullet$}node[right]{$t_4$};
\draw(2.1,0.5)node{$\Longrightarrow$};
\draw(6.5,0.5)node{$\displaystyle \begin{pmatrix} \omega_1(z_j)  \\ \omega_2(z_j)    \end{pmatrix} 
 \stackrel{(b')} {\rightarrow}  \begin{pmatrix}  1 & 1 & -1 & 0 \\
 0 & 0& 1 & -1\end{pmatrix} \times \frac{i\, dt_j}{\alpha'}  $};
\scope[yshift=-2cm]
\draw(-2.3,0.5)node{$(c') \, :$};
\draw [thick] (0,-0.1) -- (0,1.1);
\draw [thick] (0,0.5) ellipse (0.7 and 0.6);
\draw (-0.7,0.5)node{$\bullet$}node[left]{$t_1{,}t_2$};
\draw (0.6,0.8)node{$\bullet$}node[right]{$t_4$};
\draw (0.6,0.2)node{$\bullet$}node[right]{$t_5$};
\draw (0,0.5)node{$\bullet$}node[left]{$t_3$};
\draw(2.1,0.5)node{$\Longrightarrow$};
\draw(6.5,0.5)node{$\displaystyle  \begin{pmatrix} \omega_1(z_j)   \\ \omega_2(z_j) \end{pmatrix} 
 \stackrel{(c')} {\rightarrow}  \begin{pmatrix}  1& -1 & 0 & 0 \\
 0 & 1& -1 & -1\end{pmatrix} \times \frac{i\, dt_j}{\alpha'}  $};
 \endscope
\endtikzpicture 
%
 \label{dbox.8}
\end{align}
again suppresses the second term $\sim  \Delta(2,3) \Delta(4,5)$ in (\ref{dbox.3}),
and we obtain an extra minus sign in $\Delta(2,4) \Delta(3,5) \rightarrow -dt_2\ldots dt_5/(\alpha')^4$
as compared to the planar case (\ref{dbox.5}). Hence, the tropical limit of the correlator 
for diagrams ($b'$), ($c'$) is
\be
(\alpha')^4{\cal K}^{12}_{(5)}  \stackrel{(b')} {\rightarrow} -T_{12,3|4,5} \, dt_2\dots d t_5 \, , \ \ \ \ \ \
(\alpha')^4{\cal K}^{12}_{(5)}  \stackrel{(c')} {\rightarrow} -T_{12,3|4,5} \, dt_2\dots d t_5
 \label{dbox.9}
\ee
and one can read off the non-planar double-box numerators
\be
{\cal N}^{(b')}_{12,3|4,5}(\ell) =  {\cal N}^{(c')}_{12,3|4,5}(\ell) = -  T_{12,3|4,5} =  - {\cal N}^{(a')}_{12,3|4,5}(\ell)
 \label{dbox.10}
\ee
They reproduce the numerators of \cite{Mafra:2015mja} (denoted by 
${\cal N}^{(e)}_{12,3|4,5}(\ell), \,{\cal N}^{(f)}_{12,3|4,5}(\ell)$ in the reference)
and obey the color-kinematics duality when comparing with non-planar pentabox
numerators. Also note that the symmetry of ${\cal N}^{(a')}_{12,3|4,5}(\ell),\,
{\cal N}^{(b')}_{12,3|4,5}(\ell),\,{\cal N}^{(c')}_{12,3|4,5}(\ell)$ under $4\leftrightarrow 5$ 
is consistent with the vanishing of numerators associated with triangle-subgraphs.

\subsection{Assembling the supergravity amplitude}
\label{sec_assemble}

Collecting the results in the previous two subsections, we find that the field theory limit
of the genus-two scattering amplitude in Type II strings precisely produces the 
complete two-loop five-point amplitude in maximal supergravity in $D$ dimensions, 
in the double-copy representation of  \cite{Mafra:2015mja} (with the 
structure of \cite{Carrasco:2011mn}),
\begin{align}
{\cal A}^{\rm SG}_5&= \delta\Big( \sum_{j=1}^5 k_j \Big) \int_{\mathbb R^{2D}} \Big \langle
{1\over 2}  {\cal N}^{(a)}_{1,2,3|4,5}(\ell) \tilde {\cal N}^{(a)}_{1,2,3|4,5}(\ell)   I^{(a)}_{1,2,3,4,5}
+ {1\over 4}  {\cal N}^{(b)}_{1,2,3|4,5}(\ell) \tilde {\cal N}^{(b)}_{1,2,3|4,5}(\ell)  I^{(b)}_{1,2,3,4,5} \notag \\
&\phantom{=}+ {1\over 4}  {\cal N}_{1,2|4,3|5}^{(c)}(\ell,r) \tilde {\cal N}_{1,2|4,3|5}^{(c)}(\ell,r)  I^{(c)}_{1,2,3,4,5}
+ {1\over 2}  {\cal N}^{(a')}_{12,3|4,5} \tilde {\cal N}^{(a')}_{12,3|4,5}  I^{(a')}_{1,2,3,4,5} 
 \label{delttt} \\
&\phantom{=}+ {1\over 4}  {\cal N}^{(b')}_{12,3|4,5} \tilde {\cal N}^{(b')}_{12,3|4,5} I^{(b')}_{1,2,3,4,5}
+ {1\over 4}  {\cal N}^{(c')}_{12,3|4,5} \tilde {\cal N}^{(c')}_{12,3|4,5} I^{(c')}_{1,2,3,4,5}
+ {\rm sym}(1,2,3,4,5)  \Big \rangle_0 d\ell \, dr  \notag
\end{align}
Here, the symmetry factors ${1\over 2}$ and ${1\over 4}$ ensure that the sum over
$5!$ permutations of the external legs does not overcount individual diagrams. The factors 
$I_{1,2,3,4,5}^{(x)}$ are the usual products of Feynman propagators for the diagrams 
in Figure \ref{fig:1},
\begin{align}
I^{(a)}_{1,2,3,4,5}&\equiv  { 1 \over \ell^2 r^2 (\ell+r)^2 (\ell-k_1)^2 (\ell-k_{12})^2 (\ell-k_{123})^2 \, (r-k_5)^2 (r-k_{45})^2}
\notag \\
I^{(b)}_{1,2,3,4,5}&\equiv  { 1 \over \ell^2 r^2  (\ell+r)^2 (\ell-k_1)^2 (\ell-k_{12})^2 (\ell-k_{123})^2 \, (r-k_5)^2 (\ell+r+k_{4})^2}
\notag \\
I^{(c)}_{1,2,3,4,5}&\equiv  { 1 \over \ell^2 r^2  (\ell+r)^2(\ell-k_1)^2 (\ell-k_{12})^2  \, (r-k_3)^2(r-k_{34})^2 (\ell+r+k_{5})^2}
\label{allints}
\\
I^{(a')}_{1,2,3,4,5}&\equiv  { 1 \over k_{12}^2 \ell^2 r^2  (\ell+r)^2 (\ell-k_{12})^2 (\ell-k_{123})^2 \, (r-k_5)^2 (r-k_{45})^2 }
\notag \\
I^{(b')}_{1,2,3,4,5}&\equiv  { 1 \over k_{12}^2 \ell^2 r^2  (\ell+r)^2 (\ell-k_{12})^2 (\ell-k_{123})^2 \, (r-k_5)^2 (r+\ell+k_{4})^2}
\notag \\
I^{(c')}_{1,2,3,4,5}&\equiv  { 1 \over k_{12}^2  \ell^2 r^2  (\ell+r)^2(\ell-k_{12})^2 (\ell+r+k_{3})^2 \, (r-k_5)^2 (r-k_{45})^2}
\notag 
\end{align}
The zero-mode integral $\langle \ldots \rangle_0$ in (\ref{pssbracket}) yields the
components of the superspace numerators for arbitrary external states of the 
ten-dimensional Type-II multiplets, see \cite{PSSsite} for the bosonic components of
$T^m_{1,2,3|4,5}$ and $T_{12,3|4,5}$. 

\sm

The supergravity amplitude (\ref{delttt}) has been given for general spacetime dimension $D$
by considering a compactification on a $T^{10-D}$-torus and retaining only the zero-momentum
and -winding modes in the Siegel-Narain theta series (\ref{defGamma}). The superspace components
of the kinematic factors in (\ref{delttt}) can be dimensionally reduced
to any $D\leq 10$ and integrated over the loop momenta in $D<7$, where the
integrals are UV-finite. Dimensional reduction to $D=4$ does not directly reproduce
the BCJ numerators of \cite{Carrasco:2011mn} in spinor-helicity variables since their
building blocks $\gamma_{ij}$ involve certain inverse Levi-Civita invariants that are specific
to four dimensions. Still, the symmetry properties of the combinations of  $\gamma_{ij}$
in \cite{Carrasco:2011mn} match those of the superspace building blocks in  (\ref{delttt}),
see appendix D of \cite{Mafra:2015mja} for details. The difference between the amplitude 
representation in \cite{Carrasco:2011mn} and the dimensionally reduced superspace 
numerators of (\ref{delttt}) should cancel when integrating the sum over all diagrams,
for instance using the recent progress on the relevant integrals 
in \cite{Abreu:2018aqd, Chicherin:2019xeg, Abreu:2019rpt, Caron-Huot:2020vlo}.

\subsection{Comments on the Heterotic  and Type I strings}

Having correctly reproduced the two-loop integrand in maximal supergravity, one would like to also match the two-loop integrand in $\cN=4$ super-Yang-Mills theory, which is closely related to the supergravity amplitude by the double-copy prescription \cite{Carrasco:2011mn}. One possible strategy is to extract the field theory limit of the scattering amplitude of five gauge bosons in the Heterotic strings, but this would produce the integrand for half-maximal supergravity, where both vector multiplets and the gravitational multiplet propagate in the loops. While the four-point two-loop amplitude in  half-maximal supergravity is known \cite{Bern:2013qca}, this is not the case to our knowledge for the five-point amplitude. Moreover, extracting the field theory limit of Heterotic string amplitudes is  bound to be subtle, as contributions from the separating degeneration  due to the pole of $1/\Psi_{10}$ (where $\Psi_{10}$ is the genus-two Igusa cusp form of weight 10)  are known to  contribute at four points \cite{Bossard:2018rlt}, and are expected for five points as well. 

\sm

A more direct approach is to consider the oriented, open-string sector of Type I superstrings, which precisely reduces to SYM theory at low energy, without contamination from gravitational exchange.
For open superstrings, scattering amplitudes of massless gauge bosons are given by an 
integral over the moduli space of Riemann surfaces with boundaries, over the positions $z_i$ of the vertex operators along the boundaries \cite{Alvarez:1982zi}, and over loop momenta. 
Riemann surfaces with boundaries
are constructed as a quotient of a closed Riemann surface under an anti-holomorphic 
involution \cite{Bianchi:1989du}. 
As a result, the period matrix is purely imaginary, and can be parametrized by \eqref{YL123} 
for a genus-two Riemann surface with three boundaries. 
The integrand is given by the product  ${\cal K}_{(5)} \cI_{(5)} \cC_{(5)}$
where $\cC_{(5)}$ is the Chan-Paton factor, which depends only on the color indices of the 
external particles. For a five-point amplitude with gauge group $SU(N_c)$, possible choices of $\cC_{(5)}$
include a single-trace $N_c^2 \Tr( T^{a_1} T^{a_2} T^{a_3} T^{a_4} T^{a_5})$ if all 5 external particles
are attached to the same boundary and a double-trace $N_c \Tr( T^{a_1} T^{a_2} T^{a_3}) \Tr (T^{a_4} T^{a_5})$ 
if three particles are attached on one boundary and two on another (recall
that $\Tr(T^a)=0$ for a simple gauge group; the overall factors of $N_c^2$ and $N_c$ arise from $\Tr(1)$ on the boundaries which do not support any external particle). 

\sm 

At low energies, scattering amplitudes are again dominated by degenerate Riemann surfaces, with long tubes replaced by strips and closed-string vertices replaced by disks \footnote{The field theory limit of the  genus-two open-superstring partition function in a magnetic field was investigated in \cite{Magnea:2004ai, Magnea:2013lna, Magnea:2015fsa} using the Schottky representation, reproducing the Feynman diagrams contributing to the Euler-Heisenberg Lagrangian of pure Yang-Mills theory. Our interest is in scattering amplitudes in SYM theory in Minkowski background.}.
At two-loop, five points, they  can be represented by fattened versions of the graphs in Figure \ref{fig:1}, where the fattening keeps track of the position of the vertex operators. For the pentabox diagrams $(a,b,c)$, the same 
computations as in subsection \ref{sec:62} apply, and reproduce the field theory integrands in 
color-kinematics dual form. Double-box diagrams, however, arise in a different fashion than for closed
strings, since the rules \eqref{dbox.1}, \eqref{dbox.2} for contact diagrams no longer apply. Instead, 
kinematic poles only arise from prime forms involving 
pairs of neighbouring punctures on the same boundary,
\be
\partial_{z_i} \ln E(z_i,z_{i\pm 1})  \sim \frac{1}{z_i {-}z_{i\pm 1}} 
\rightarrow \mp \frac{   \delta(z_i,z_{i\pm 1}) }{s_{i(i\pm 1)}}
\label{altdbox}
\ee
Therefore, the coefficient of a single-trace Chan-Paton factor $\sim N_c^2 {\rm tr}(T^{a_1} T^{a_2} T^{a_3}T^{a_4}T^{a_5})$ exhibits kinematical  poles of the form $1/s_{12}$, $1/s_{23}$, $1/s_{34}$, $1/s_{45}$, $1/s_{51}$,
while a double-trace Chan-Paton factor $\sim N_c {\rm tr}(T^{a_1} T^{a_2} T^{a_3}) 
 {\rm tr}(T^{a_4}T^{a_5})$ is accompanied by poles of the form $1/s_{12}, 1/s_{23}, 1/s_{31}$. The numerators can be extracted in the same way as before, and turn out to match with the prescription
 of \cite{Carrasco:2011mn}, after converting color-ordered traces into the color factors associated to the cubic graphs in Figure \ref{fig:1}. All cubic graphs are accessible from
 the partial amplitudes $\sim N_c^2 {\rm tr}(T^{a_1} T^{a_2} T^{a_3}T^{a_4}T^{a_5})$ and 
 $\sim N_c {\rm tr}(T^{a_1} T^{a_2} T^{a_3}) {\rm tr}(T^{a_4}T^{a_5})$ since the
$N_c^{-2}$-suppressed single-trace contribution $\sim \Tr( T^{a_1} T^{a_2} T^{a_3} T^{a_4} T^{a_5})$ 
is expressible in terms of permutations of the former \cite{Edison:2011ta} 
(see \cite{Bern:1997nh} for the $N_c^{-2}$-suppressed four-point single-trace amplitude).

\newpage

\section{Conclusion and future directions}
\setcounter{equation}{0}
\label{sec:7}

In this work, we have proposed a spacetime supersymmetric expression for the chiral 
two-loop five-point amplitude relevant to massless states of Type II, Heterotic, and Type I superstring theories.
The construction of the chiral amplitude is driven by the BRST cohomology of vertex operators in
the pure spinor formalism and the constraints from homology invariance in the chiral splitting procedure.
The main result in (\ref{simp.7}) and (\ref{simp.8}) is written in pure spinor superspace and therefore
allows to address arbitrary combinations of massless external states in the gauge and gravity
supermultiplets.

\sm

The key result of this work is to obtain the full $\alpha'$ dependence of the two-loop five-point amplitudes, including the contributions to the correlators beyond the OPE analysis and the low energy limit of Type~I and Type~II amplitudes in \cite{Gomez:2015uha}. In doing so we provide the starting point for a systematic study of the low energy expansion of Type~II string amplitudes beyond leading order, and comparison with predictions from string dualities, which will be the subject of  a companion paper \cite{compone}.  Our result will be further validated by a  derivation from first principles in the RNS formalism of the  chiral amplitude for external NS bosons and even spin structure to be given in another companion paper \cite{comptwo}.  
 
\sm 

We have also extracted the loop integrands for two-loop five-point amplitudes of super-Yang--Mills 
and maximal supergravity in $D\leq 10$ dimensions: The worldline limit of the string amplitudes
in this work reproduce the representation of the field theory amplitudes proposed in \cite{Mafra:2015mja}.
This form of the super-Yang--Mills and supergravity amplitudes features the color-kinematics duality and 
double-copy structure  \cite{Bern:2008qj, Bern:2010ue, Bern:2019prr}. Therefore, our work is yet another showcase that hidden relations between gauge and gravity amplitudes  may be conveniently studied from a string-theory perspective.

\sm

Our methods should be useful to determine and organize chiral two-loop amplitudes for higher numbers
of massless states. The explicit construction of the kinematic factors will require further cohomology
studies in pure spinor superspace as for instance done at genus one \cite{Mafra:2014gsa, Mafra:2018nla}. 
The decomposition (\ref{simp.7}) of the chiral amplitude into a basis of differential forms 
is easily extended to higher multiplicity: At six points for instance, the problem reduces to 
constructing 14 sub-correlators along with the basis forms that are individually homology-invariant
functions of the punctures related by permutations of the external legs.

\sm

Given that the chiral correlators in (\ref{simp.8}) have no explicit $\alpha'$ dependence, our results 
may also be exported to the pure spinor incarnation of the ambi-twistor string \cite{Berkovits:2013xba, Adamo:2015hoa}, and
should pave the way towards obtaining five-point supergravity amplitudes from correlators on the bi-nodal sphere using the techniques of \cite{Geyer:2016wjx, Geyer:2018xwu}.

\newpage

\appendix

\section{Clifford-Dirac algebra and pure spinor identities}
\setcounter{equation}{0}
\label{sec:A}

Weyl spinors in the {\bf 16} and {\bf 16'} representations of the Lorentz group $SO(10)$ in ten-dimensional space-time  $\RR^{10}$  will be denoted with an upper and a lower index, respectively, such as $\xi^\a$ and $\chi_\a$ where $\a = 1, \cdots , 16$.  The Clifford-Dirac matrices $(\gamma ^m) _{\a \b}$ and $(\gamma ^ m)^{ \a \b}$ acting on Weyl spinors in the {\bf  16} and {\bf 16'} respectively  satisfy the Clifford algebra,
\bea
(\gamma ^m)_{\a \b} \, (\gamma ^n)^{ \beta \gamma} + (\gamma ^n) _{\a \b} \,  (\gamma ^m)^{ \beta \gamma} = 2 \eta ^{mn} \delta _\a {} ^\gamma
\eea 
where $\eta^{mn}$ is the flat Minkowski metric on $\RR^{10}$ and $m,n=1,\cdots, 10$. The summation convention over pairs of repeated upper and lower vectorial or spinorial indices is adopted throughout. We shall often be led to complexifying the momenta and polarization data of the fields, in which case space-time is $\CC^{10}$, the Lorentz group is  $SO(10;\CC)$, and the metric $\eta^{mn}$ is the Kronecker $\delta^{mn}$, and all formulas in this section continue to hold as stated.

\subsection{Basic identities}

The anti-symmetric tensor $\gamma$-matrices are defined by,
\bea
(\gamma ^{mn})_\a {} ^\b & = & 
{1 \over 2!} ( \gamma ^m)_{\a \gamma} (\gamma ^n)^{\gamma \beta} - \hbox{ 1 permutation of } m,n
\no \\
(\gamma ^{mnp})_{\a \b} & = & 
{1 \over 3!}  ( \gamma ^m)_{\a \gamma} (\gamma ^n)^{\gamma \delta} (\gamma ^p)_{ \delta \beta}  \pm \hbox{ 5 permutation of } m,n, p
\eea
and so on for $\gamma ^{mnpq}, \gamma ^{mnpqr}$, and similarly for the $\gamma$-matrices with reversed spinor indices such as $(\gamma ^{mn})^\a {}_\b$. We shall not need $\gamma$-matrices of rank 6 or higher which are related to $\gamma$-matrices of lower rank by Poincar\'e duality. The $\gamma$-matrices have the following  symmetry properties, 
\begin{align}
\label{sym}
(\gamma ^m )_{\a \b} & =  + (\gamma ^m )_{\b \a} 
& (\gamma ^{mn} )^\a {}_\b & = - (\gamma ^{mn})_\b {}^\a
\no \\
(\gamma ^{m n p} )_{\a \b} & =  - (\gamma ^{m n p} )_{\b \a} 
& (\gamma ^{mnpq} )^\a {}_\b & = +(\gamma ^{mnpq})_\b {}^\a
\no \\
(\gamma ^{m n  p q r} )_{\a \b} & =  + (\gamma ^{m n p q r} )_{\b \a}
\end{align}
satisfy the following product identities,
\bea
\label{gam2}
\gamma _{mn} \gamma _s & = & \gamma _{mns} + \gamma _m \eta _{ns} - \gamma _n \eta _{ms}
\no \\
\gamma _{mnp} \gamma _s & = & \gamma _{mnps} + \gamma _{mn} \eta _{ps} - \gamma _{mp} \eta _{ns} + \gamma _{np} \eta _{ms}
\no \\
\gamma _{mnpq} \gamma _s & = & \gamma _{mnpqs} + \gamma _{mnp} \eta _{qs} - \gamma _{mnq} \eta _{ps} + \gamma _{mpq} \eta _{ns} - \gamma _{npq} \eta _{ms}
\eea
as well as the following contraction identities,
\bea
\label{gam3}
\gamma ^m \gamma _{m\, n_1\,  \cdots \,  n_p } & = & (10-p)  \gamma _{n_1 \,  \cdots \, n_p }
\no \\
\gamma ^m \gamma _{n_1 \cdots n_p} \gamma _m & = &  (10-2p) (-)^p \gamma _{n_1 \cdots n_p}
\eea
As an immediate consequence for arbitrary commuting or anti-commuting spinors $\xi ^\a, \psi ^\a$, we have the following decomposition formulas, 
\bea
\label{Fierz3}
\xi ^\a \psi ^\b + \xi ^\b \psi ^\a & = & { 1 \over 8} \, (\xi \gamma _m \psi) (\gamma ^m )^{\a \b}
+ { 1 \over 16 \cdot 5!} \, (\xi \gamma _{mnpqr} \psi) (\gamma ^{mnpqr} )^{\a \b}
\no \\
\xi ^\a \psi ^\b - \xi ^\b \psi ^\a & = & { 1 \over 8 \cdot 3!} \, (\xi \gamma _{mnp} \psi) (\gamma ^{mnp} )^{\a \b}
\eea
For an arbitrary commuting Weyl spinor $\xi$, combining the first equation of (\ref{Fierz3}) with the second equation of (\ref{gam3}) we obtain,
\bea
\label{Fierz2}
(\gamma ^m \xi )_\a (\gamma _m \xi )_\b = - \half (\gamma _m)_{\a\b} (\xi \gamma ^m \xi)
\eea
Finally, we have the following Fierz identity,
\bea
\label{Fierz1}
8 \delta _\beta {} ^\gamma \delta _\a {}^\delta & = &
4 (\gamma ^m)_{\a \b} (\gamma _m) ^{\gamma \delta} -  (\gamma ^{mn} )_\a {}^\gamma 
(\gamma _{mn} )_\b {} ^\delta - 2 \delta _\a {}^\gamma \delta _\b {}^\delta
\eea
and the famous supersymmetry Fierz identity,
\bea
 \label{Fierz4}
0 & = & (\gamma ^m)_{\a \b} (\gamma _m) _{\gamma \delta} + 
(\gamma ^m)_{\b \gamma} (\gamma _m) _{\a \delta} + 
(\gamma ^m)_{\gamma \a} (\gamma _m) _{\b \delta} 
\eea

\subsection{Identities involving pure spinors}

A commuting pure Weyl spinor $\lambda$ is defined to satisfy (\ref{pure}), namely $(\lambda \gamma ^m \lambda)=0$.
Combining (\ref{pure}) with (\ref{Fierz2}) and with the last equation of (\ref{gam2}) respectively, we see that an arbitrary commuting pure spinor satisfies the following fundamental identities,
\bea
\label{Fierz5}
(\lambda \gamma ^m)_\a (\lambda \gamma _m )_\b & = & 0
\no \\
(\lambda \gamma _{mnpqr} \lambda) (\lambda \gamma ^m )_\a & = & 0
\eea
The tensor product of two identical pure Weyl  spinors has the following decomposition, 
\bea
\lambda ^\a \lambda ^\b & = & { 1 \over 32 \cdot 5!} \, (\lambda \gamma _{mnpqr} \lambda) (\gamma ^{mnpqr} )^{\a \b}
\eea
The following identity holds for the tensor product of three identical  pure Weyl spinors,
\bea
\label{Fierz6}
(\lambda \gamma _{[mnpqr} \lambda ) (\lambda \gamma _{s]} )_\a=0
\eea
where the anti-symmetrization bracket is applied to all six indices. The identity may be proven as follows. The symmetric tensor product of three arbitrary Weyl spinors in the {\bf 16} is reducible by contracting two of the Weyl spinors with a $\gamma$-matrix. However, this contraction vanishes for pure spinors by (\ref{pure}) and hence the symmetrized tensor product of three pure Weyl  spinors is irreducible. Its further tensor product with a {\bf 16} is readily shown not to contain an anti-symmetric rank 6 tensor, which is Poincar\'e dual to an anti-symmetric rank 4 tensor, which proves the identity.

\newpage

\section{Functions and differentials on Riemann surfaces}
\setcounter{equation}{0}
\label{sec:C}

In this appendix, we review the basic holomorphic and meromorphic functions, differentials, and Green functions on a compact Riemann surface $\Sigma$ of genus $h$ from which all string correlators needed here can be constructed. Standard references are \cite{DHoker:1988pdl,fay,Verlinde:1986kw}.

\subsection{Homology and modular transformations}

A canonical basis for the homology group $H_1(\Sigma, \ZZ)$ consists of 1-cycles $\mA_I$ and $\mB_I$ with  $I=1, \cdots, h$ and canonical intersection pairing $\mJ$,
\begin{align}
\label{1b1}
\mJ (\mA_I, \mA_J) & = \, \mJ (\mB_I, \mB_J)  \, = 0 
\no \\
 \mJ (\mA_I, \mB_J) & =  - \mJ (\mB_I, \mA_J)  =  \delta _{IJ}
\end{align}
Different canonical bases $(\mA_I, \mB_I)$ and $(\tilde \mA_I, \tilde \mB_I)$  are related by linear transformations represented by a matrix $M$  with integer entries, 
\bea
\label{1b2}
 \left ( \begin{matrix} \tilde \mB \cr \tilde \mA \cr \end{matrix} \right ) 
 = M \left ( \begin{matrix} \mB \cr \mA \cr \end{matrix} \right )
\eea
Here, $\mA$ and $\mB$ stand for the column matrices with entries $\mA_I$ and $ \mB_I$, respectively, and $M$ is an element of  the group $Sp(2h,\ZZ)$ of modular transformations, which preserve the canonical intersection matrix $\mJ$, 
\bea
\label{1b2}
 M^t \mJ M = \mJ 
 \hskip 0.7in 
\mJ = \left ( \begin{matrix}0 & -I_h \cr I_h & 0 \cr \end{matrix} \right )
 \hskip 0.7in
M = \left ( \begin{matrix} A & B \cr C & D \cr \end{matrix} \right )
\eea
 where $A,B,C,D$ are $h \times h$ matrices with integer entries. An important subgroup of $Sp(2h,\ZZ)$ is the group $Gl(h,\ZZ)$ which  consists of those modular transformations $M$ which transform $\mA$-cycles into linear combinations of $\mA$-cycles and $\mB$-cycles into linear combinations of $\mB$-cycles. It is obtained by setting $B=C=0$ and $D= (A^t)^{-1}$.

\subsection{Holomorphic 1-forms and the period matrix}

A canonical basis of the cohomology group $H^{(1,0)} (\Sigma, \ZZ)$ consists of holomorphic $(1,0)$-forms  $\om_I$ with $I=1,\cdots, h$ whose periods on the homology basis $(\mA_I, \mB_I)$ are given by,\footnote{For our conventions and notations for integrals of $(1,0)$ forms see footnote 2.}
\bea
\label{1b5}
\oint _{\mA_I} \om_J = \delta _{IJ} \hskip 1in \oint _{\mB_I} \om_J = \Omega _{IJ}
\eea
The $\mA$-periods fix the canonical normalization of $\om_I$, while the $\mB$-periods give the period matrix $\Omega $, which is symmetric by the Riemann bilinear relations, and for which the matrix,
\bea
\label{Y}
Y= \Im \, \Omega
\eea 
is positive definite. Under modular transformations $M \in Sp(2h,\ZZ)$, whose  parametrization in terms of $h \times h$ matrices $A,B,C,D$ is given in (\ref{1b2}), the matrix of holomorphic Abelian differentials $\om$, the period matrix $\Omega$, its imaginary part $Y$, and the determinant thereof  $\det Y$ transform as follows,
\bea
\tilde \om & = & \om (C \Omega+D)^{-1}
\no \\
\tilde \Omega & = & (A \Omega +B) (C \Omega +D)^{-1} 
\no \\
\tilde Y & = & (\Omega C^t+D^t)^{-1} Y (C\Omega^* +D)
\no \\
\det \tilde Y & = & |\det (C \Omega +D)|^2 \, \det Y
\eea

\subsection{The Abel map and Jacobi $\tet$-functions}

The Jacobian of the surface $\Sigma$ is the Abelian variety defined by,
\bea
J(\Sigma) = \CC^h / \{ \ZZ^h + \Omega \ZZ^h\}
\eea
Given a base point $z_0 \in \Sigma$, the Abel map sends a divisor $D$ of $n$ points $z_i \in \Sigma$ with weights $q_i \in \ZZ$ for $i=1,\cdots,n$, formally denoted by  $D=q_1 z_1 + \cdots  q_n z_n$, into $\CC^h$ by,
\bea
\label{abelmap}
q_1 z_1 + \cdots + q_n z_n 
\equiv 
\sum _{i=1} ^n q_i \int _{z_0} ^{z_i} (\omega _1, \cdots , \omega _h)
\eea
where the $h$-tuple $(\omega _1, \cdots , \omega _h)$ stands for the vector of holomorphic $(1,0)$-forms $\om_I$.
The Abel map into $\CC^h$ is multiple valued, but it is single valued as a map into $J(\Sigma)$. 

\sm

The Jacobi $\tet$-functions with characteristics $\kappa$ are defined on $\zeta = (\zeta _1, \cdots, \zeta _h)^t \in \CC^h$ by, 
\bea
\label{deftheta}
\tet [\kappa] (\zeta| \Omega) 
\equiv  
\sum _{n \in {\bf Z}^h } 
\exp \Big (i \pi (n + \kappa ') ^t \Omega (n+ \kappa ') + 2\pi i (n+\kappa ') ^t  (\zeta + \kappa '') \Big ) 
\eea
Here, $\kappa  = \left ( \kappa ' | \, \kappa '' \right )$ is a general characteristic, where $\kappa ', \ \kappa '' \in \mathbb C^h$ are both written as a column vector. Henceforth, we shall assume that $\kappa$ corresponds to a spin structure, and thus be valued in $\kappa ', \kappa '' \in ( \mathbb Z/2 \mathbb Z) ^h$. The parity of the spin structure is determined by the parity of the $\tet$-functions which satisfy,
\bea
\tet [\kappa ] (- \zeta | \Omega ) = (-1) ^{4 \kappa ' \cdot \kappa ''}
\tet [\kappa ](\zeta |\Omega)
\eea
According to whether $4\kappa ' \cdot \kappa ''$ is even or odd,  $\kappa$ is referred to as an {\sl even or odd spin structure}. Upon shifting by full periods $M, \ N \in \ZZ^h$,
\bea
\tet [\kappa] (\zeta + M + \Omega N| \Omega ) =
\exp \Big (-i \pi N^t \Omega N - 2 \pi i N^t (\zeta + \kappa ') + 2
\pi i M^t \kappa '' \Big ) \tet [\kappa] (\zeta | \Omega)
\quad
\eea
Under a modular transformation $M \in Sp(2h,\ZZ)$ as given in (\ref{1b2}), the characteristic $\kappa = (\kappa ' | \, \kappa'' )$ transforms as (see for example \cite{fay,igusa})
\bea
\left ( \begin{matrix} \tilde \kappa ' \cr \tilde \kappa '' \end{matrix} \right )
=
\left ( \begin{matrix} D & -C \cr -B & A \end{matrix}  \right )
\left ( \begin{matrix} \kappa ' \cr \kappa '' \end{matrix}  \right )
+ \half {\rm diag} \left ( \begin{matrix} C\, D^t \cr A\, B^t \end{matrix}  \right ) 
\eea
The $\tet$-function transforms as follows, 
\bea
\tet [\tilde \kappa ] \Big ((\Omega C^t +D^t)^{-1}  \zeta \big | (A \Omega +B)(C\Omega +D)^{-1} \Big ) =
\epsilon (\kappa, M) \Big ( \det (C \Omega + D) \Big ) ^{\half} \tet [\kappa ](\zeta |\Omega)
\qquad
\eea
where $\epsilon(\kappa,M)$ is an eighth root of unity satisfying $\epsilon^8=1$. Its explicit form is given in \cite{fay,igusa} but will not be needed here.

\subsection{The prime form \label{sec_prime}}

The prime form is constructed as follows \cite{fay}. For any odd spin structure $\nu$, the $2h-2$ zeros of the holomorphic $(1,0)$-form, 
\bea
\label{hnudef}
h_\nu^2 (z) = \sum _I \p^I \tet [\nu ](0|\Omega) \omega _I(z)  \hskip 1in \p^I = { \p \over \p \zeta _I}
\eea
 are double and the form admits a unique (up to an overall sign) square root $h_\nu (z)$ which is a holomorphic $(1/2,0)$ form. The prime form is a $(-1/2,0)$ form in $z,w$, living in the covering space of $\Sigma$, defined by
\bea
\label{Edef}
E(z,w|\Omega) = {\tet [\nu ] (z-w| \Omega) \over h_\nu (z) \, h_\nu (w)}
\eea
where the argument $z-w$ of the $\tet$-functions stands for the Abel map of (\ref{abelmap}) with $z_1=z$, $z_2=w$ and $q_1=-q_2=1$. The form $E(z,w|\Omega)$ defined in (\ref{Edef})  is independent of $\nu$, holomorphic in $z$ and $w$, odd under swapping $z$ and $ w$,  and has a unique simple zero at $z=w$. It is single valued when $z$ is moved around $\mA_I$ cycles, but has non-trivial monodromy around a $\mB_I$ cycle,
\bea
E(z+\mB_I ,w|\Omega) = - \exp \biggl ( -i \pi \Omega _{II} - 2 \pi i \int ^z _w \! \omega _I \biggr ) E(z,w|\Omega) 
\eea
In terms of the first derivatives, we have,
\bea
\p_z \ln E(z+\mB_I,w) & = & \p _z \ln E(z,w) - 2 \pi i \om_I(z)
\no \\
\p_z \ln E(z,w+\mB_I) & = & \p _z \ln E(z,w) + 2 \pi i \om_I(z)
\label{mondE}
\eea
The combination $\p_z \p_w \ln E(z,w|\Omega)$ is a single valued meromorphic differential with one double pole at
$z=w$ and no single poles. Its integrals around homology cycles are given by,
\bea
\label{prf}
\oint _{\mA_I} \! dz \p_z \p_w \ln E(z,w|\Omega) & = & 0
\nonumber \\
\oint _{\mB_I} \! dz \p_z \p_w \ln E(z,w|\Omega) & = & 2 \pi i \omega _I(w)
\eea
and will be of use throughout.

\newpage

\section{Chiral splitting and loop momenta}
\setcounter{equation}{0}
\label{sec:D}

In this appendix, we review chiral splitting for the $x^m$-field in 10-dimensional space-time on a compact Riemann surface of arbitrary genus $h$. The functional integrals of interest may be obtained through a generating functional which includes both the contributions from the Koba-Nielsen factor and from multi-linear insertions of the current $\p x_m$ required in the vertex operators, and is given by (\ref{genM}).   

\sm

The worldsheet field contents of the pure spinor string has been arranged so that their combined Weyl and holomorphic anomalies cancel. Omitting the contribution to these anomalies from the $x$-field by itself, its Gaussian functional integral evaluates to, 
\bea
\cJ =  (2 \pi )^{10} \, \delta (k) \, { |Z |^{-20} \over (\det 2 Y)^5}  \, \exp   \Big \{ \sum _{i,j=1}^N \cE_{ij} \Big \} 
\hskip 1in k = \sum _{i=1}^N k_i
\eea
Here, the determinant is taken of the matrix $Y$ with components $Y_{IJ} = \Im \Omega _{IJ}$,  while $Z$ is the chiral scalar partition function which is holomorphic in moduli, and  $\cE_{ij}$ is given by,
\bea
\cE_{ij} & = & 
- \half k_i \cdot k_j \, G(z_i,z_j) + i k_i \cdot  \ep _j \,  \p_{z_j} G(z_i,z_j) 
+   i k_i \cdot  \bar \eta _j \,   \p_{\bar z_j} G(z_i,z_j) \no \\ &&
+ \half  \ep  _i  \cdot \ep  _j \, \p_{z_i} \p_{z_j} G(z_i,z_j)
+ \half \bar \eta _i \cdot \bar \eta _j \, \p_{\bar z_i} \p_{\bar z_j} G(z_i,z_j)
+  \bar \eta _i \cdot  \ep _j \, \p_{\bar z_i} \p_{z_j} G(z_i,z_j)
\eea
The Green function $G$ is given in (\ref{Green}), but may equivalently be replaced by the Arakelov Green function of (\ref{GreenA}). We split $\cE_{ij}$ into a part which involves only the holomorphic prime form $E(z_i,z_j)$, another part which involves its complex conjugate, and a part which involves the holomorphic Abelian differentials and $Y^{IJ}$,
\bea
\cE_{ij} = \cE_{ij} ^+ + \cE_{ij} ^- + \cE_{ij} ^0
\eea
The individual contributions are given as follows, 
\bea
\cE_{ij} ^+ & = &  \half k_i \cdot k_j \, \ln E(z_i,z_j) - i k_i \cdot  \ep _j  \,  \p_{z_j} \ln E(z_i,z_j) 
- \half  \ep _i  \cdot \ep _j \p_{z_i} \, \p_{z_j} \ln E(z_i,z_j)
\no \\
\cE_{ij} ^- & = &  \half k_i \cdot k_j \, \ln \overline{E(z_i,z_j)} - i k_i \cdot \bar \eta _j \,  \p_{\bar z_j} \ln \overline{E(z_i,z_j)} 
- \half  \bar \eta _i \cdot \bar \eta _j \, \p_{\bar z_i} \p_{\bar z_j} \ln \overline{E(z_i,z_j)} 
\eea
and the sum of $\cE_{ij}^0$ is given by,
\bea
\sum _{i,j=1}^N \cE_{ij}^0 & = &    { \pi \over 2}  Y^{IJ} \left ( \zeta _I - \tilde \zeta _I \right ) \cdot
\left ( \zeta _J - \tilde \zeta  _J \right )
\eea
where we have defined,
\bea
\zeta ^m _I & = & 
\sum_{j=1}^N \Big (  \ep ^m _j \om _I(z_j) + i k^m _j \, \int ^{z_j} _{z_0} \om _I  \Big )
\no \\ 
\tilde \zeta ^m _I & = & 
\sum_{j=1}^N \Big ( \bar \eta ^m _j \, \overline{\om} _I(z_j) +  i k^m _j \, \int ^{z_j} _{z_0} \overline{\om} _I  \Big )
\eea
Next, we shall represent the combination of the $(\det Y)$-denominator and the exponential of the sum of $\cE^0_{ij}$ by an integral over loop momenta $p^m _I \in \RR$,
\bea
{ \exp  \left \{ \sum _{i,j} \cE_{ij}^0 \right \} \over (\det 2Y)^5}  
= \int _{\RR^{10h}}  dp \, \exp  \left\{ - 2 \pi  Y _{IJ} \, p ^I \cdot p ^J 
+  2 \pi  p^I \cdot  (\zeta _I  -  \tilde \zeta _I) \right \}
\eea
The full generating function is then given as follows, 
\bea
\label{genJ}
\cJ =  \delta  (k) \int_{\RR^{10h}}  dp \,  \cB (z_i,  \ep_i, k_i, p^I |\Omega ) 
\,  \overline{\cB (z_i,\eta_i ,  - k_i^*, -p^I |\Omega )} 
\eea
where the chiral amplitude is given by, 
\bea
\cB (z_i, \ep_i, k_i,  p^I |\Omega ) & = &  Z ^{-10}  \exp \Bigg \{ 
 i \pi  \Omega _{IJ} \, p ^I \cdot p ^J 
+ \sum _{i}  2 \pi  p ^I \cdot  \Big ( \ep _i \, \om_I (z_i) + i k_i   \int ^{z_i} _{z_0} \om _I   \Big ) 
\no \\ && \hskip 0.7in
- \half  \sum _{i \not=  j} \Big ( i k_i + \ep _i \p_{z_i} \Big )  \Big ( i k_j + \ep _j \p_{z_j} \Big )
\ln E(z_i,z_j)  \Bigg \} 
\eea
and similarly for its conjugate chiral amplitude.
The chiral amplitude may be recast in the form of a chiral correlator, 
\bea
\cB (z_i, \ep_i, k_i,  p^I |\Omega ) & = &  
Z ^{-10} \exp \left \{   i \pi \Omega _{IJ} p ^I  \cdot p ^J 
+  \sum _{i}  2 \pi i p^I \cdot k_i  \int ^{z_i} _{z_0} \om _I   \right \}
\\ && \times 
\left \< \exp \sum _i \left \{  \ep _i \cdot \Big ( \p_z x_+ (z_i) +    2 \pi  p^I \om_I(z_i) \Big )   
+ i k _i \cdot  x_+(z_i) \right  \} \right \>
\no \eea
The effective rule for the Wick contraction of the chiral bosonic field $x_+$ is given by (\ref{wick}).
We have grouped together the various terms involving the polarization vectors, which make it clear that the effective rule for the insertion of the derivatives in the formulation with loop momenta is given by the following substitution,
\bea
\p x^m (z) \quad \longrightarrow \quad \p x_+ ^m (z) +   2 \pi  (p^I)^m \om _I (z)
\eea
It is this effective rule of which we shall make use here when applying chiral splitting.

\newpage

\section{Operator product expansions}
\setcounter{equation}{0}
\label{sec:B}

The short-distance behavior of the physical canonical fields is given by the following OPEs,
\bea
\label{D1}
x^m(z) x^n (y) & \sim & - \eta ^{mn} \ln (z-y)
\no \\
p_\a (z) \theta ^\b (y) & \sim & { \delta _\a {}^\b \over z-y}
\eea
As a result, the OPEs of the composite matter fields $d_\a, \Pi^m$ defined in (\ref{dpi}) may be deduced from the OPEs of the physical canonical fields, 
\begin{align}
\label{D2}
d_\a (z) \, f \big (x(y), \theta (y) \big ) & \sim  { D_\a f \over z-y}
&
d _\a (z) \, d_\b (y) & \sim  - { \gamma ^m _{\a \b} \, \Pi_m \over z-y}
\no \\
\Pi _m (z) \, f \big (x(y), \theta (y) \big ) & \sim  - { \p_m f \over z-y}
&
d_\a (z) \, \Pi ^m (y) & \sim  { \gamma ^m _{\a \b} \, \p \theta ^\b \over z-y}
\end{align}
where $D_\a$ is the superspace derivative defined in (\ref{superD}), 
from which the BRST transformations of the matter fields in (\ref{BRST}) may be evaluated.
The OPEs of the ghost fields are given by, 
\bea
w_\a (z) \, \lambda ^\beta (y) & \sim & 
{ \delta _\a {}^\b + (\gamma ^m \lambda )_\a \Lambda _m^\beta   \over z-y} 
\no \\
\bar w ^\a (z) \, \bar \lambda _\b (y) & \sim & 
{ \delta ^\a {} _\b + (\gamma _m \bar \lambda )^\a \bar \Lambda^m_{ \beta}  
-  (\gamma _m r)^\a \phi^m_{\beta} \over z-y} 
\no \\
s^\a (z) \, r_\b (y) & \sim & { \delta ^\a {}_\b + (\gamma_m \bar \lambda )^\a \psi^m_{\b} \over z-y}
\eea
 The presence of the functions $\Lambda _m^\beta, \bar \Lambda^m_{\b}, \phi^m_{\beta} , \psi ^m_\b$ is required in order for the OPEs to be compatible with the pure spinor constraints (\ref{pure}), and specifically to cancel the singularities in the OPE of the fields $w_\a, \bar w_\a, s^\a$ with the pure spinor constraints of (\ref{pure}).  To do so, $\Lambda _m ^\beta$ and $\psi_m ^\beta$ must satisfy,
\bea
\label{Lambeta}
(\gamma ^n \lambda) _\a (\Lambda _n \gamma ^m \lambda ) + ( \gamma ^m \lambda )_\a & = & 0
\no \\
(\gamma ^n \bar \lambda) ^\a (\psi _n \gamma ^m \bar \lambda ) + ( \gamma ^m \bar \lambda )^\a & = & 0
\eea  
while $\bar \Lambda^m_{\beta}$ and $\phi^m_{\beta}$ must satisfy the following set of coupled equations, 
\bea
(\gamma _m \bar \lambda )^\a + (\gamma ^n  \bar \lambda )^\a (\bar \Lambda _n \gamma _m \bar \lambda ) - (\gamma^n r)^\a (\phi _n \gamma _m \bar \lambda) & = & 0
\no \\
(\gamma ^n r )^\a + (\gamma ^p  \bar \lambda )^\a (\bar \Lambda _p \gamma ^n r ) - (\gamma^p r)^\a (\phi_p \gamma ^n r) & = & 0
\eea
Note that the functions $\Lambda _m ^\b, \bar \Lambda^m_{\beta},\psi^m_{\beta}$ are commuting, while $\phi^m_{ \beta}$ is anti-commuting.  The solutions  to these equations are not unique as there are non-trivial kernels. For example,  we cannot solve them simply by setting $(\Lambda _n \gamma ^m \lambda )= - \delta _n{}^m$ since this would be inconsistent with the constraint $\lambda \gamma ^m \lambda =0$. Similarly for the other equations and their solutions.

\sm

The contributions from $\Lambda _m, \bar \Lambda _m , \phi_m, \psi^m_\b$ will cancel out of the OPEs of the composites $N_{mn}, J, T_\lambda$, and their analogues for the ghosts $\bar w^\a$ and $s^\a$. Their  OPEs with  $\lambda ^\a$ are given by the corresponding linear transformations on $\lambda^\a$,
\bea
N_{mn} (z) \, \lambda ^\a (y) & \sim & \half { (\gamma _{mn} \lambda )^\a \over z-y}
\no \\
J_\lambda (z) \, \lambda ^\a (y) & \sim & { \lambda ^\a \over z-y}
\no \\
T_\lambda (z) \, \lambda ^\a (y) & \sim & { \p \lambda ^\a \over z-y}
\eea
while their OPEs with $w_\a$ are subject to extra terms due to the constraints (\ref{pure}) and will not be needed here.
The OPEs of the currents are more complicated because of the constraints, and we quote here only the relevant results, 
\bea
N_{mn} (z) \, N_{pq}(y) & \sim &
{ \eta _{np} N_{mq} - \eta _{mp} N_{nq} - \eta _{nq} N_{mp} + \eta_{mq} N_{np} \over z-y}
\no \\ &&
  -3 \, { \eta _{mq} \eta _{np} - \eta _{mp} \eta _{nq} \over (z-y)^2}
\eea

\end{document}